\documentclass{hch}

\usepackage{graphicx}
\usepackage{amssymb}

\catcode`\ä = \active
\catcode`\ö = \active
\catcode`\ü = \active
\catcode`\Ä = \active
\catcode`\Ö = \active
\catcode`\Ü = \active
\catcode`\ß = \active
\catcode`\é = \active
\catcode`\è = \active
\catcode`\ë = \active
\catcode`\ô = \active
\catcode`\ê = \active
\catcode`\ø = \active
\defä{\"a}
\defö{\"o}
\defü{\"u}
\defÄ{\"A}
\defÖ{\"O}
\defÜ{\"U}
\defß{\ss}
\defé{\'{e}}
\defè{\`{e}}
\defë{\"{e}}
\defô{\^{o}}
\defê{\^{e}}
\defø{\o}

\newcommand{\subrm}[1]{\ensuremath{_{\mathrm{#1}}}}
\newcommand{\mst}[1]{ \ensuremath{|m\subrm{F} = #1\rangle}  }
\newcommand{\fmst}[2]{|F = #1, m_F = #2\rangle}

\newcommand{\vr}{\vec{r} }
\newcommand{\afz}{ \langle F_z \rangle }
\newcommand{\lastate}[3]{#1, \ldots #2 \ldots; #3}
\newcommand{\sqw}{S(\vec{q}, \omega)}
\newcommand{\sq}{S(\vec{q})}
\newcommand{\twovec}[2]{ \left( \begin{array}{c} #1 \\ #2\end{array} \right) }
\newcommand{\bfo}{\hat{\Psi}} 
\newcommand{\bfod}{\bfo^\dagger} 

\begin{document}

\title{Spinor condensates and light scattering from Bose-Einstein condensates}

\author{Wolfgang Ketterle}

\address{Department of Physics and Research Laboratory of Electronics, Massachusetts
Institute of Technology, Cambridge, MA  02139, USA\\E-mail: ketterle@mit.edu}

\authorsup{Dan M. Stamper-Kurn\inst[Norman Bridge
Laboratory of Physics, California Institute of Technology, Pasadena, CA  91125, USA]{1}, Wolfgang
Ketterle\inst[Department of Physics and Research Laboratory of Electronics, Massachusetts Institute
of Technology, Cambridge, MA  02139, USA]{2}}

\runningtitle{Spinor condensates and light scattering}

\maketitle

\begin{abstract}
These notes discuss two aspects of the physics of atomic
Bose-Einstein condensates:  optical properties and spinor
condensates.  The first topic includes light scattering
experiments which probe the excitations of a condensate in both
the free-particle and phonon regime.  At higher light intensity, a
new form of superradiance and phase-coherent matter wave
amplification were observed. We also discuss properties of spinor
condensates and describe studies of ground--state spin domain
structures and dynamical studies which revealed metastable excited
states and quantum tunneling.
\end{abstract}

\section{Introduction}

The possibility of creating optical fields with many photons in a
single mode of a resonator was realized with the creation of the
laser in 1960.  The possibility of creating a matter-wave field
with many atoms in a single mode of an atom trap (which is the
atomic equivalent of an optical resonator) was realized with the
achievement of Bose-Einstein condensation (BEC) in 1995. Because
of the wealth of new phenomena which the condensates display, and
the precision and flexibility with which they can be manipulated,
interest in them has grown explosively in the communities of
atomic physics, quantum optics, and many-body physics.  At least
twenty groups have created condensates, and the publication rate
on Bose-Einstein condensation has soared following the discovery
of the gaseous condensates.

Although atomic condensates and laser light share many properties, they also differ fundamentally:
atoms interact readily, while photons do not.  As a result, the atomic condensates constitute a
novel class of many-body systems that provide a new laboratory for many-body physics.  They have
already yielded discoveries such as stability and collapse of condensates with attractive
interactions, multi-component condensates, Feshbach resonances and novel optical properties, and
have led to advances in many-body theory. Furthermore, because atoms interact, atom optics is
inherently non-linear optics. Consequently, nonlinear effects such as four-wave mixing that were
first achieved with light only with difficulty, occur almost automatically with coherent matter
waves.

These lecture notes will focus on two aspects of Bose-Einstein condensation: light scattering from
a Bose-Einstein condensate and spinor condensates. Our lectures at Les Houches covered a broader
range of topics, including trapping techniques, methods to probe the condensate and studies of
sound and condensate formation. For those topics we refer to our Varenna Summer School Notes which
give a comprehensive discussion of experimental techniques, static and dynamic properties,
coherence and optical trapping of condensates \cite{kett99var}.  This paper and some other recent
review papers summarize the state of the field \cite{ingu99,dalf99rmp,park98}.

Research on gaseous BEC can be divided into two areas:  In the
first,  which could be labeled ``the atomic condensate as a
coherent gas'' or ``atom lasers,'' one would like to have as
little interaction as possible between atoms --- almost like
photons in an optical laser. Thus the experiments are
preferentially done at low densities. The Bose-Einstein condensate
serves as an intense source of ultracold coherent atoms for
experiments in atom optics, in precision studies or for
explorations of basic aspects of quantum mechanics. The second
area could be labeled as ``BEC as a new quantum fluid'' or ``BEC
as a many-body system.''  The focus here is on the interactions
between the atoms which are most pronounced at high densities.

The topics covered in these notes illustrate both aspects of BEC.
Spinor condensates realize a new class of quantum fluids. Coherent
matter wave amplification is at the heart of atom lasers. Our
studies of light scattering from a Bose condensate link both
aspects together: light scattering was used to imprint phonons
into the condensate, but also to measure the coherence of an atom
laser and to realize a matter wave amplifier.

Chapters 2 and 4 of this review are based on the thesis of one of the authors \cite{stam99thesis}.
An abbreviated version of chapter 2 will appear in Ref.\ \cite{stamp00mbx}.

\section{Optical properties of a Bose-Einstein condensate}

What does a trapped Bose-Einstein condensate look like?  More precisely, how does it interact with
light, and does this differ fundamentally from what one would naively expect from a similar
collection of very cold atoms?  In the early 1990s, before Bose-Einstein condensation was realized
in atomic gases, there were lively debates about how a condensate could be observed. Some
researchers thought it would absorb all light and would therefore be ``pitch black,'' some
predicted it would be ``transparent'' (due to superradiant line-broadening \cite{java94}), others
predicted that it would reflect light due to polaritons \cite{svis90,poli91} and be ``shiny'' like
a mirror.

All the observations of Bose condensates have employed scattering
or absorption of laser light. These observations were either done
on ballistically expanding dilute clouds or with
far--off--resonant light. Under those circumstances, a Bose
condensate scatters light as ordinary atoms do. On resonance, the
condensate strongly absorbs the light, giving rise to the
well-known ``shadow pictures'' of expanding condensates where the
condensate appears black.  For off-resonant light, the absorption
can be made negligibly small, and the condensate acts as a
dispersive medium bending the light like a glass sphere. This
regime has been used for non-destructive in-situ imaging of
Bose-Einstein condensates.

\begin{figure}
    \begin{center}
    \includegraphics[height=2.8 in]{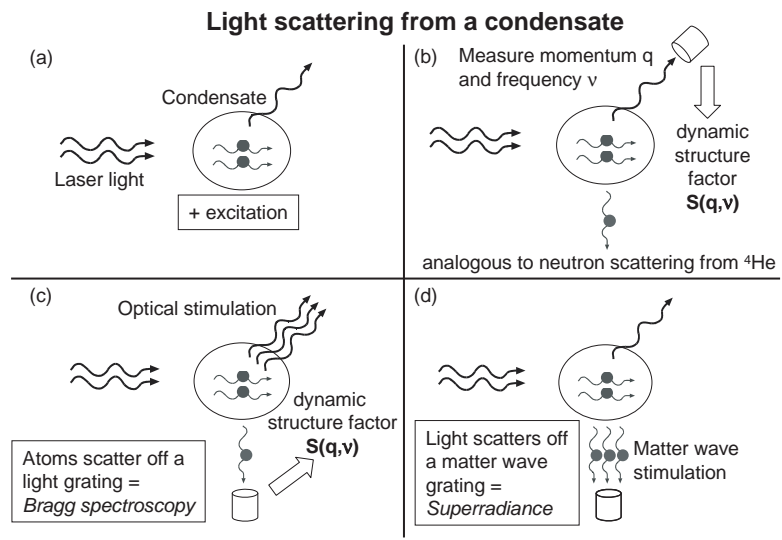} \caption[Light scattering from a
    Bose-Einstein condensate]{
Light scattering from a Bose-Einstein condensate.  When a photon is scattered, it transfers
momentum to the condensate and creates an excitation (upper left). Therefore, an analysis of the
scattered light allows the determination of the dynamic structure factor, in close analogy to
neutron scattering experiments with superfluid helium (upper right).  For sufficiently large
momentum transfer the excitation leads to an atom scattered out of the condensate.  The signal is
greatly increased by stimulating the light scattering by a second laser beam and detecting the
scattered atoms (lower left)---this is the scheme for Bragg spectroscopy. Light scattering can also
be stimulated by adding a coherent atomic field (lower right). This led to superradiant scattering
of light and atoms.
    \label{light_scatt}}
    \end{center}
\end{figure}

Our group has recently looked more closely at how coherent, weakly--interacting atoms interact with
coherent light.  Light scattering imparts momentum to the condensate and creates an excitation in a
many--body system (Fig.\ \ref{light_scatt}). Consequently, the collective nature of excitations and
the coherence of the condensate can affect its optical properties. Thus light scattering can be
used to illuminate properties of the condensate.

\subsection{Light scattering from a Bose-Einstein condensate}

\subsubsection{Elastic and inelastic light scattering}
\label{sec:scattering_basics}

Let us begin by considering the effect on a single atom of a single light scattering event. The
initial state of the atom--light system is $|N_{k}, \ldots 0_{l} \ldots ; i \rangle$ where $N_{k}$
photons are in an incident beam with wavevector $k$, no photons are in other photon modes (such as
mode $l$), and the atom is in state $|i\rangle$ which can be either a trapped or untrapped state.
After adiabatically eliminating the excited atomic state, the coupling between the atom and the
light is described by the operator
\begin{equation}\label{eq:laint}
  \mathcal{H^\prime} = C \sum_{k,l,m,n} \hat{c}^\dagger_{l}
  \hat{a}^\dagger_n \hat{c}_k \hat{a}_m \delta_{l+n-k-m}
\end{equation}
Here $\hat{c}_k$ ($\hat{c}^\dagger_k$) is the destruction (creation) operator for optical waves,
and $\hat{a}_k$ ($\hat{a}^\dagger_k$) is the destruction (creation) operator for atomic waves of
wavevector $k$.  The strength of atomic resonances and the detuning of the light determines the
strength of the coupling, summarized in the coefficient $C$.

Due to the coupling, light is scattered from the incident beam to wavevector $l = k - q$ with a
cross--section proportional to\footnote{Our discussion is limited to low intensity of the probe
light, allowing us to neglect the Mollow triplet fluorescence spectrum observed at high
intensities.}
\begin{eqnarray}\label{eq:scatxsection}
  \frac{d \sigma}{d \Omega} & \propto & |C|^2 \times \\
  & & \left( \left| \langle
  \lastate{N_k - 1}{1_l}{i} | \sum_m \hat{c}^\dagger_{k-q}
  \hat{a}^\dagger_{m+q} \hat{c}_k \hat{a}_m | {\lastate{N_k}{0_l}{i}} \rangle
  \right|^2 \right. \nonumber \\
  & & \left. + \sum_{j \neq i} \left| \langle
  \lastate{N_k - 1}{1_l}{j} | \sum_m \hat{c}^\dagger_{k-q}
  \hat{a}^\dagger_{m+q} \hat{c}_k \hat{a}_m | {\lastate{N_k}{0_l}{i}} \rangle
  \right|^2 \right) \nonumber
\end{eqnarray}

Two contributions to the light scattering are separated in the above expression. The first part
describes ``coherent'' light scattering, i.e.\ diffraction and refraction, in which the atomic
system is left in its original state, and the scattered light has the same frequency as the
incident light. The coupling shifts the phase of the elastically scattered light by an amount
proportional to $\langle \lastate{N_k - 1}{1_l}{i} | \mathcal{H}^\prime | {\lastate{N_k}{0_l}{i}}
\rangle$. This phase shift can be used to spatially image an atomic cloud by a dispersive imaging
technique, such as dark--ground or phase--contrast imaging.  For a dilute cloud of size $d$, the
coherent scattering is limited to the diffraction angle $\lambda /d$.  When the cloud is much
larger than an optical wavelength $\lambda$, only small--angle forward scattering is coherent.

The second part of the scattering cross--section describes ``incoherent'' light scattering in which
the state of the atom is changed.  For the case considered here, $d \gg \lambda$, this occurs when
light  is scattered outside the diffraction angle. Incoherent, or inelastic light scattering is
used for absorption imaging where the light which is transmitted by the cloud is collected, and the
amount of probe light scattered out of the imaging system is determined.

One can learn more from inelastic scattering than just by observing absorption, i.e.\ counting the
total sum of scattered photons. An inelastically scattered photon is shifted in frequency from the
incident photon. Further, the outgoing angle of the scattered photon determines the momentum $\hbar
\vec{q}$ which is imparted to the sample.  Thus, a spectroscopic analysis of inelastically
scattered photons at a given angle from the incident light beam determines the response of the
atomic sample to a given energy and momentum transfer (Fig.\ \ref{light_scatt}).

Analyzing photons scattered by a Bose--Einstein condensate from a single beam would be a difficult
task. Alkali Bose--Einstein condensates are currently produced with $\lesssim 10^7$ atoms. If one
would scatter light from a small fraction of these over a $4 \pi$ solid angle, only a few photons
would be collected and one would need to determine their frequency amidst a large background of
incident and scattered light.

Instead, we have adopted a different approach.  Rather than
detecting {\it spontaneous} scattering from a single beam, we
study light scattering as a {\it stimulated} process, called Bragg
scattering, induced by two laser beams which illuminate the atomic
sample (Fig.\ \ref{light_scatt}). The momentum and energy transfer
is {\it pre--determined} by the angle and frequency difference
between the incident beams, respectively, rather than {\it
post--determined} by the position of a photo--detector and by a
difficult frequency measurement. The quantities of interest are
matrix elements which characterize the response of the condensate,
and they are the same for spontaneous and stimulated scattering.
Furthermore, since the momentum transfer can be much greater than
the momentum spread of the sub--recoil atomic sample, and since
stimulated light scattering can be made to dominate over
spontaneous scattering, the response of the system can be assessed
by the nearly background--free detection of recoiling atoms.

We have studied Bose--Einstein condensates by the spectroscopic measurement of the Bragg scattering
resonance. In this paper, we describe two applications of Bragg spectroscopy to study excitations
of a Bose--Einstein condensate in either the free--particle \cite{sten99brag} (large momentum
transfer) or the phonon \cite{stam99phon} (small momentum transfer) regime.  The discussion
includes a description of the dynamic structure factor of a Bose--Einstein condensate which leads
to the interpretation of our measurements as an observation of the zero--point momentum
distribution of trapped condensates, as a measurement of the energies of free--particle and phonon
excitations, and as evidence for correlations in the many--body condensate wavefunction introduced
by interatomic interactions.

\subsubsection{Light scattering from atomic beams and atoms at rest}
\label{sec:beam scattering}

The interaction of a neutral atomic beam with an optical standing
wave was studied by several groups in the early 1980's.
Quantitative studies which focused on the effect of conservative
optical potentials were performed by Pritchard and collaborators
\cite{goul86,mart88bragg}.  In two different experiments, a
collimated atomic beam of sodium was incident upon a standing
light wave formed by a retro--reflected laser beam.  Two regimes
of scattering were identified: Kapitza--Dirac scattering from a
thin optical grating (tightly focused beams) which is
non--specific in the angle between the incident atomic beam and
the standing wave\cite{goul86}, and Bragg scattering from a thick
grating (loosely focused beams) which occurs only at specific
resonant angles\cite{mart88bragg}.  In these atomic beam
experiments, the kinetic energy of the atoms, and thus the
magnitude  of their momentum, is unchanged by scattering off the
stationary optical field. Thus, scattering can occur only if the
optical field contains photons propagating in a direction so that
the atomic momentum can be rotated by the absorption and
stimulated emission of photons from the standing wave.  A thin
optical grating contains photons propagating over a wide angular
range, and thus scattering is not limited to specific angles.  In
the case of a broad focus, the angular divergence of the photons
is too small to allow scattering except at specific incident
angles of the atomic beam to the standing wave (the so-called
Bragg angles).

Kapitza--Dirac \cite{ovch99} and Bragg \cite{kozu99bragg} scattering of Bose--Einstein condensates
have also been demonstrated.  These experiments were performed by exposing the nearly stationary
atomic sources to a pulse of two intersecting laser beams which had a variable differential
detuning $\omega$.  Such an experimental situation is identical to the aforementioned atomic beam
experiments when viewed in the frame of reference of the atoms: the duration of the optical pulse
corresponds to the width of the optical grating, and a differential detuning between the optical
beams is equivalent to an atom crossing an optical grating at an angle which introduces opposite
Doppler shifts to the two counter--propagating laser beams. Kapitza--Dirac scattering occurs for
short pulses which contain frequency components necessary to excite the atom to an energy of $\hbar
\omega_q^0 = \hbar^2 q^2 / 2 m$ where $\hbar q$ is the momentum recoil due to a single scattering
event. Similarly, the condition for Bragg scattering becomes a resonance condition for exciting an
atom to an excited momentum state: $\omega = \omega_q^0$ (for first order scattering).

The Bragg resonance condition is sensitive to the motion of the atom with respect to the optical
standing wave orientation. By simple energy and momentum conservation, the energy transferred to an
atom with initial velocity $\vec{v}_i$ by a momentum kick of $\hbar \vec{q}$ is
\begin{equation}\label{eq:doppersens}
  \hbar \omega = \frac{ (\hbar \vec{q}  + m \vec{v}_i)^2}{2 m} - \frac{m v_i^2}{2} = \frac{\hbar^2
  q^2}{2 m} + \hbar \vec{q} \cdot \vec{v}_i
\end{equation}

Thus, the Bragg resonance is Doppler sensitive and can be used to determine spectroscopically the
velocity distribution of an atomic sample.  It has been used previously to determine the
temperature of laser-cooled atoms \cite{cour94}.  Here we extend the method to the determination of
the zero-point motion of a condensate.

\subsubsection{Relation to the dynamic structure factor of a many--body system}

\label{sec:meassqw}

Inelastic scattering has long been used to probe the properties of
condensed--matter systems.  In the case of liquid helium, both
neutron and light scattering were used to determine the elementary
excitations of this system
\cite{hove54,grey78,soko95,nozi90,grif93}.  A theoretical
discussion of the spectrum of inelastically scattered light from a
Bose--Einstein condensate has been presented by a number of authors
\cite{java95spec,java95scat,grah96scat}.   The dilute atomic
condensates are particularly simple examples for the general
scattering theory since the scattering can be treated in an atomic
basis. Following the experimental studies of inelastic light
scattering which are summarized in this review, a thorough
interpretation of light scattering from an inhomogeneous
Bose--Einstein condensate was presented \cite{zamb00}.

Let us discuss how Bragg scattering is used to probe the properties of a many--body system.  An
atomic sample is exposed to two laser beams, with wavevectors $\vec{k}_1$ and $\vec{k}_2$ and a
frequency difference $\omega$ which is generally much smaller than the detuning  $\Delta$ of the
beams from an atomic resonance. The two laser beams interfere to form a ``walking'' wave intensity
modulation $I\subrm{mod}(\vec{r},t) = I \cos (\vec{q} \cdot \vec{r} - \omega t)$ where $\vec{q} =
\vec{k}_1 - \vec{k}_2$.  Due to the ac Stark effect\cite{cohe92}, atoms exposed to this intensity
modulation experience a conservative optical potential with a spatial modulation of $V\subrm{mod} =
(\hbar \Gamma^2 / 8 \Delta) \cdot (I\subrm{mod}/I\subrm{sat})$ where $\Gamma$ is the line width of
the atomic resonance and $I\subrm{sat}$ the saturation intensity.

To determine the Bragg scattering response of a many--body system, we express the modulated
potential in second quantized notation as
\begin{equation}\label{eq:vmod}
  V\subrm{mod} = \frac{V}{2} \left( \hat{\rho}^\dagger( \vec{q}) e^{-i \omega t} + \hat{\rho}^\dagger(
-\vec{q} ) e^{+i \omega t} \right)
\end{equation}
where $\hat{\rho}(\vec{q}) = \sum_m \hat{a}^\dagger_{m+q} \hat{a}_m$ is the Fourier transform of
the atomic density operator at wavevector $\vec{q}$.  Equivalently, $V\subrm{mod}$ is found by
isolating those terms in $\mathcal{H}^\prime$ (Eq.\ \ref{eq:laint}) which involve the
macroscopically occupied optical modes at wavevectors $\vec{k}_1$ and $\vec{k}_2$, and replacing
the photon creation and destruction operators with $c$--numbers proportional to the electric field
strength of the Bragg scattering laser beams.

We may then determine the Bragg scattering rate using Fermi's golden rule.  Considering scattering
out of the many--body ground state $|g \rangle$, we neglect the counter--rotating term in
$V\subrm{mod}$ and obtain the excitation rate per particle as
\begin{equation}\label{eq:braggrate}
  \frac{W}{N} = \frac{2 \pi}{N \hbar}
  \left(\frac{V}{2}\right)^{\,2} \sum_f | \langle f |
  \hat{\rho}^\dagger(\vec{q}) | g \rangle |^2 \delta(\hbar \omega
  - (E_f - E_g)) \equiv 2 \pi \omega_R^2 \sqw
\end{equation}
Here $N$ is the number of atoms in the system, and the sum is
performed over all final excited states $|f\rangle$ with energy
$E_f$.  We have introduced the dynamic structure factor $\sqw$
which is the Fourier transform of density--density fluctuations in
state $|g\rangle$ with spatial and temporal frequencies of
$\vec{q}$ and $\omega$, respectively \cite{nozi90,grif93}. The
dynamic structure factor generally characterizes the response of
the system to longitudinal perturbations of any source, not solely
to optical excitation. The density fluctuation spectrum is
directly determined by the Bragg scattering response, normalized
by the two--photon Rabi frequency $\omega_R = V / 2 \hbar$.
Integrating over all frequencies $\omega$ one obtains the static
structure factor $\sq = \langle g | \hat{\rho}(\vec{q})
\hat{\rho}^\dagger(\vec{q}) | g \rangle$ which is equivalent to
the line strength of the Bragg resonance.

\subsection{The dynamic structure factor of a Bose--Einstein condensate}

\label{sec:dynstrucsection}

In this section, we use the theory of the weakly--interacting Bose--Einstein condensate to predict
the dynamic structure factor, first for a homogeneous condensate and then for the situation of
experimental relevance, an inhomogeneous condensate confined by a harmonic trapping potential.

\subsubsection{The homogeneous condensate}

\label{sec:sqwbec}

\begin{figure}
    \begin{center}
    \includegraphics[height=3 in]{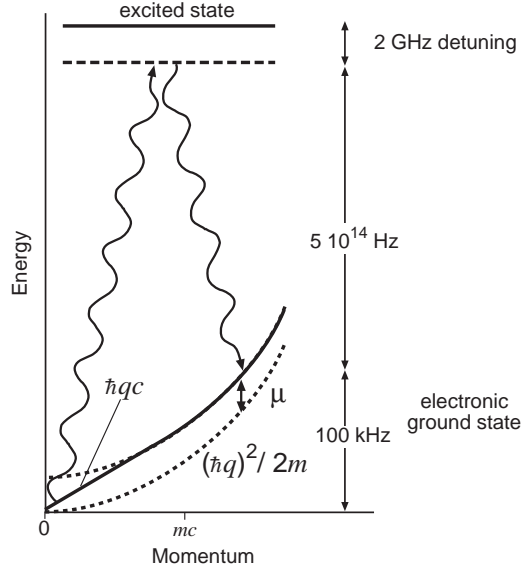}
    \caption[Dispersion relation of a Bose-Einstein condensate]{
Probing the dispersion relation of a Bose-Einstein condensate by off-resonant light scattering. The
microscopic excitation spectrum in a homogeneous weakly--interacting Bose--Einstein condensate is
given by the Bogoliubov dispersion relation (solid line).  For small momenta $\hbar q$, such that
$\hbar q \ll m c$ the dispersion relation is phonon--like (linear). Here $m$ is the mass, and the
speed of sound $c$ is related to the interaction energy $\mu$ by $\mu=m c^2$. For large momenta
($\hbar q \gg m c$) it is particle--like (quadratic) offset from the energy of a free--particle
excitation by a mean--field shift of $\mu$ ($ \approx$ 5 kHz for our experiments). Excitations can
be created optically by stimulated light scattering using two laser beams which are both far
detuned (about 2 GHz) from the atomic resonance.  Momentum and energy are provided by absorption of
one photon to a virtual excited level, followed by stimulated emission of a second, lower energy
photon.
    \label{bogoliubov}}
    \end{center}
\end{figure}

A Bose--Einstein condensate is quite different from other fluids in that the {\it microscopic}
(i.e.\ single--atom) excitations of the system become manifest as {\it macroscopic} density
fluctuations due to interference with the macroscopic wavefunction. Considering density
fluctuations in a homogeneous Bose--Einstein condensate, we may approximate
\begin{equation}\label{eq:excinbec}
 |e\rangle = \frac{1}{\sqrt{N}} \hat{\rho}^\dagger(\vec{q}) |g\rangle
  = \frac{1}{\sqrt{N}} \sum_m \hat{a}^\dagger_{m+q}
  \hat{a}_m |g \rangle
   \simeq \frac{(\hat{a}^\dagger_q \hat{a}_0 + \hat{a}^\dagger_0
  \hat{a}_{-q}) |g\rangle}{\sqrt{N}}
  = |e^+\rangle + |e^-\rangle
\end{equation}
Here, the macroscopic occupation of the zero--momentum state picks out two terms in the sum.
Following Bogoliubov\cite{bogo47}, we identify $\hat{a}^\dagger_0 = \hat{a}_0 = \sqrt{N_0}$ and
transform to Bogoliubov operators by substituting $\hat{a}_k = u_k \hat{b}_k - v_k
\hat{b}^\dagger_{-k}$.  The operators $\hat{b}^\dagger_{k}$ and $\hat{b}_{k}$ are creation and
destruction operators for the proper microscopic quasi--particle excitations of the condensate,
with $u_k = \cosh \phi_k$, $v_k = \sinh \phi_k$ and $\tanh 2\phi_k = \mu / (\hbar \omega_k^0 +
\mu)$. Here, $\mu$ is the chemical potential, and, again, $\hbar \omega_k^0 = \hbar^2 k^2 / 2 m$ is
the free recoil energy at wavevector $\vec{k}$.  The Bogoliubov quasi--particle spectrum is given
as $\hbar \omega_k^B = \sqrt{\hbar \omega_k^0 (\hbar \omega_k^0 + 2 \mu)}$ (see Fig.\
\ref{bogoliubov}).

The many--body wavefunction of the condensate $|g \rangle$ corresponds to the quasi--particle
vacuum defined by the relations $\hat{b}_k |g\rangle = 0, \forall k$. Thus, we find
\begin{equation}\label{eq:excinbec2}
  \hat{\rho}^\dagger(\vec{q}) |g\rangle \simeq \sqrt{N_0} (u_q - v_q)
  \hat{b}^\dagger_{q} |g\rangle
\end{equation}
From this, it follows that for a homogeneous Bose--Einstein condensate $\sqw \simeq (u_q - v_q)^2
\delta(\omega - \omega_q^B)$, and $\sq = (u_q - v_q)^2 = \omega_q^0 / \omega_q^B$.

Thus, we expect Bragg scattering from an interacting, homogeneous
Bose--Einstein condensate to differ from the Bragg scattering of a
non--interacting condensate in two ways.  First, the Bragg
resonance occurs at the Bogoliubov quasi--particle energy which is
higher than the free--particle recoil energy, i.e.\ the Bragg
resonance line at a momentum transfer of $\hbar \vec{q}$ is
shifted upwards in frequency from the free particle resonance by
$\Delta \omega = \omega_q^B - \omega_q^0$. Second, the Bragg
scattering response is weakened relative to that of free particles
by a factor $\sq = (u_q - v_q)^2 < 1$.  In other words, light
scattering from a Bose--Einstein condensate is suppressed by the
presence of repulsive interatomic interactions. Indeed, these two
statements are equivalent: by the $f$--sum rule which states that
$\int \omega \sqw d \omega = \omega_q^0$, an increase in the
resonance frequency implies a decrease in the excitation
strength\cite{nozi90}.

Let us evaluate the structure factor in two limits of the Bragg scattering wavevector $\vec{q}$.
The wavevector which corresponds to the interaction energy is the inverse healing length $\xi^{-1}
= \sqrt{2} m c/ \hbar$ where $c$ is the speed of Bogoliubov sound. For large scattering wavevectors
($q \gg \xi^{-1}$), Bragg scattering occurs in the free--particle regime. The Bragg resonance is
shifted upwards by the chemical potential $\Delta \omega \simeq \mu / \hbar$, and the line strength
tends to $\sq \rightarrow 1 - \mu / \hbar \omega_q^0$. Thus, by measuring the frequency shift of
the Bragg scattering resonance in the free--particle regime, one can directly measure the chemical
potential \cite{sten99brag}. For small wavevectors ($q \ll \xi^{-1}$), the Bose--Einstein
condensate responds to optical excitation collectively with the creation of phonons.  The static
structure factor tends to $\sq \rightarrow \hbar q / 2 m c$ and vanishes in the long wavelength
limit, as required of a zero--temperature system with finite compressibility \cite{pric54}.

\subsubsection{Bragg scattering as a probe of pair correlations in the condensate}

\label{sec:probecorr}

It is interesting to re--examine these modifications to light scattering from a Bose--Einstein
condensate in terms of the structure of the many--body condensate wavefunction.  What is it about a
dilute, weakly--interacting Bose--Einstein condensate that suppresses light scattering compared to
non--interacting atoms? As discussed above, the static structure factor $\sq$ is the magnitude of
the state vector $|e\rangle \simeq |e^{+}\rangle + |e^{-}\rangle$ (Eq.\ \ref{eq:excinbec}). The
states $|e^{+}\rangle$ and $|e^{-}\rangle$ represent two means by which momentum is imparted to the
condensate: either by promoting a zero-momentum particle to momentum $\hbar {\vec{q}}$, or else by
demoting a particle from momentum $- \hbar {\vec{q}}$ to zero momentum.

If correlations could be neglected, the total rate of excitation would simply be the sum of the
independent rates for these two processes, proportional to $\langle e^{+}|e^{+} \rangle = \langle
N_{q}^{0}\rangle + 1 = u_{q}^{2}$ and $\langle e^{-}|e^{-} \rangle = \langle N_{-q}^{0} \rangle =
v_{q}^{2}$ where $\langle N_{k}^{0} \rangle$ is the expected number of atoms of momentum $\hbar
{\vec{k}}$ in the condensate.  Indeed, a simple--minded rate equation leads to an expression for
the total scattering rate which is proportional to $N_0 (\langle N_{q}^{0}\rangle +1) + \langle
N_{-q}^{0} \rangle (N_0 +1)$.  This expression is the sum of the rates of the two processes which
can transfer momentum $\hbar {\vec{q}}$ to the system and scatter particles into or out of the
condensate.  Each rate is proportional to the number of particles in the initial state and, because
of bosonic stimulation, to the number of atoms in the final state plus one. This would apply, for
example, to a condensate in a pure number state, or to an ideal gas condensate with a thermal
admixture of atoms with momenta $\pm \hbar {\vec{q}}$, and would always lead to $S({\vec{q}}) > 1$.

Yet, for the many-body ground state of the interacting Bose gas, the behavior is dramatically
different. Interactions between zero-momentum atoms admix into the condensate pairs of atoms at momenta
$\pm \hbar {\vec{q}}$ the population of which comprises the quantum depletion\cite{huan87}. As a
result, the two momentum transfer mechanisms described above produce indistinguishable final
states, and the total rate of momentum transfer is given by the interference of two amplitudes, not
by the sum of two rates. Pair excitations in the condensate are correlated so as to minimize the
total energy, and this results in a relative phase between the two amplitudes which produces
destructive interference: $S({\vec{q}}) = (u_{q} - v_{q})^{2} < 1$. For high momentum, $\langle
N_{q}^{0}\rangle \ll 1$ and the interference plays a minor role. In the phonon regime, while the
independent rates $u_{q}^{2}$ and $v_{q}^{2}$ (and hence $\langle N_{\pm q}^{0} \rangle$) diverge
as $1/q$, the correlated quantum depletion extinguishes the rate of Bragg excitation.

It is interesting to note that the extinction of Bragg scattering at low momentum, i.e. the minus
sign in the amplitude $u_{q} - v_q$ of state $|e \rangle$, arises from minimizing the energy of an
interacting Bose gas with positive scattering length.  One might wonder whether a negative sign of
the scattering length would lead to an enhancement of scattering. However, homogeneous condensates
with negative scattering length are unstable.  In a trap, they are only stable when their size is
smaller than the healing length, effectively cutting off the long--wavelength phonon modes.
Therefore, a condensate with negative scattering length should show essentially the same behavior
as a similar--sized condensate of non-interacting atoms.

\subsubsection{Mean--field theory determination of $\sqw$}

These same results are obtained also in a self--consistent mean--field approach. Considering again
the homogeneous case, we introduce the perturbation
\begin{equation}\label{eq:vmodsp}
  V\subrm{mod} = \frac{V}{2} \left( {e^{i \vec{q} \cdot \vec{r} - i \omega t}} + {e^{- i \vec{q} \cdot \vec{r} + i \omega t}} \right)
\end{equation}
to the time--dependent Gross--Pitaevskii equation\cite{gros61,pita61}, and use a condensate
wavefunction of the form
\begin{equation}\label{eq:perturbedpsi}
  \psi(\vec{r},t) = e^{- i \mu t / \hbar} \left( \psi_0(\vec{r},t) + u(t) {e^{i \vec{q} \cdot \vec{r} - i \omega t}} + v^*(t) {e^{- i \vec{q} \cdot \vec{r} + i \omega t}}
\right)
\end{equation}
Here,  $\mu = n g$ is the chemical potential where $n$ is the condensate density, $g = 4 \pi
\hbar^2 a / m$ and $a$ is the $s$--wave scattering length. In the absence of the perturbation, the
ground--state condensate wavefunction is $\psi_0(\vec{r},t) = \sqrt{n}$.

The weak potential $V\subrm{mod}$ introduces the small perturbations $u(t) {e^{i \vec{q} \cdot
\vec{r} - i \omega t}}$ and $v(t) {e^{- i \vec{q} \cdot \vec{r} + i \omega t}}$ where the
amplitudes $u(t)$ and $v(t)$ are slowly varying functions of time. Using the Gross--Pitaevskii
equation and isolating terms proportional to ${e^{i \vec{q} \cdot \vec{r} - i \omega t}}$ and
${e^{- i \vec{q} \cdot \vec{r} + i \omega t}}$, we obtain the set of equations
\begin{eqnarray}\label{eq:excgrowth1}
  \hbar \omega u + i \hbar \frac{d u}{d t} & = & \left(
  \frac{\hbar^2 q^2}{2 m} + n g \right) u + n g v +
  \frac{V}{2} \psi_0 \\
  -\hbar \omega v - i \hbar \frac{d v}{d t} &  =&  \left(
  \frac{\hbar^2 q^2}{2 m} + n g \right) v  + n g u + \frac{V}{2}
  \psi_0 \label{eq:excgrowth2}
\end{eqnarray}
Let us now apply the Bogoliubov transformation, and write
\begin{equation}\label{eq:bogtrans}
  \twovec{u(t)}{v(t)} = \alpha(t) \twovec{u_q}{- v_q} + \beta(t)
  \twovec{v_q}{-u_q}
\end{equation}
The two component vectors $\twovec{u_q}{-v_q}$ and $\twovec{v_q}{-u_q}$ are solutions of the
equations \cite{bogo47,pita61}
\begin{equation}\label{eq:bdg}
  \left[ \begin{array}{c c}
  \left(\hbar \omega_q^0 + \mu \right) & \mu \\
  -\mu & - \left(\hbar \omega_q^0 + \mu \right) \\
  \end{array} \right] \, \twovec{u}{v} = \hbar \tilde{\omega}
  \twovec{u}{v}
\end{equation}
with frequencies $\tilde{\omega} = \omega_q^B$ and $\tilde{\omega} = - \omega_q^B$, respectively.
The negative frequency solution corresponds to excitations in the $-\vec{q}$ direction. With this
substitution, Eqs.\ \ref{eq:excgrowth1} and \ref{eq:excgrowth2} are decoupled:
\begin{eqnarray}\label{eq:excgrowth3}
  \hbar \omega \alpha + i \hbar \frac{d \alpha}{d t} & = & \hbar
  \omega_q^B \alpha + (u_q - v_q) \frac{V}{2} \psi_0 \\
  \label{eq:excgrowth4}
  \hbar \omega \beta + i \hbar \frac{d \beta}{d t} & = & - \hbar
  \omega_q^B \beta + (u_q - v_q) \frac{V}{2} \psi_0
\end{eqnarray}
These equations are identical to those of first--order perturbation theory for a single--particle
Schr\"{o}dinger equation, and thus the rates of growth of $|\alpha|^2$ and $|\beta|^2$ can be
evaluated using Fermi's golden rule. $|\alpha|^2$ and $|\beta|^2$ are the probabilities for
creating quasiparticles with momentum $\hbar \vec{q}$ and $-\hbar \vec{q}$, respectively.

The response of the condensate to Bragg scattering can be evaluated by calculating the momentum
imparted to the condensate
\begin{equation}\label{eq:mom1}
  \langle \psi(\vec{r}, t) | \hat{p} | \psi(\vec{r}, t) \rangle = \hbar \vec{q} \times \left( |\alpha|^2
  - |\beta|^2 \right) = \hbar \vec{q} \times \left( |u|^2 - |v|^2 \right)
\end{equation}
Normalizing by the Bragg scattering momentum $\hbar q$, we obtain the Bragg excitation rate per
particle as
\begin{displaymath}
\frac{d}{dt} \left(\frac{\langle \psi(\vec{r}, t) | \hat{p} | \psi(\vec{r}, t) \rangle}{\hbar q
  \, N_0}\right) = 2 \pi \omega_R^2 (u_q - v_q)^2 \left[\delta(\hbar \omega - \hbar \omega_q^B)
  - \delta(\hbar \omega + \hbar \omega_q^B) \right]
\end{displaymath}
\begin{equation}\label{eq:momperparticle}
 = 2 \pi \omega_R^2 \left[ \sqw - S(
  -\vec{q}, -\omega) \right]
\end{equation}
where, again, $\omega_R$ is the two--photon Rabi frequency.

This treatment reveals two important points.  First, the mean--field treatment reproduces the
suppression of the structure factor, even though, as discussed in Sec.\ \ref{sec:probecorr}, this
suppression is indicative of correlations in the many--body condensate wavefunction.  In the
mean--field theory, correlations in the condensate wavefunction are explicitly neglected by the use
of a Hartree wavefunction.  Nevertheless, when the excitation of the condensate is also treated in
a Hartree approximation (i.e.\ assuming all particles are in the same, albeit time--varying,
single--particle wavefunction), the dynamic response is correctly obtained.

Second, this treatment takes into account both the positive and
negative frequency terms in $V\subrm{mod}$, unlike in Sec.\
\ref{sec:meassqw} where we chose to consider only the positive
frequency part for simplicity. One finds that the momentum
imparted by stimulated scattering from the two Bragg beams
measures the difference $\sqw - S( -\vec{q}, -\omega)$. This is
important when one considers Bragg scattering from a non--zero
temperature system which has thermally excited states and undergo
anti-Stokes scattering. The dynamic structure factor for a
non--zero temperature Bose--Einstein condensate is given as
\cite{nozi90,grif93}
\begin{equation}\label{eq:sqwfiniteT}
  \sqw = (u_q - v_q)^2 \left[ ( \langle N_q^B \rangle + 1)
  \delta(\hbar \omega - \hbar \omega_q^B) + \langle N_{-q}^B
  \rangle \delta(\hbar \omega + \hbar \omega_q^B) \right]
\end{equation}
where $\langle N_q^B \rangle$ is the thermal population of {\it quasi--particles}.  Light
scattering is thus quite sensitive to the presence of excitations in the condensate, as confirmed
by the recent observation of superradiant light scattering from a condensate which is due to the
buildup of excited particles in a preferred mode\cite{inou99super}.  However, inserting Eq.\
\ref{eq:sqwfiniteT} into Eq.\ \ref{eq:momperparticle} shows that the effects of thermally--excited
particles are cancelled out in the Bragg scattering response, and thus one measures the
zero--temperature structure factor even in a finite--temperature sample.

\subsubsection{The inhomogeneous condensate}

The Bose--Einstein condensates realized experimentally differ from the homogeneous condensates
considered above due to their confinement. This confinement changes the Bragg scattering resonance
from that predicted for a homogeneous Bose--Einstein condensate by introducing an inhomogeneous
density distribution and by introducing Doppler broadening due to the zero--point momentum
distribution.

The confining potential of magnetic or even optical traps is typically harmonic, taking the form
$V(\vec{r}) = (m/2) (\omega_x^2 x^2 + \omega_y^2 y^2 + \omega_z^2 z^2)$, where $\omega_x$,
$\omega_y$, and $\omega_z$ are the trap frequencies. The condensate wavefunction $\psi(\vec{r})$
can be determined by the mean--field Gross--Pitaevskii equation\cite{gros61,pita61}. For large
condensates for which the interaction energy is much larger than the kinetic energy, the
wavefunction is given by the Thomas--Fermi solution as\cite{dalf99rmp}
\begin{equation}
|\psi(\vec{r})|^2 = n \, \max\left( 1 - \left(\frac{x}{x_{c}}\right)^2 -
\left(\frac{y}{y_{c}}\right)^2 - \left(\frac{z}{z_{c}}\right)^2,\; 0\right) \label{eq:condintf}
\end{equation}
where the Thomas--Fermi radii are defined as $x_{c}^2 = 2 \mu / m \omega_x^2$ (similar for $y_c$
and $z_c$), and $\mu = g n$ where $n$ is the maximum condensate density and $g = 4 \pi \hbar^2 a /
m$ with $a$ being the $s$--wave scattering length.

Let us first consider the two effects of confinement separately:

\begin{itemize}
  \item [a)]{\bf Mean--field shift and broadening:} The effects of interactions
  on Bragg scattering from an inhomogeneous density distribution can be accounted for using a local
density approximation (discussed further in \cite{zamb00}).
Using the density distribution of a condensate in the
Thomas--Fermi regime (Eq.\ \ref{eq:condintf}) and the predicted
$\sqw$ for a homogeneous condensate at the local value of the
density, the Bragg resonance line shape is calculated to be
\cite{stam99phon}
\begin{equation}
    I_{\mu}(\omega) \, d\omega = \frac{15}{8} \frac{\omega^{2} -
    {\omega_{q}^{0}}^{2}}{\omega_{q}^{0} (\mu / \hbar)^{2}}
    \sqrt{1 - \frac{\omega^{2} -
    {\omega_{q}^{0}}^{2}}{2 \omega_{q}^{0} \mu / \hbar}} \, d\omega
    \label{eq:lineshape}
\end{equation}

The line strength $S({\vec{q}})$ and center frequency can be obtained from Eq.\ \ref{eq:lineshape}
by integration. Explicitly, the static structure factor of a harmonically--confined Bose--Einstein
condensate is given by
\begin{equation}\label{eq:explicitsq}
  S({\vec{q}}) = \frac{15 \eta}{64} (y + 4 \eta - 2 y \eta^2 + 12 \eta^3 - 3 y \eta^4)
\end{equation}
where $\eta^2 = \hbar \omega_q^0 / 2 \mu$ and $y = \pi - 2 \arctan ((\eta^2 - 1)/2 \eta)$.  The
line strength has the limiting values of $S({\vec{q}}) \rightarrow 15 \pi/32 \, (\hbar
\omega_{q}^{0}/ 2 \mu)^{1/2}$ in the phonon regime and $S({\vec{q}}) \rightarrow 1 - 4 \mu / 7
\hbar \omega_{q}^{0}$ in the free--particle regime. In accordance with the $f$--sum rule, the
center frequency $\bar{\omega}$ is given as $\omega_{q}^{0} / S({\vec{q}})$. In the free--particle
regime ($\hbar \omega_q^0 \gg \mu$), the line center is shifted upwards from the free--particle
resonance frequency by $4 \mu / 7 \hbar$, and broadened to an rms--width of
\begin{equation}\label{eq:mfwidth}
  \Delta \omega_\mu = \sqrt{\frac{8}{147}} \frac{\mu}{\hbar}
\end{equation}

  \item [b)]{\bf Doppler broadening:} Doppler broadening arises due to the initial momentum distribution
of the condensate. The momentum distribution of a trapped
Bose--Einstein condensate, assuming its full coherence, is given
by the Fourier transform of the condensate wavefunction (Eq.\
\ref{eq:condintf}). Thus, neglecting the mean--field shift (the
impulse approximation as discussed in \cite{zamb00}), the Bragg
excitation rate $I_D(\omega)$ from a Bose--Einstein condensate at
a frequency difference of $\omega$ between the two Bragg beams is
\begin{eqnarray}\label{eq:dopplerline2}
  \lefteqn{I_D(\omega) d\omega }  \\
  & &  \propto \int d^3\vec{k} \ \delta \! \left( \frac{\hbar \vec{k} \cdot
  \vec{q}}{m} - (\omega - \omega_q^0)\right) \ \left| \int d^3\vec{r} \ e^{-i \vec{k} \cdot
  \vec{r}} \psi(\vec{r}) \right|^2 d\omega  \nonumber \\
  & & \propto  \int dx_1 \, dx_2 \, e^{-i k(\omega)
  \cdot (x_2 - x_1)} \int dy \, dz \, \psi^*(x_1, y, z) \psi(x_2, y, z) \, d\omega   \nonumber
\end{eqnarray}
where $\hbar \vec{q}$ is the Bragg scattering momentum and $\psi({\vec{r}})$ is the condensate wave
function. In the last line, a coordinate system is chosen so that $\hbar \vec{q}$ lies in the
$\hat{x}$--direction, and $k(\omega) = m (\omega - \omega_q^0) / \hbar q$ . In the Thomas--Fermi
regime, for Bragg scattering along one of the principal axes of the harmonic trap, the Doppler line
shape is then
\begin{eqnarray}\label{eq:dopplertf}
  \lefteqn{ I_D(\omega) d\omega \propto } \\
  & & \frac{2 (4 + \kappa^2) J_1(\kappa) J_2(\kappa) +
  \kappa J_0(\kappa) [5 \kappa J_1(\kappa) - 16 J_2(\kappa) + 3 \kappa J_3(\kappa)]}{\kappa^3}
   \,  d\omega \nonumber
\end{eqnarray}
where $\kappa = k(\omega) x_{c}$, and $J_i$ are Bessel functions.  This line shape is similar to a
Gaussian, but its rms--width is undefined.  We therefore fitted a Gaussian function to the line
shape and extracted an effective rms--width of
\begin{equation}\label{eq:dopplerwidth}
  \Delta \omega_D \simeq 1.58 \frac{\hbar q}{m x_{c}}
\end{equation}

\end{itemize}

\subsubsection{Relevance of Doppler broadening}

Thus, the Bragg scattering resonance for a trapped Bose--Einstein
condensate is sensitive both to the velocity (Doppler shift) and
the density (mean--field shift) of the atomic sample.  The above
treatments, in which we considered the effects of each of these
shifts separately, are valid predictions in two limiting cases:
for large condensates so that $\mu / \hbar \gg \hbar q / m x_c$
one can neglect Doppler broadening, while for small condensates
where $\mu / \hbar \ll \hbar q / m x_c$ one can neglect
mean--field broadening.  However, in our experiments on Bragg
scattering in the free--particle regime\cite{sten99brag}, the
Doppler and mean--field widths were comparable, and one must
consider both effects simultaneously.

We now show with the aid of simple sum rules that it is correct to
add the Doppler broadening (Eq.\ \ref{eq:dopplerwidth}) and the
mean--field broadening (Eq.\ \ref{eq:mfwidth}) in quadrature to
obtain the total rms line width which can then be compared to
experiments.  A unified approach toward determining fully the
dynamic structure factor in the presence of both Doppler and
mean--field shifts has been presented recently by Zambelli
\emph{et al.}\ and was compared favorably with experimental data
\cite{zamb00}.

The condensate wavefunction $|g\rangle$ determined by the Gross--Pitaevskii equation is a solution
of the equation $\mathcal{H}_0 |g\rangle = \mu |g\rangle$ where
\begin{equation}\label{eq:condensateh0}
  \mathcal{H}_0 = \frac{\hat{p}^2}{2 m} + U(\vec{r}) + g n(\vec{r})
\end{equation}
In the regime $\hbar \omega_q^0 \gg \mu$, the excitations relevant to Bragg scattering are well
described as free--particle excitations which obey
\begin{equation}
    \hbar \omega_{f} |f \rangle = \left( \frac{{\hat{p}}^{2}}{2 m} + U({\vec{r}}) + 2 g n(\vec{r})
 \right) |f \rangle = {\mathcal H}\subrm{exc} |f \rangle
\end{equation}
This can be seen, for example, by considering Eq.\ \ref{eq:bdg} in the free--particle regime where
$v_q \rightarrow 0$.  Thus, the excitations are eigenfunctions of the Hamiltonian
$\mathcal{H}\subrm{exc} = \mathcal{H}_0 + g n(\vec{r})$ which is different from the Hamiltonian
$\mathcal{H}_0$ which gives the condensate wavefunction \cite{dalf97coll}.  The extra term $g
n(\vec{r})$ represents the repulsion of excitations from the condensate which gives rise to the
mean--field shift in the free--particle regime.

An exact determination of the resonance line shape using Fermi's golden rule requires detailed
knowledge of the excitation wavefunctions. Such an explicit calculation has been
performed\cite{kill99spec} in the context of the two--photon optical excitation from the $1S$ to
the $2S$ state in hydrogen which has been employed to probe properties of a hydrogen Bose--Einstein
condensate \cite{kill98,frie98}.  However, even without this exact description, moments of the
spectral line can be easily determined. The first moment of the spectral line is given by
\begin{eqnarray}
    \hbar \bar{\omega} & = &
        \frac{
            \int_{0}^{\infty} d\omega \, \omega S({\vec{q}}, \omega)}{ \int_{0}^{\infty} d\omega \,
     S({\vec{q}}, \omega)} \\ & = & \langle e^{i \vec{q} \cdot \vec{r}}\, \mathcal{H}\subrm{exc} \,
     e^{-i \vec{q} \cdot \vec{r}}
      - {\mathcal H}_{0}\rangle_{c}\\
    & = & \frac{\hbar^{2} q^{2}}{2 m} + g \langle n \rangle_{c}
\end{eqnarray}
where $\langle X \rangle_{c} = \langle g | X | g \rangle$, and
$\langle \vec{p} \rangle_{c} = 0$ in the ground state.  Thus the
Bragg resonance line for a harmonically trapped condensate is
shifted from the free--particle resonance by $g \langle n
\rangle_c = 4 \mu / 7 \hbar$.

The rms--width of the line $\Delta \omega = \sqrt{\overline{\omega^{2}} - \overline{\omega}^{2}}$
is calculated using
\begin{equation}
    \hbar^{2} \overline{\omega^2} = \langle e^{i \vec{q} \cdot \vec{r}} {\mathcal H}^2_{exc} e^{-i
    \vec{q} \cdot \vec{r}}
     - {\mathcal H}_{0}^{2} \rangle_{c}
\end{equation}
by which one obtains
\begin{equation}\label{eq:totalwidth}
    \hbar^{2} (\Delta \omega)^{2} = \left< \left( \frac{\hbar \vec{q} \cdot \vec{p}}{m} \right)^{2}
    \right>_{c} + g^{2} \left( \langle n^{2} \rangle_{c} - \langle n\rangle_{c}^{2} \right)
\end{equation}
Thus the total rms--width of the line is the sum of two widths in quadrature: the Doppler width due
to the finite size of the condensate, and the line broadening due to the inhomogeneous condensate
density.

\subsection{Experimental aspects of Bragg spectroscopy}

For the experimental studies of light scattering from
Bose--Einstein condensates, condensates of atomic sodium were
produced as in our previous experiments
\cite{mewe96bec,kett99var}. These condensates typically contained
$10^7$ atoms, and were held in a magnetic trap which provided a
cylindrically--symmetric cigar--shaped harmonic confinement.  The
strength of the confining field could be easily modified by
adjusting the currents through the magnetic trapping coils,
allowing us to vary the peak condensate density in the range (0.5
-- 6) $\times 10^{14} \, \mbox{cm}^{-3}$ and the radial
half--width of the condensate ($x_c$) between 7 and 15 $\mu$m.

The two laser beams used for Bragg scattering were derived from a
common source, and had a detuning to the red of the $3S_{1/2},
|F=1\rangle \rightarrow 3P_{3/2}, |F^\prime = 0,1,2\rangle$
optical transitions of about $\Delta = 1.7$ GHz (similar
parameters were used for experiments at NIST, Gaithersburg
\cite{ovch99,kozu99bragg,hagl99coh,deng99}). A small frequency
difference $\omega$ between the counter--propagating beams, on the
order of $2 \pi \times 100$ kHz, was introduced by using two
independently--controlled acousto--optic modulators (AOM's) to
control the frequency of two independent beams\footnote{In our
study of the free--particle regime using counter--propagating
Bragg beams, we alternately used a single AOM to generate a single
beam with two different frequencies which was retro--reflected.
This was done to minimize the deleterious effects of vibrations of
optical components}.

\begin{table}
    \begin{center}
        \begin{tabular}{r|c|c|c}
        & Phonon \cite{stam99phon} & BEC & Free--particle \cite{sten99brag} \\
        \hline
        Velocity & $\hbar q / m = 7$ mm/s & $c = 10$ mm/s &
        $\hbar q / m = 60$ mm/s \\
        Energy & $\hbar \omega_q^0 = h \times 1.5$ kHz & $\mu = h
        \times 6$ kHz & $\hbar \omega_q^0 = h \times 100$ kHz\\
        \end{tabular}
        \caption[Conditions for Bragg scattering in the phonon and the free--particle regimes]{
Conditions for Bragg scattering in the phonon and the free--particle regimes.  The two regimes can
be distinguished by comparing the Bragg scattering velocity $\hbar q / m$ to the speed of
Bogoliubov sound $c = \sqrt{\mu / m}$, or, similarly, the Bragg scattering kinetic energy $\hbar
\omega_q^0 = \hbar^2 q^2 / 2 m$ to the chemical potential $\mu$.  Bragg scattering using beams
inclined by 14$^\circ$ to one another accessed the phonon regime, while the use of
counter--propagating beams accessed the free--particle regime.
        \label{tab:braggcomp}}
    \end{center}
\end{table}

To determine the Bragg scattering response, the laser beams were pulsed on for about 400 $\mu$s.
After the optical excitation, the Bose--Einstein condensate was allowed to freely expand for 20 --
70 ms time of flight, after which the atoms were imaged by absorption imaging.  Bragg scattered
atoms were clearly distinguished by their displacement in the time--of--flight images (see Figs.\
\ref{fig:bragg}, \ref{fig:braggphonon}).  The Bragg resonance spectrum for both trapped and
untrapped Bose--Einstein condensates was determined by scanning the frequency difference $\omega$
and determining the fraction of Bragg scattered atoms.

As discussed above, there are two different regimes of excitations
described by the Bogoliubov theory for the zero--temperature,
weakly--interacting Bose--Einstein condensate: the phonon and the
free--particle regimes. These two regimes can be distinguished in
a number of equivalent ways.  In the phonon regime, the wavevector
of the excitation $q$ is smaller than the inverse healing length
$q \ll \xi^{-1} = \sqrt{2 m \mu / \hbar^2}$.  Equivalently, the
free recoil velocity $\hbar q / m$ is smaller than the speed of
sound $c$. One can also distinguish the free--particle from the
phonon regime by comparing the free recoil energy $\hbar^2 q^2 / 2
m =  \hbar \omega_q^0$ to the condensate interaction energy $\mu$:
$\hbar \omega_q^0 \ll \mu$ in the phonon regime, and $\hbar
\omega_q^0 \gg \mu$ in the free--particle regime.

Either of these types of excitations can be accessed by Bragg
scattering.  In an $N$-th order Bragg scattering event induced by
laser beams of wavevectors $\vec{k}_1$ and $\vec{k}_2$
($|\vec{k}_1| \simeq |\vec{k}_2| = k$), $N$ photons are
transferred from one incident beam to another, imparting a
momentum $\hbar q = 2 N \hbar k \sin(\theta / 2)$, where $\theta$
is the angle between $\vec{k}_1$ and $\vec{k}_2$ . Thus,
small--angle Bragg scattering can be used to excite low--momentum
phonon excitations, while large--angle or high--order Bragg
scattering can excite high--momentum free--particle excitations.

The conditions in our two experiments are shown in Table
\ref{tab:braggcomp}. In our first experimental study of Bragg
scattering, excitations in the free--particle regime were studied
by using two counter--propagating Bragg beams \cite{sten99brag}.
As shown in Table \ref{tab:braggcomp}, the recoil velocity and
energy for an excitation with a momentum of two photon recoils
($\hbar q = 2 \hbar k$) were clearly in the free--particle regime.
In a second experiment, we accessed the phonon regime by using
Bragg beams which were inclined at a small angle of about
14$^\circ$ with respect to one another. This yielded a recoil
velocity and energy which implied that Bragg scattering in a
trapped Bose--Einstein condensate would occur in the phonon
regime.

\subsection{Light scattering in the
free--particle regime}

\subsubsection{Measurement of line shift and line broadening}

\begin{figure}
    \begin{center}
    \includegraphics[height=2.7 in]{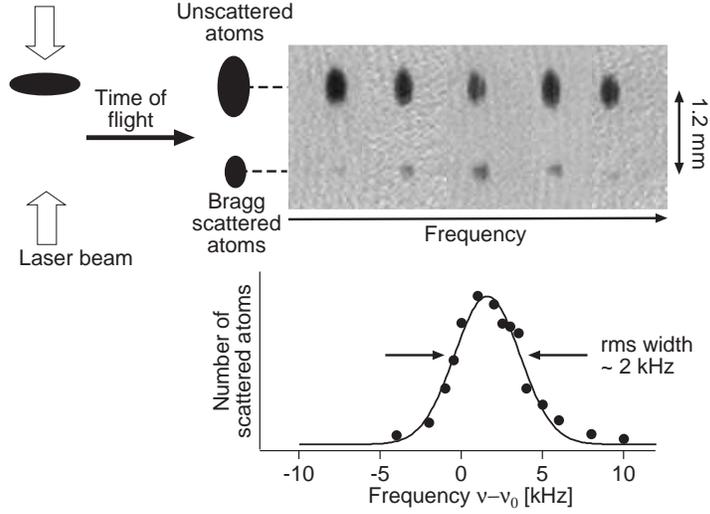}
    \caption[Measuring the momentum distribution of a condensate]{
Measuring the momentum distribution of a condensate\protect\cite{sten99brag}. Atoms were stimulated
by two counter--propagating laser beams to absorb a photon from one beam and emit it into the other
beam, resulting in momentum transfer to the atoms, as observed in ballistic expansion after 20 ms
time of flight.  The number of Bragg scattered atoms showed a narrow resonance when the difference
frequency between the two laser beams was varied.
    \label{fig:bragg}}
    \end{center}
\end{figure}

Fig.\ \ref{fig:bragg} illustrates our experimental method for
probing the free--particle regime.  Counter--propagating Bragg
beams were incident along a radial direction
(vertical in image) of the cigar--shaped condensate.  After
probing, the condensate was allowed to freely expand, allowing
Bragg scattered atoms to separate spatially from the unscattered
atoms, as shown in the figure.  During free expansion, a
cigar--shaped condensate expands primarily radially, with a
maximum radial velocity of $v_r = \sqrt{2 \mu / m} = \sqrt{2} c$.
Thus, since the Bragg scattered atoms are clearly separated from
the remaining condensate during the time of flight, $\hbar q / m
\gg c$ and thus Bragg scattering at a momentum of $\hbar q = 2
\hbar k$  produces free--particle excitations.

\begin{figure}
    \begin{center}
    \includegraphics[height=1.8 in]{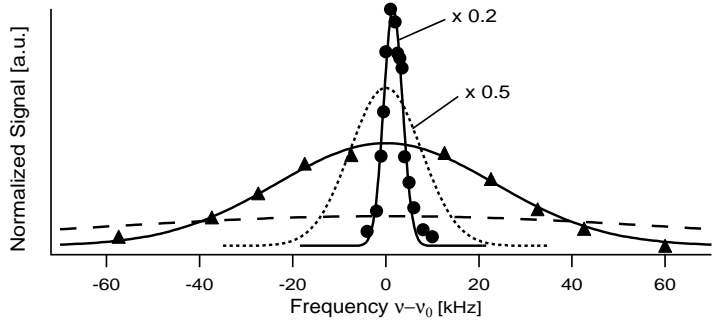}
   \caption{
Bragg resonances for a trapped condensate (circles) and after 3 ms time of flight (triangles).  For
comparison, the momentum distributions of the ground state of the trapping potential (dots) and of
a 1 $\mu$K cold, thermal cloud (dashes) are indicated.
    \label{fig:fpbraggcurves}}
    \end{center}
\end{figure}

\begin{table}
    \begin{center}
        \begin{tabular}{c|c|c|c|c}
        & thermal & ideal & interacting & time--of-- \\
        & cloud & BEC & BEC & flight \\
       \hline
        $\Delta x$ & $a\subrm{osc} \sqrt{k_B T/ \hbar \omega}$ & $a\subrm{osc}$ &
        $a\subrm{osc} \xi$ & \\
        $\Delta p$ & $ (\hbar/ a\subrm{osc}) \sqrt{k_B T/ \hbar \omega}$  & $ \hbar / a\subrm{osc} $ &
        $ (\hbar / a\subrm{osc}) (1/\xi)$ & $ (\hbar/ a\subrm{osc}) \xi$ \\
        $\Delta x \Delta p / \hbar$ & $ k_B T/ \hbar \omega $ & 1 & 1 &
        \end{tabular}
        \caption[Spatial width and momentum width of thermal clouds and Bose-Einstein condensates ]{
Scaling of spatial and momentum widths of thermal clouds and
Bose-Einstein condensates in an isotropic harmonic trap with
frequency $\omega$. The parameter $\xi=\sqrt{\mu / \hbar \omega}$
denotes the strength of the mean-field interaction, and $a\subrm{osc}=\sqrt{\hbar/m \omega}$ the oscillator length.
        \label{tab:widths}}
    \end{center}
\end{table}

The use of Bragg spectroscopy allowed the true momentum distribution of a gaseous Bose--Einstein
condensate to be observed for the first time.  Previous to this work, the onset of Bose--Einstein
condensation had been observed by the observation of a bimodal density distribution either \emph{in
situ}, or else in time--of--flight.  In both cases, the condensate can be distinguished from the
thermal cloud because its energy $\mu$, which is typically dominated by interaction energy, is
smaller than the thermal energy $k_B T$.  \emph{In situ}, a separation between the two components
of the gas occurs due to the inhomogeneous quadratic trapping potential, while in time--of--flight
images, the separation is seen in velocity space after the explosive conversion of the interaction
energy of the condensate to kinetic energy of ballistic expansion; in both cases, the spatial
extent of the gas scales as the square root of its energy. As a result, both in time--of--flight
and in the spatial domain, is the onset of BEC indicated by a narrowing of the density distribution
relative to that of the thermal component by a factor of $\sqrt{\mu /k_B T}$ (Table \ref
{tab:widths}). Indeed, we used Bragg spectroscopy to observe the expanding condensate in momentum
space \emph{after} its interaction energy had been converted to kinetic energy during its time of
flight (Fig.\ \ref{fig:fpbraggcurves}). In a time--of--flight analysis, the condensate distribution
is just $\approx$ 3 times narrower than that of the thermal cloud (corresponding to the factor
$\sqrt{\mu /k_B T}$), but wider than the momentum distribution of the harmonic oscillator ground
state.

\begin{figure}
    \begin{center}
    \includegraphics[height=3.5 in]{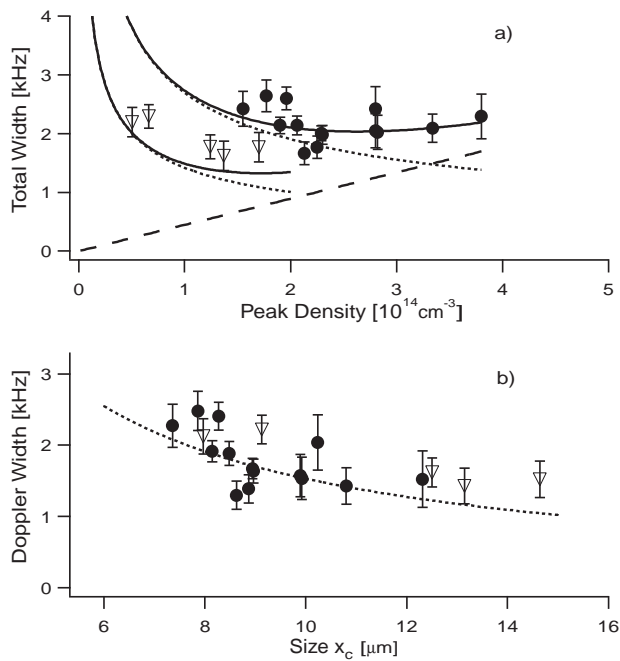}
    \caption[Total width and Doppler width]{
Line width in Bragg spectroscopy.  The widths are shown for various densities and sizes of the
condensate using two different radial trapping frequencies, $\nu_r = (195\pm 20)$ Hz (circles), and
$\nu_r = (95\pm 20)$ Hz (triangles). The lines in a) show the contributions of the mean--field
energy (dashed), due to the finite size (dotted, for both trapping frequencies), and their
quadrature sum (solid lines). Fig.\ b) displays the Doppler width, obtained by subtracting the
contributions of the mean--field and the finite pulse duration from the total width. The observed
Doppler width of the resonance is proportional to the condensate's momentum uncertainty $\Delta p$,
and was consistent with the Heisenberg limit $\Delta p \approx \hbar / x_c $ as the condensate
radius $x_c$ was varied. Thus, the coherence length of the condensate is equal to its physical
size, i.e.\ the condensate is one ``coherent matter wave.'' An analogous measurement in the time
domain has been done in Gaithersburg\protect\cite{hagl99coh}.  The error bars are 1$\sigma$ errors
of the Gaussian fits to the data.
    \label{fig:rmswidths}}
    \end{center}
\end{figure}

However, the \emph{in situ} momentum distribution of the trapped
condensate is much narrower than in ballistic expansion, by a
factor $\xi ^2 = {\mu/ \hbar \omega}$ (Table \ref {tab:widths}).
It is also narrower than the momentum distribution of the harmonic
oscillator ground state.  This could be directly determined by the
use of Bragg spectroscopy \emph{before} releasing the condensate
from its trap. As shown in Fig.\ \ref{fig:fpbraggcurves}, the
momentum distribution of a trapped Bose--Einstein condensate is
significantly narrower than that observed in time--of--flight; for
the data shown, the \emph{in situ} momentum width is 10 times
narrower than in time--of--flight. Thus, the distinction between
the thermal cloud and the condensate in momentum space is much
more stark than in coordinate space and emphasizes the original
description by Einstein of Bose--Einstein condensation as a
condensation in momentum space.

We obtained Bragg resonance spectra for trapped condensates over a range of condensate densities
and radial widths, and determined the shifts of the line center (Fig.\ \ref{fig:s-q}) and the
rms--widths (Fig.\ \ref{fig:rmswidths}) of the resonances.  The shifts of the line center for
trapped condensates from the free--particle resonance frequency were compared with measurements of
the chemical potential determined from the width of condensates in time--of--flight images
\cite{kett99var}.  The measured shift of $(0.54 \pm 0.07) \mu / \hbar$ was in agreement with the
predicted $0.57 \mu / \hbar$.  In related work, mean--field line shifts measured by two--photon
spectroscopy of atomic hydrogen were used as a measurement of the $a_{1S,2S}$ scattering length for
collisions between atoms in the $1S$ and $2S$ states \cite{kill98}.

The rms line widths showed effects of both mean--field broadening
and Doppler broadening: mean--field broadening caused the line
width to increase as the condensate number, and hence its size and
density, were increased, while Doppler--broadening caused the
rms--width to increase as the radial width of the condensate was
decreased. Experimental Bragg spectra for conditions at the
cross--over between mean--field dominated (local density
approximation) and Doppler dominated (impulse approximation)
behaviour are shown in Fig.\ \ref{fig:rmswidths}.  These were
analyzed in Ref.\ \cite{zamb00} and compared with a theoretical
treatment based on an eikonal approximation which accounts for
both effects.

The Doppler line widths were extracted from Bragg spectra by subtracting the broadening due to the
mean--field and the finite pulse duration. Doppler widths were determined for both trapped
condensates and condensates released from the trap. The width for freely--expanding condensates
during the conversion of their interaction energy into kinetic energy was in full agreement with
mean--field theory \cite{cast96}.  Thus, both in momentum space and in coordinate
space\cite{kett99var}, the predicted behaviour of freely--expanding Bose--Einstein condensates has
been quantitatively verified, nicely confirming the mean--field description of large--amplitude
dynamics of a condensate.

\subsubsection{A measurement of the coherence length of a Bose--Einstein condensate}

\label{sec:cohlength}

The Doppler width of the Bragg scattering resonance of trapped Bose--Einstein condensates measured
their momentum distribution and thus measured their coherence length. By observing momentum
distributions which were equal to the Heisenberg--limited momentum distribution determined by the
condensate size (Fig.\ \ref{fig:rmswidths}), our measurements showed that the coherence length of
the condensate was no smaller than the radial width of the condensate, or simply that condensates
can be described by a single macroscopic wavefunction.

\begin{figure}
    \begin{center}
    \includegraphics[height=1.25in]{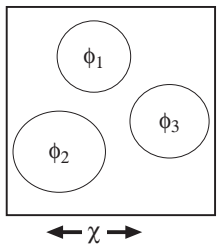}
    \caption[Pictorial depiction of a quasi--condensate]{
A quasi--condensate can be depicted pictorially as being composed of many smaller condensates, each
with a well--defined phase which is unrelated to the phase of the other small condensates. The
range of phase coherence is the coherence length $\chi$.
    \label{fig:quasicond}}
    \end{center}
\end{figure}

To make this statement more quantitative, consider the possibility that the condensate is not fully
coherent. As shown pictorially in Fig.\ \ref{fig:quasicond}, consider a situation in which the
trapped Bose--Einstein condensate is actually composed of many smaller coherent condensates of
typical size $\chi$ with no phase relation between them.  Formally, one considers
\begin{equation}\label{eq:coherencelength}
  \left< \bfod(\vec{r}_1) \bfo(\vec{r}_2) \right> = \left<\bfod(\vec{r}_1) \right> \, \left<
  \bfo(\vec{r}_2) \right> \, g^{(1)}(|\vec{r}_1 - \vec{r}_2|)
\end{equation}
where $g^{(1)}(r)$ is the first--order coherence function which decays from $g^{(1)}(0) = 1$ to
$g^{(1)}(r \gg \chi) = 0$.  The Doppler line shape of the Bragg resonance, shown in Eq.\
\ref{eq:dopplerline2} for the case of full coherence, now becomes
\begin{eqnarray}\label{eq:dopplerincoh}
  \lefteqn{I_D(\omega) d\omega \propto}\\
  & &  \int dx_1 \, dx_2 \, e^{-i k(\omega) (x_2 - x_1)}
  \int dy dz \, \left< \bfod(x_1, y, z) \bfo(x_2, y, z) \right>  \,
  d\omega = \nonumber \\
  & &  \int dx_1 \, dx_2 \, e^{-i k(\omega) (x_2 - x_1)} g^{(1)}(|x_1 - x_2|)
  \int dy dz \, \psi^*(x_1, y, z) \psi(x_2, y, z)   \, d\omega
  \nonumber
\end{eqnarray}

\begin{figure}
    \begin{center}
    \includegraphics[height=1.5in]{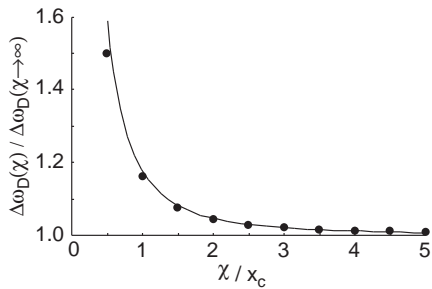}
    \caption[Doppler width of the Bragg resonance as a function of the coherence length]{
Doppler width of the Bragg resonance as a function of the
coherence length $\chi$. Shown is the ratio of the Doppler width
of a partly coherent condensate to that of a fully coherent
condensate, i.e. one for which $\chi \rightarrow \infty$. The
coherence length is given in units of the Thomas--Fermi radius
$x_{c}$.  The results of a numerical calculation of Eq.\
\ref{eq:dopplerincoh} using a Thomas--Fermi wavefunction and
$g^{(1)}(r) =  \exp(-r^2 / 2 \chi^2)$ (points) are well--described
by the approximation of Eq.\ \ref{eq:dopincohtf} (line).
    \label{fig:doppler}}
    \end{center}
\end{figure}

The effect of a limited coherence length is to increase the momentum distribution and thereby
broaden the Doppler width of the Bragg resonance.  For example, consider a condensate wavefunction
which is Gaussian $\psi(\vec{r}) = \exp(-\sum x_i^2 / 2 \sigma_i^2)$ with rms--widths of
$\sigma_i$, ($i = {x,y,z}$), and a first--order correlation function of the form $g^{(1)}(r) =
e^{-r^2 / 2 \chi^2}$.  For Bragg scattering along the $\hat{x}$--direction, the Doppler line width
becomes $\Delta \omega_D(\chi) = \hbar q / m \times \sqrt{ \sigma_x^{-2} + \chi^{-2} }$.  Fig.\
\ref{fig:doppler} shows the Doppler line width $\Delta \omega_D(\chi)$ calculated using the
Thomas--Fermi condensate wavefunction for different values of the coherence length $\chi$. The
calculations agree well with an approximation of the form
\begin{equation}\label{eq:dopincohtf}
  \Delta \omega_D(\chi) \simeq  \sqrt{ [\Delta
  \omega_D(\chi \rightarrow \infty)]^2 + \left[ \frac{\hbar q}{m \chi} \right]^2}
\end{equation}
This relation can be used to extract a coherence length from the Doppler widths for trapped Bose--Einstein
condensates. Taking the average for all data points shown in Fig.\
\ref{fig:rmswidths}b (assuming the ratio of the coherence length to
the condensate size is constant), one finds an average value for
the ratio of the measured Doppler width to that predicted for a
fully coherent condensate of $\Delta \omega_D(\chi) / \Delta
\omega_D(\chi \rightarrow \infty) = 1.09 \pm 0.11$.  This
corresponds to a determination of the coherence length in the
range $\chi =1.4^{+ \infty}_{-0.5} \times x_{c}$, i.e.\ the
measurements place a lower bound on the coherence length of $\chi
\simeq x_{c}$ and are consistent with full coherence ($\chi
\rightarrow \infty$).

The coherence of a condensate was first demonstrated by Andrews {\it et al.}\ by the observation of
matter--wave interference when two independent Bose--Einstein condensates were overlapped \cite{andr97int}. The
regular high--contrast fringes indicated, at least qualitatively, that the condensates possessed
long--range first--order coherence. Our measurements using Bragg scattering provide a more
quantitative measure of this property. Recently, two other groups have made measurements of the
coherence length\cite{hagl99coh,bloc00coh}. These measurements give evidence of a decay of the
coherence length over two length scales.  In particular, the Munich group ascribes the short--range
decay of the first--order coherence function to the presence of thermal excitations.  Instead of
light, one could also use the scattering of fast atoms to determine the coherence length of a
condensate, as suggested recently \cite{kukl99}.

\subsection{Light scattering in the phonon regime}

\subsubsection{Experimental study}

\label{sec:braggphonon}

\begin{figure}
    \begin{center}
    \includegraphics[height=1.3 in]{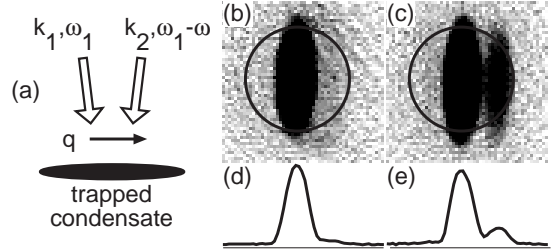}
    \caption[Bragg scattering in the phonon regime]{
Observation of momentum transfer by Bragg scattering. (a) Atoms were exposed to laser beams with
wavevectors $\vec{k_1}$ and $\vec{k_2}$ and frequency difference $\omega$, imparting momentum
$\hbar \vec{q}$ along the axis of the trapped condensate. The Bragg scattering response of trapped
condensates (b,d) was much weaker than that of condensates after a 5 ms free expansion (c,e).
Absorption images (b,c) after 70 ms time of flight show scattered atoms distinguished from the
denser unscattered cloud by their axial displacement.  Curves (d,e) show radially averaged
(vertically in image) profiles of the optical density after subtraction of the thermal
distribution. The Bragg scattering velocity is smaller than the speed of sound in the condensate
(position indicated by circle). Images are 3.3 $\times$ 3.3 mm. Figure taken from Ref.\
\cite{stam99phon}.
    \label{fig:braggphonon}}
    \end{center}
\end{figure}

A second experiment explored the use of Bragg spectroscopy of a Bose--Einstein condensate in the
phonon regime \cite{stam99phon}. Laser beams separated by an angle of about 14$^\circ$ were
directed at a condensate so that the Bragg scattering momentum $\hbar \vec{q}$ was directed along the
condensate axis (Fig.\ \ref{fig:braggphonon}).  Orienting the Bragg scattering momentum in the
axial direction was necessary to allow for a clear distinction between the Bragg scattered atoms
and the unscattered condensate in time--of--flight images, since the condensate expands radially in
at a velocity $v_r = \sqrt{2} c$ which is greater than the recoil velocity $\hbar q / m$ in the
phonon regime ($c > \hbar q / m$). The axial expansion is slower than the radial expansion by a
factor which is $2/\pi$ times the aspect ratio of the cylindrical condensate \cite{cast96}, and was
negligible for our experimental conditions.

We could directly compare Bragg scattering in the phonon and
free--particle regimes by using the identical optical setup to
optically excite either a trapped Bose--Einstein condensate, or a
condensate that had been allowed to freely expand for 5 ms before
the Bragg excitation. As discussed above, excitation in the
trapped condensate occurred in the phonon regime.  As for the
expanded cloud, its density was reduced by a factor of 23 and thus
the speed of sound by a factor of 5 from that of a trapped
Bose--Einstein condensate. Therefore, excitations in expanded
cloud occurred in the free--particle regime.

\begin{figure}[ht]
    \begin{center}
    \includegraphics[height=1.5 in]{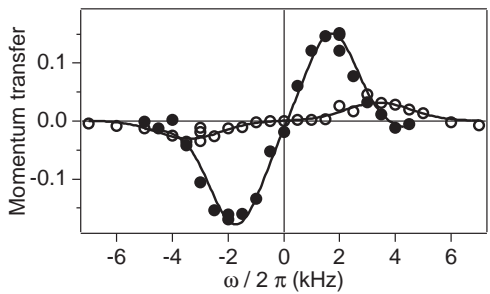}
    \caption{
Bragg scattering of phonons and of free particles. Momentum transfer per particle, in units of
$\hbar q$, is shown vs.\ the frequency difference $\omega / 2 \pi$ between the two Bragg beams.
Open symbols represent the phonon excitation spectrum for a trapped condensate at a chemical
potential $\mu / h = 9.2$ kHz (compared to the free recoil shift of $\approx 1.4$ kHz). Closed
symbols show the free-particle response of an expanded cloud. Lines are fits to the difference of
two Gaussian line shapes representing excitation in the forward and backward directions.}
    \label{fig:phononspectra}
    \end{center}
\end{figure}

\begin{figure}
    \begin{center}
    \includegraphics[height=2.5 in]{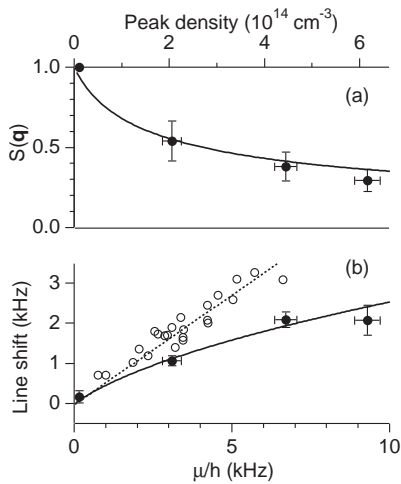} \caption[Line strengths and line shifts]{
Line strengths and line shifts of Bragg resonances. Shown are (a)
the static structure factor $S(\vec{q})$ and (b) the shift of the
line center from the free-particle resonance. $S(\vec{q})$ is the
ratio of the line strength at a given chemical potential $\mu $ to
that observed for free particles. As the density and $\mu$
increase, the structure factor is reduced, and the Bragg resonance
is shifted upward in frequency. Solid lines are predictions of a
local-density approximation for light scattering by 14$^\circ$.
The dotted line indicates a mean-field shift of $4 \mu / 7 h$ as
measured in the free-particle regime using a scattering angle of
180 degrees. Figure taken from Ref.\ \cite{stam99phon}.
    \label{fig:s-q}}
    \end{center}
\end{figure}

As shown in Figs.\ \ref{fig:braggphonon} and
\ref{fig:phononspectra}, the Bragg scattering response in the
phonon regime was significantly weaker in strength and shifted
upwards in frequency from that of free particles as predicted by
Bogoliubov theory (Sec.\ \ref{sec:dynstrucsection}). The line
strength and center shift were measured for condensates at various
densities, as presented in Fig.\ \ref{fig:s-q}.   As the
condensate density is increased, the speed of sound increases and,
at a constant Bragg scattering momentum $\hbar q$, one pushes
further into the phonon regime where the line strength decreases.
We could directly compare our experimental results to the
predictions of the local density approximation (Eq.\
\ref{eq:lineshape}) since the Doppler shift was negligible at
small momentum transfer.  The striking difference from the
free--particle regime demonstrates the collective nature of low
momentum excitations.

The results summarized in Fig.\ \ref{fig:s-q} show that we have confirmed the Bogoliubov dispersion
relation (Fig.\ \ref{bogoliubov}) in both the small-- and large--momentum regime.  Data were taken
at only two values of the momentum transfer, but by varying the condensate density, the ratio of
the recoil energy to the mean--field energy could be smoothly varied.

\subsubsection{Suppression of light scattering from a Bose--Einstein condensate}

The diminished line strength for stimulated light scattering observed in the phonon regime shows
that inelastic spontaneous light scattering (Rayleigh scattering) from a Bose--Einstein condensate
is different from the light scattering from an equal number of non--interacting atoms.  Consider
Rayleigh scattering from a homogeneous Bose--Einstein condensate with a chemical potential $\mu$.
Light scattered at an angle $\theta$ from an incident beam of light with wavevector $\vec{k}$
imparts a momentum of magnitude $\hbar q = 2 \hbar k \sin(\theta/2)$.  The intensity of light
scattered at this angle is diminished by $S(q) = \omega_q^0 / \omega_q^B$. Integrating over all
possible scattering angles $\theta$ and accounting for the dipolar emission pattern, we find that
Rayleigh scattering from a homogeneous interacting Bose--Einstein condensate is suppressed by a
factor
\begin{eqnarray}\label{eq:suppfac}
  F & = & \frac{3}{8 \pi}
  \int d\Omega \, \left( \cos^2 \theta + \sin^2 \theta \sin^2 \phi \right)
  \frac{2 k \sin(\theta/2)}{\sqrt{(2 k \sin(\theta/2))^2 + 2
  \xi^{-2}}}   \nonumber \\
  & = & \frac{1}{64 x^3 \sqrt{1 + 2 x}} \left[ \sqrt{2 x} \left( 15 + 46 x + 64 x^2 + 64 x^3 \right)
   \right.  \nonumber\\
  & & \left.- 3
  \sqrt{1 + 2 x} \left( 5 + 12 x + 16 x^2 \right) \tanh^{-1}
  \left( \sqrt{\frac{2 x}{1 + 2 x}} \right) \right]
 \end{eqnarray}
where $x = \hbar \omega_{k}^{0} / \mu$.

\begin{figure}
    \begin{center}
    \includegraphics[height=1.5in]{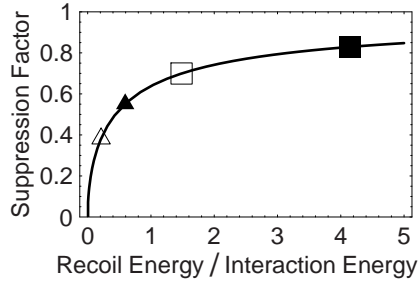}
    \caption[Suppression of Rayleigh scattering from a Bose--Einstein condensate]{
Rayleigh scattering from a homogeneous interacting Bose--Einstein
condensate is suppressed with respect to scattering from free
atoms.  The suppression factor is plotted against the ratio of the
recoil energy $\hbar \omega_k^0 = \hbar^2 k^2 / 2 m$ to the
interaction energy $\mu$, where $k$ the wavevector of the incident
light. Points indicate the factor by which the scattering of
near--resonant light at $\lambda = 589$ nm (closed symbols) and
infrared light at $\lambda = 985$ nm (open symbols) is suppressed
at typical condensate densities attained in our magnetic ($n = 4
\times 10^{14} \, \mbox{cm}^{-3}$, squares) and optical ($n = 3
\times 10^{15} \, \mbox{cm}^{-3}$, triangles) traps.
    \label{fig:suppress}}
    \end{center}
\end{figure}

This suppression in Rayleigh scattering should be observable in current Bose--Einstein condensation
experiments (Fig.\ \ref{fig:suppress}). For example, Rayleigh scattering of near--resonant light
from a sodium condensate at a density of $3 \times 10^{15} \, \mbox{cm}^{-3}$, which is the maximum
density of condensates which has been obtained in optical traps \cite{stam98odt}, should be reduced
by a factor of two.  It would also be interesting to measure this suppression of light scattering
at a Feshbach resonance \cite{inou98}, where the chemical potential can be made quite large by
tuning the scattering length $a$ using magnetic fields. In such experiments, light scattering may
allow one to study dynamically how the pair correlations in a condensate are established in
response to a sudden increase in the interaction strength.

\section{Amplified scattering of light}

\subsection{Introduction}
In the previous section, we discussed the elementary process in
which a photon is inelastically scattered from a condensate and
creates a quasi-particle. We focused on the limit of low intensity
of the probe light which means that each photon is assumed to
probe the ``original condensate.'' This is valid as long as the
quasi-particles created by photon scattering decay or decohere
fast enough so that the optical properties of the condensate don't
change.

However, the low temperature and coherent nature of the condensate
imply very long decoherence times, leading to long-lived
quasi-particles which can affect the scattering of subsequent
photons. The presence of long-lived excitations introduces strong
correlations between successive Rayleigh scattering events
\cite{sait99}.  This provides a positive feedback mechanism and
leads to directional amplified Rayleigh scattering.

This non-linear behavior happens already at very low laser intensity. The amplification of the
scattering process is stimulated by the matter-wave field analogous to stimulated emission in the
optical laser. When this amplification process starts spontaneously, a new form of superradiance is
realized (Sect.\ \ref{sec:superradiance}).  When it is initiated by seeding the condensate with an
input matter wave, it leads to coherent atom amplification (Sect.\ \ref{sec:amplification}).

\subsection{Superradiant Rayleigh scattering}
\label{sec:superradiance}

\subsubsection{Semiclassical derivation of the gain mechanism}
\label{sec:semiclassical derivation}

The gain mechanism for Rayleigh scattering from a condensate can
be derived semi-classically.  When a condensate of $N_{0}$ atoms
is exposed to a laser beam with wavevector $\vec{k}_{0}$ and
scatters a photon with wavevector $\vec{k}_{i}$, an atom (or
quasi-particle, also called momentum side-mode in
\cite{moor99pra}) with recoil momentum $\hbar \vec{K}_{j}= \hbar
(\vec{k}_{0}-\vec{k}_{i})$ is generated. Since light propagates at
a velocity about ten orders of magnitude greater than the atomic
recoil velocity (3 cm/s for sodium), the recoiling atoms remain
within the volume of the condensate long after the photons have
left and affect subsequent scattering events. They interfere with
the condensate at rest to form a moving matter wave grating of
wavevector $\vec{K}_{j}$, which diffracts the laser beam into the
phase-matching direction $\vec{k}_{i} \;
(=\vec{k}_{0}-\vec{K}_{j})$. This diffraction is a self-amplifying
process because every diffracted photon creates another recoiling
atom which further increases the amplitude of the matter wave
grating.

When $N_j$ recoiling atoms interfere with $N_0$ condensate atoms, the density modulation comprises
$N\subrm{mod}=2 \sqrt{N_0 N_{j}}$ atoms.  The light scattered by these atoms interferes
constructively in the phase-matching direction with a total power $P$ of

\begin{eqnarray}
    P &=& \hbar \omega \;f_{j} \;R \; \frac{N\subrm{mod}^{2}}{4} \; ,
    \label{power}
    \\ f_{j} &=& \frac{\sin^{2} \theta_{j}}{8\pi/3} \;\Omega_{j} \; .
    \label{eq:super semiclass}
\end{eqnarray}

Here, $R$ is the rate for single-atom Rayleigh scattering which is proportional to the laser
intensity, and $\omega$ is the frequency of the radiation. The angular term in Eq.\ (\ref{eq:super
semiclass}) reflects the dipolar emission pattern with $\theta_{j}$ being the angle between the
polarization of the incident light and the direction of emission. Due to the finite size of the
sample, the phase matching condition is fulfilled over the solid angle $\Omega_{j}\sim
\lambda^2/A$, where $A$ is the cross-sectional area of the condensate perpendicular to the
direction of the light emission and $\lambda$ the optical wavelength.  More rigorously,
$\Omega_{j}$ is given by the usual phase-matching integral for superradiance in extended samples
\cite{rehl71}:

\begin{equation}
    \Omega_{j} = \int d \Omega ({\vec{k}})
    \left|
    \int \tilde{\rho}({\vec{r}})
    \exp^{i ({\vec{k}}_{i}- {\vec{k}}) \cdot {\vec{r}}} d{\vec{r}}
    \right|^{2} \; ,
\end{equation}

where $|\vec{k}|=|\vec{k}_{i}|$, and $\tilde{\rho}({\vec{r}})$ is the normalized density
distribution of the condensate ($\int \tilde{\rho}({\vec{r}}) d{\vec{r}} =1$).

Since each scattered photon creates a recoiling atom, we obtain the growth rate for $N_{j}$ from
$P/\hbar \omega$:

\begin{equation}
    \dot{N_{j}}= G_j N_{j}  \; .\label{eq:class}
\end{equation}

This equation predicts exponential growth of $N_{j}$ with the
small-signal gain $G_j = R N_0 f_j \sim R\, \sin^{2} \theta_{j}
\,D_j$, where  $D_j \sim \rho_0 \, \lambda^2 \,l_{j}$ is the
resonant optical density for a condensate with an atomic density
$\rho_0$ and a length $l_{j}$ along the axis of emission.
Therefore, for an anisotropic Bose condensate, the gain is largest
when the light is emitted along its longest axis (the ``end-fire
mode'' \cite{dick64}).  Eq.\ (\ref{eq:class}) is valid in the
absence of decoherence and predicts the build-up of highly
anisotropic Rayleigh scattering from a non-spherical sample of
atoms.

The built-up of a matter-wave grating dramatically changes the optical properties of the condensate
--- the condensate becomes ``reflective.''  This property is only due to the coherent nature of the
condensate.  This phenomenon is completely different from the
reflectivity caused by polaritons which depends on the
dipole-dipole interaction of excited atoms and requires high
atomic densities \cite{poli91,svis90}.

\subsubsection{Four-wave mixing of light and atoms}

Further insight is obtained by describing the scattering process
fully quantum-mechanically.  As discussed in Sect.\
\ref{sec:scattering_basics}, the Hamiltonian $\mathcal{H}^\prime$
describes light scattering as a coupling (i.e.\ a four--wave
mixing) between two Schr\"odinger (matter) waves and two
electromagnetic waves. It is convenient to rewrite
$\mathcal{H}^\prime$ by choosing to quantize the recoiling atoms
 in the volume of the condensate and use plane
waves for the light.  The important terms in $\mathcal{H}^\prime$
include the macroscopically occupied initial atom field
$\hat{a}_0$(the condensate) and electromagnetic field $\hat{c}_0$
(the incident laser beam with wavevector $\vec{k}_0$), i.e.\
$\mathcal{H}^\prime \simeq C_{i,j} \hat{c}^\dagger_i
\hat{a}^\dagger_j \hat{c}_0 \hat{a}_0$ where $\vec{k}_i$ is the
wavevector of the outgoing light, and $\hat{a}^\dagger_j$ is the
creation operator for an atom in the final state $j$.

The square of the matrix element which describes the creation of
recoiling atoms is proportional to $N_j+1$ reflecting bosonic
stimulation in the atomic field, where $N_j$ is the number of
atoms in the final state $j$. We use Fermi's Golden Rule to sum
over all final states of the electromagnetic field which are
assumed to be initially empty (spontaneous scattering) and obtain
the growth rate of $N_{j}$ as
\begin{eqnarray}
    \dot{N_{j}} &=& G_j (N_{j}+1) \label{eq:quant}
    \\ \nonumber &=& R \; N_{0} \;\frac{\sin^{2} \theta_{j}} {8\pi/3} \;\Omega_{j} \; (N_{j}+1)\;,
\end{eqnarray}
which is identical to the classical treatment of Sect.\ \ref{sec:semiclassical derivation} except
that $N_{j}$ is replaced by $N_{j}+1$.  The solid angle factor $\Omega_j$ now reflects the number
of electromagnetic plane wave modes $i$ which are excited together with a quasiparticle in state
$j$ localized in the volume of the condensate. Energy conservation requires that the frequency of
the scattered light be red-shifted with respect to the incident radiation by the kinetic energy of
the recoiling atoms. Eq.\ (\ref{eq:quant}) describes both normal Rayleigh scattering at a constant
total rate $\Sigma \dot{N}_j = R N_{0}$ when $N_{j} \ll 1$, and exponential gain of the $j$-th
recoil mode due to bosonic stimulation once $N_{j}$ becomes non-negligible.  Initially, the angular
distribution of the scattered light follows the single-atom spontaneous (dipolar) emission pattern,
but can become highly anisotropic when stimulation by the atomic field becomes important.

The choice of plane waves for the scattered light is arbitrary. Another basis set are spherical
waves \cite{dick64}.  For the situation considered here it is more convenient to choose orthogonal
modes for the different directions of light emission which have diffraction limited beam waists
$A_j$ identical to the cross section of the condensate perpendicular to the axis of light emission
(see Ref.\ \cite{dick64}, page 45). Therefore, for scattering into a specific direction, one has to
consider only the longitudinal ``beam waist modes'' along the direction of scattering, and their
wavevector is determined by energy conservation. Of course, the final result is the same (Eq.\
\ref{eq:quant}). The solid angle factor $\Omega_{j} \propto 1/A_j$ appears in this derivation
because the energy density of a single photon in the beam waist is inversely proportional to $A_j$.

One can account for decoherence and losses by adding a damping
term to Eq.\ \ref{eq:quant}.
\begin{equation}
    \dot{N_{j}} = G_{j} (N_{j}+1) -D_j N_j \; .
    \label{eq:loss}
\end{equation}
As we discuss below, this damping term reflects the finite linewidth of the Bragg resonance and is
necessary to account for the observed threshold behavior.

\subsubsection{Bosonic stimulation by scattered atoms or scattered light?}
\label{sec:cavity}

In the previous discussion, we evaluated coupling matrix elements
assuming that the atoms scatter light always into empty modes of
the electromagnetic field, and thus found that bosonic stimulation
by the occupied atomic recoil mode was responsible for
superradiant light scattering.  If the time between light
scattering events from the condensate is much longer than the time
it takes the scattered photons to leave the condensate, this
assumption is clearly justified.  But suppose that the Rayleigh
scattering rate were increased so that the mean number of
scattered photons in the condensate were non--zero.  What effect
would this have on the observed superradiance?  Does the number of
photons occupying the scattered light mode lead to additional
bosonic stimulation?

To answer this question and, in particular, to clarify the
importance of the decay times of the atom and optical modes, let
us consider light scattering into just one selected
electromagnetic mode which is defined as the mode of a cavity
built around the condensate. In the so-called bad cavity limit, we
will obtain the results of the previous section.

The two modes of the light and the atoms are labeled as before and
coupled by the operator
\begin{equation}\label{eq:couplingagain}
  \mathcal{H}^\prime = C \hat{c}^\dagger_{i} \hat{a}^\dagger_j
\hat{c}_0 \hat{a}_0 + {\rm h.c.}
\end{equation}
where $C$ denotes the strength of the coupling. In the interaction
picture, we obtain coupled equations of motion for the mode
operators
\begin{eqnarray}
    \frac{\mathrm{d}{\hat{a}_{j}}} {\mathrm{d} t} &=& - \frac{D_a}{2} \; \hat{a}_j -
    (\mathrm{i} C \hat{a}_0 \hat{c}_0) \hat{c}_i^\dagger
    \label{eq:coupled eq1}
    \\ \frac{\mathrm{d}\hat{c}_i^\dagger} {\mathrm{d} t} &=&
    (\mathrm{i} C \hat{a}_0^\dagger \hat{c}_0^\dagger) \hat{a}_j -
    \frac{D_c}{2} \; \hat{c}_i^\dagger
    \label{eq:coupled eq2}
\end{eqnarray}
$D_a$ and $D_c$ are the damping rates for the number of atoms in
mode $j$ and for the number of photons in mode $i$. Similar
coupled equations were considered in Refs.\
\cite{law98amp,moor99pra}. If the scattered light is strongly
damped we can adiabatically eliminate it by setting the l.h.s.\@
of Eq.\ \ref{eq:coupled eq2} to zero
\begin{equation}
    \hat{c}_i^\dagger = \frac{2 \mathrm{i} C}{D_c} \; \hat{a}_0^\dagger \hat{c}_0^\dagger \hat{a}_j
    \label{eq:ad elim}
\end{equation}
Inserting this expression into Eq.\ \ref{eq:coupled eq1} leads to
a gain equation for the atomic field operator
\begin{equation}
    \frac{\mathrm{d}{\hat{a}_{j}}} {\mathrm{d} t} =
    \frac{1}{2} \left[ \frac{4|C|^2}{D_c} \; N_0 n_0 -D_a \right] \hat{a}_j
    \label{eq:gain-eq}
\end{equation}

The field operators for the initial states were replaced by the square root of the occupation
numbers $N_0$ and $n_0$ for the atoms in the condensate and the photons in the laser beam which
leads to an expression for the gain $G = 4(|C|^2/D_c) N_0 n_0$.

Alternatively, we can find the eigenvalues $\lambda _{1,2}$ of the matrix in Eqs.\ \ref{eq:coupled
eq1} and \ref{eq:coupled eq2} as
\begin{equation}
    \lambda _{1,2}= -\frac{D_a + D_c}{4} \pm \sqrt{ {\left( \frac{D_a - D_c}{4} \right)}^2 + C^2 N_0
    n_0}
    \label{eq:eigenvalues}
\end{equation}

For $C=0$, the two eigenvalues are $-D_a/2$ and  $-D_c/2$.  When
$D_c\gg D_a,~G$, the two eigenvalues are $-D_c/2$ and $\frac{1}{2}
\left[ (4|C|^2/D_c) \; N_0 n_0 - D_a \right]$; this is the result
obtained in Eq.\ \ref{eq:gain-eq}.

Eqs.\ \ref{eq:coupled eq1} and \ref{eq:coupled eq2} can be applied to the free--space situation
discussed in the previous section where the optical modes $i$ are the beam-waist modes quantized in
a longitudinal box of length $L$ and mode density $L/c$. The rate of spontaneous emission into the
beam waist mode is given by Fermi's golden rule and is proportional to the square of the matrix
element times the mode density $|C|^2 N_0 n_0 (L/c)$ and is equivalent to Eq.\ \ref{eq:quant}.

On resonance, a cavity of finesse $F$ enhances the density of states by the factor $8 F/\pi$
\cite{moi83}.  The cavity walls create mirror images of the radiating atoms which enhance the
scattering rate.  The number of effective mirror images is proportional to the finesse of the
cavity. Therefore, the gain in Eq.\ \ref{eq:gain-eq} is proportional to the rate of spontaneous
emission into the solid angle $\Omega_j$ and to the finesse of the cavity.
\begin{equation}
    G=R \; N_{0} \;\frac{\sin^{2} \theta_{j}} {8\pi/3} \;\Omega_{j} \; \frac{8 F}{\pi}
    \label{eq:cavity gain}
\end{equation}
The free-space limit is retrieved by setting the finesse to about unity ($F=\pi/8$) which cancels
the last factor in Eq.\ \ref{eq:cavity gain}.  Or equivalently, free space corresponds to the limit
of a cavity with a damping time of about $L/c$.

This result is useful to discuss bosonic stimulation by the matter wave field and the optical
field.  The operator describing light scattering (Eq.\ \ref{eq:couplingagain}) has a matrix element
squared of $|C|^2 (N_j+1) (n_i+1) N_0 n_0$. Since this expectation value is proportional to
$(N_j+1) (n_i+1)$, it seems that there is bosonic stimulation by both the final atomic state
(occupied by $N_j$ atoms) and the final optical state (occupied by $n_i$ photons). However, Eqs.\
\ref{eq:gain-eq} and \ref{eq:cavity gain} show that this notion is incorrect. After adiabatic
elimination of the rapidly--damped scattered light we have only stimulation by the atomic field.
The superradiant gain $G$ is increased by increasing the finesse of the optical cavity, thus
increasing the decay time of the scattered light.  However, the gain does not explicitly depend on
the number of photons in this mode.

Since the coupled equations have complete symmetry between light and atoms, one can also discuss
the complementary case where the atoms are rapidly damped and are adiabatically eliminated. In this
case, one obtains a gain equation for the scattered light showing optical stimulation, but not
stimulation by the matter wave field in the final state.  This situation applies to the optical
laser and also to stimulated Brioullin scattering (see also Sect.\ \ref{sec:non-linear phenomena}).

On first sight it appears counterintuitive that  bosonic stimulation is not multiplicative if we
have a large occupation in both the final photon state and the final atomic state.  However, when
we have occupancy in all final states, the scattering can go both ways.  Thus, let us consider the
situation in which there are $N_j$ recoiling atoms and $n_i$ scattered photons in the volume of the
condensate, and look at time scales much longer than the coherence time,  i.e.\ a situation in
which rate equations apply.  The process $|n_0, n_i ; N_0, N_j\rangle \rightarrow | n_0-1, n_i+1;
N_0-1, N_j+1 \rangle$ by which atoms scatter out of the condensate into the recoil mode $j$ occurs
with a rate proportional to the square of the matrix element  $|C|^2 (N_j+1) (n_i+1) N_0 n_0$. The
process $|n_0, n_i; N_0, N_j\rangle \rightarrow |n_0+1, n_i-1; N_0+1, N_j-1 \rangle$ by which atoms
scatter \emph{back into} the condensate has a rate proportional to $|C|^2 N_j n_i (N_0+1) (n_0+1)$.
The \emph{net} rate of scattering atoms from the condensate into mode $j$ is the difference of the
two partial rates. If we assume $N_0, N_j, n_0 \gg 1$ and also much larger than $n_i$, since the
scattered light was assumed to suffer the strongest damping, we obtain the leading term in the net
scattering rate to be proportional to $|C|^2 N_j N_0 n_0$ as before. The bosonic stimulation by the
\emph{least} populated field (mode $i$) in the matrix element dropped out when the net rate was
calculated.

This finding should be generally applicable.  For example, when two condensed atoms form molecules
by photoassociation or near a Feshbach resonance, the coupled equations involve a matter wave field
for the atoms and for the molecules \cite{wyna00,abel99} which leads to bosonic stimulation in the
coupling matrix element and therefore in the partial rate.  However, molecules are the most rapidly
damped mode.  Therefore the net rate of molecule formation is independent of bosonic stimulation by
the particles in the final state and can not distinguish between a coherent molecular matter wave
field and an incoherent classical ensemble of molecules in the final state.

Eq.\ \ref{eq:coupled eq1} also describes Bragg scattering where a condensate is exposed to two
strong laser beams in modes $0$ and $i$.  This results in Rabi oscillations of the atoms between
states $0$ and $j$.  For short times $t$, one has $N_j \propto N_0 n_0 n_i t^2$.  The scattering
rate $\dot{N_j}$ is now proportional to $N_j^{1/2}$ which is incompatible with the simple concept
of bosonic stimulation. The simple picture of bosonic stimulation is only obtained when we are in
the regime where rate equations apply. This example shows that one has to be careful not to
over-interpret the intuitive concept of bosonic stimulation.  It can be misleading to use this
concept in the case of coherently coupled fields.  It can be applied to the situation where one of
the fields (either the atomic or the optical one) is strongly damped, but then there is bosonic
stimulation of the net scattering rate \emph{only} by the long-lived field.

\subsubsection{Observation of directional emission of light and atoms}

\begin{figure}
    \begin{center}
    \includegraphics[height=2.5 in]{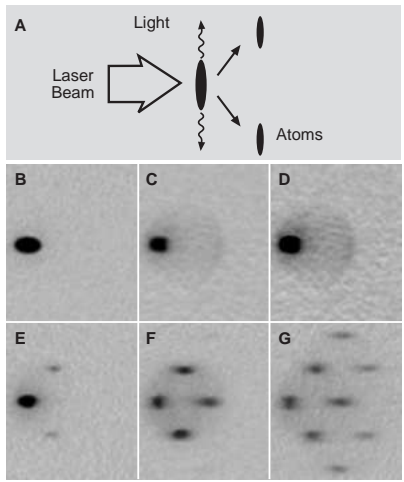} \caption[Observation of superradiant Rayleigh scattering]{
Observation of superradiant Rayleigh scattering. ({\bf A}) An elongated condensate is illuminated
with a single off-resonant laser beam. Collective scattering leads to photons scattered
predominantly along the axial direction, and atoms at 45 degrees. ({\bf B-G}) Absorption images
after 20 ms time-of-flight show the atomic momentum distribution after their exposure to a laser
pulse of variable duration. When the polarization was parallel to the long axis, superradiance was
suppressed, and normal Rayleigh scattering was observed (B-D). For perpendicular polarization,
directional superradiant scattering of atoms was observed (E-G), and evolved to repeated scattering
for longer laser pulses (F,G). The pulse durations were 25 (B), 100 (C,D), 35 (E), 75 (F), 100
$\mu$s (G). The field of view of each image is 2.8 $\times$ 3.3 mm. The scattering angle appears
larger than 45 degrees due to the angle of observation. All images use the same gray scale except
for (D), which enhances the small signal of Rayleigh scattered atoms in (C). Reprinted with
permission from Ref.\ \protect\cite{inou99super}, copyright 1999 American Association for the
Advancement of Science.
    \label{fig:super1}}
    \end{center}
\end{figure}

For the experimental study of directional Rayleigh scattering, elongated Bose--Einstein condensates
were prepared in a magnetic trap \cite{mewe96bec}. The trapped condensates were approximately $20
\,\mu{\rm m}$ in diameter $d$ and $200 \,\mu{\rm m}$  in length $l$ and contained several million
sodium atoms in the $F=1$ hyperfine ground state.  The condensate was exposed to a single
off-resonant laser pulse which was red-detuned by 1.7 GHz from the $3S_{1/2}, F=1 \rightarrow
3P_{3/2}, F=0,1,2 $ transition. The beam had a diameter of a few millimeters, propagated at an
angle of 45 degrees to the vertical axis and intersected the  elongated condensate perpendicular to
its long axis (Fig.\ \ref{fig:super1}). Typical laser intensities were between 1 and  $100\,{\rm
mW/cm}^{2}$ corresponding to Rayleigh scattering rates of 45 to $4500\ {\rm s}^{-1}$, and the pulse
duration between 10 and 800 $\mu$s. In order to probe the momentum distribution of scattered atoms,
the magnetic trap was suddenly turned off immediately after the light pulse, and the ballistically
expanding cloud was imaged after 20 to 50 ms time-of-flight using resonant probe light propagating
vertically onto a CCD camera.

The momentum distributions of atoms after light scattering (Fig.\ \ref{fig:super1} \ B-G) showed a dramatic
dependence on the polarization of the incident laser beam. For polarization parallel to the long
axis of the elongated condensate ($\theta_{j} =0$), light emission into the end-fire mode was
suppressed, and the distribution of atoms followed the dipolar pattern of normal Rayleigh
scattering. For perpendicular polarization ($\theta_{j} = \pi/2$), photons were predominantly
emitted along the long axis of the condensate where the superradiant gain was the largest. The
recoiling atoms appeared as highly directional beams propagating at an angle of 45 degrees with
respect to this axis.

\begin{figure}
    \begin{center}
    \includegraphics[height=1.8 in]{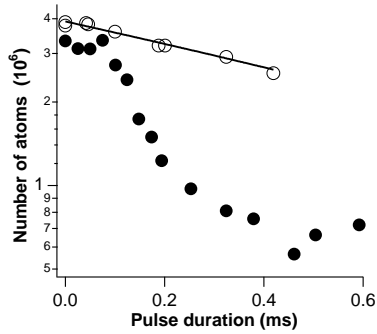} \caption[Exponential and superradiant decay]{
Exponential and superradiant decay. The decay of atoms in the condensate at rest showed the normal
exponential decay for parallel polarization (open circles) and faster superradiant decay for
perpendicular polarization (full circles).  The laser intensities (13 ${\rm mW/cm}^{2}$) and
oscillator strengths were equal in both cases. Reprinted with permission from Ref.\
\protect\cite{inou99super}, copyright 1999 American Association for the Advancement of Science.
    \label{fig:super2}}
    \end{center}
\end{figure}

A characteristic feature of superradiance is an accelerated decay of the initial state. In our
experiment, normal exponential decay and the superradiant decay could be directly compared by
tracing the number of atoms remaining in the condensate at rest after exposure to light of
different polarizations. For parallel polarization, we observed a simple exponential decay with the
expected Rayleigh scattering rate (Fig.\ \ref{fig:super2}). For perpendicular polarization, the end-fire mode was
active and the condensate decayed non-exponentially with a strongly accelerated superradiant rate.

\begin{figure}
    \begin{center}
    \includegraphics[height=3.5 in]{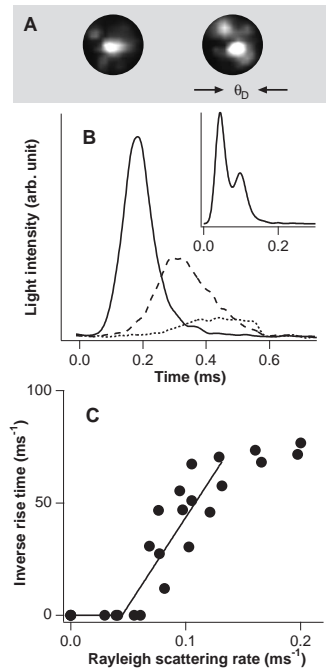}
    \caption[Observation of directional emission of light]{ Observation of directional emission of
light. ({\bf A}) The angular pattern of the emitted light along the axial direction showed a few
bright spots with an angular width $\theta_{D}$ ($1/e^{2}$ diameter) of $107 \pm 20$  mrad,
corresponding to the diffraction limited angle of an object of $\sim$ 14 $\mu$m in diameter. The
images were integrated over the entire duration of the light pulse. ({\bf B}) The temporal
evolution of the light intensity showed a strong initial increase characteristic of a stimulated
process. For higher laser power, the pulse was shorter and more intense. The laser intensities were
3.8 (solid line), 2.4 (dashed line), and ${\rm 1.4\, mW/cm^{2}}$ (dotted line), and the duration
was $550\, \mu$s. The inset shows a double-peak in the temporal signal when the laser intensity was
about $15\,{\rm mW/cm}^{2}$, which was above the threshold for sequential superradiant scattering.
The photomultiplier recorded the light over an angle of 200 mrad around the axial direction. ({\bf
C}) The dependence of the inverse initial rise time on the Rayleigh scattering rate shows a
threshold for the stimulated process. The solid curve is a straight-line fit. Reprinted with
permission from Ref.\ \protect\cite{inou99super}, copyright 1999 American Association for the
Advancement of Science.
    \label{fig:super3}}
    \end{center}
\end{figure}

The directional scattering of atoms implies that the light is scattered preferentially into the
axial direction.  This was verified by observing the scattered light with a CCD camera (Fig.\
\ref{fig:super3} A). The camera was positioned out of focus of the imaging system, so that the
images represent the angular distribution of photons emitted around the axial direction. The images
consisted of bright spots with angular widths equal to the diffraction limit for a source with  a
diameter $\sim 14 \, \mu {\rm m}$. Typical images showed more than one such spot, and their pattern
changed randomly under the same experimental conditions.  The observation of a few spots is
consistent with a Fresnel number $F=\pi d^2/ 4 l \lambda$ slightly larger than one, implying that
the geometric angle $d/l$ is larger than the diffraction angle $\lambda/d$.  $F>1\;$ leads to
multimode superradiance \cite{vreh82} since there is now more than one end-fire mode.

By replacing the camera with a photomultiplier, a time-resolved measurement of the scattered light
intensity was obtained (Fig.\ \ref{fig:super3} B). Simple Rayleigh scattering would give a constant signal during
the square-shaped laser pulse. Instead, we observed a fast rise and a subsequent decay consistent
with a stimulated process.  Measurements at variable laser intensities showed a threshold for the
onset of superradiance, and a shorter rise time for higher laser intensities. This behavior can be
accounted for by Eq.\ \ref{eq:loss}.

We determined the exponential rate $(G_{j} - D_{j})$ by fitting the initial rise in the light
intensity. At early times the depletion of the condensate is negligible and $G_{j}$ and $D_{j}$ are
constants. Fig.\ \ref{fig:super3} C shows the inverse rise time $\dot{N}/N$ vs.\ the Rayleigh scattering rate $R$,
which was measured by ``switching off'' the superradiance using parallel polarization. The slope
gives $G_{j}/R$ and the offset determines the loss $L_{j}$. The agreement between the calculated
value for $G_{j}/R \sim 890$ (using $\Omega_j \sim 1.9 \times 10^{-4}$ and $N_0=4.7 \times 10^{6}$)
and the result of the simple linear fit (790) is better than the uncertainty in the Rayleigh
scattering rate (40\%). The offset in Fig.\ \ref{fig:super3} C determines the threshold for superradiance and
yields $1/D_{j}= 35 \,\mu s$.

The rate of decoherence $D_{j}$  for the superradiance indicates the decay of the matter wave
interference. This has been studied separately using stimulated Rayleigh scattering (or Bragg
spectroscopy) \cite{sten99brag}, where the linewidth of the Bragg resonance resulted from Doppler and
mean-field broadening. The observed FWHM of approximately 5 kHz yields a decoherence time of
$32\,\mu$s, in good agreement with the value shown above.

For higher laser powers, a distinct change in both the momentum pattern of the atoms (Fig.\
\ref{fig:super1} F,G) and in the photomultiplier traces (Fig.\ \ref{fig:super3}B) was observed.  The atomic
pattern showed additional momentum peaks which can be explained as a sequential scattering process
where atoms in the initial momentum peak undergo further superradiant scattering. These processes
are time-delayed with respect to the primary process, and showed up as a second peak in the
time-resolved photomultiplier traces (Fig.\ \ref{fig:super3}B).  Those higher-order peaks may also be
affected by stimulated Raman scattering and four-wave mixing between matter waves \cite{deng99}
which couple the different recoil modes.

\subsubsection{Relation to other non-linear phenomena}
\label{sec:non-linear phenomena} The directional emission of light
and atoms described in the previous sections is analogous to the superradiance discussed by Dicke
\cite{dick64}.  He considered an elongated radiating system of incoherently electronically excited
atoms, and showed that this system realizes ``a laser which does not employ mirrors in order to
produce feedback amplification.'' The amplification of this ``Coherence brightened laser'' is
provided by electronic coherence: ``The memory of the previously emitted electromagnetic field is
burned into the radiating system rather than being sent back into the radiating system by the use
of mirrors \cite{dick64}.'' The key feature of superradiance (or
superfluorescence)\cite{skri73,boni75,shuu81,gros82,vreh82} is that spontaneous emission is not a
single-atom process, but a collective process of all atoms, leaving the atoms in a coherent
superposition of ground and excited states \cite{dick54}. The condensate at rest ``pumped'' by the
off-resonant laser corresponds to the electronically excited state in the Dicke case. It can decay
by a spontaneous Raman process to a state with photon recoil (corresponding to the ground state).
The rate of superradiant emission in Dicke's treatment is proportional to the square of an
oscillating macroscopic dipole moment.  In the present case, the radiated intensity is proportional
to the square of the contrast of the matter wave interference pattern between the condensate and
the recoiling atoms.  In both cases, the initial emission of light shows the single atom dipole
pattern. Quantum noise and spontaneous emission create spatial coherence and lead to directional
emission different from forward scattering.

The situation of an atom cloud with a small excited state admixture (``dressed'' condensate) nicely
demonstrates the analogy between the optical laser and the atom laser.  If the emitted light is
allowed to build up in a cavity, an optical laser is realized (called the Coherent Atomic Recoil
Laser (CARL) \cite{boni94,moor98,berm99}. If the matter wave field builds up in a cavity, it
realizes an atom laser by optical pumping \cite{olsh96,wise95atla,jani96}. In our experiment, the
matter wave field builds up in a traveling wave even without a cavity because of the slow motion of
the recoiling atoms.

For smaller momentum transfer, the nature of the quasi-particles in the condensate will change from
recoiling atoms to phonons. The observed phenomenon is therefore related to stimulated Brillouin
scattering which was called ``phonon maser action'' \cite{chia64} due to the amplification of sound
waves.   However, the analogy is closest when the roles of light and atoms are exchanged.  In
Brillouin scattering, the optical field builds up, whereas the phonon field generated by the
momentum transfer is usually rapidly damped and adiabatically eliminated from the equations. In our
case, the matter wave field builds up and the optical field can be eliminated due to the fact that
light propagates at a velocity which is ten orders of magnitude faster than the atomic recoil
velocity.

The observed phenomenon differs from other stimulated processes
like Brillouin scattering and Dicke superradiance in two regards.
In those cases, the electromagnetic field is treated either
classically,  or spontaneous emission (``quantum noise'') is
needed only to  initiate the process, which then evolves
classically after many photons have been accumulated.  In our
case, all of the light is emitted  spontaneously, i.e.\ without
\emph{optical} stimulation. Furthermore, our system gives rise to
a ``cascade'' of superradiant scattering processes which does not
exist in the two-level superradiance systems studied so far.

More generally, one can regard the observed superradiance as an instability against spatial pattern
formation triggered by noise.  The spatial pattern is the periodic density modulation, the noise is
the quantum noise which leads to spontaneous scattering.  Spontaneous pattern formation is an
interesting non-linear phenomenon studied in fluid mechanics, optics and chemical reactions
\cite{cros93}.

Superradiance is based on the coherence of the emitting system,
but it does not require quantum degeneracy. The condition for
superradiance is that the gain exceed the losses, or that the
superradiant decay time be shorter than any decoherence time.
Above the BEC transition temperature $T_c$, thermal Doppler
broadening results in a thirty times shorter decoherence time than
for a condensate. Further, the larger size of the thermal cloud
reduces the solid angle $\Omega_j$ and therefore the gain by
another factor of ten.  Therefore, the threshold for superradiance
in a thermal cloud is several orders of magnitude higher than for
a condensate.  No signs of superradiant scattering were observed
above $T_c$; rather, the sudden appearance of superradiant
emission was a sensitive indicator for reaching the phase
transition.

\subsection{Phase-coherent amplification of matter waves}
\label{sec:amplification}

Atom amplification differs from light amplification in one important aspect. Since the total number
of atoms is conserved (in contrast to photons), the active medium of a matter wave amplifier has to
include a reservoir of atoms. One also needs a coupling mechanism which transfers atoms from the
reservoir to an input mode while conserving energy and momentum. The gain mechanism which was
explained above can act as a matter wave amplifier. The momentum required to transfer atoms from
the condensate at rest to the input mode is provided by light scattering.  Refs.\
\cite{law98amp,moor99pra} discussed that a condensate pumped by an off-resonant laser beam acts as
a matter wave amplifier which can amplify input matter waves within the momentum range which can be
reached by scattering a single pump photon.

The inversion in this matter wave amplifier is most apparent in the dressed atom picture where the
condensate at rest and the pump light field are treated as one system.  An atom in the dressed
condensate can now spontaneously decay into a recoiling atom and a scattered photon which escapes.
Inversion is maintained since the photons escape and thus the inverse process of combining
recoiling atom and emitted photon into a dressed atom at rest is not possible.  Thus, in principle,
a complete transfer of the condensate atoms into the recoil mode can occur.

\begin{figure}[btp]
    \begin{center}
    \includegraphics[height=2in]{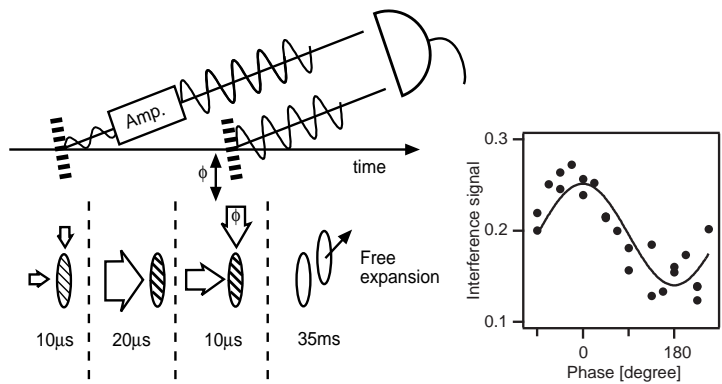}
    \caption{ Experimental scheme for observing phase coherent matter wave amplification. A
small-amplitude matter wave was split off the condensate by applying a pulse of two off-resonant
laser beams (Bragg pulse). This input matter wave was amplified by passing it through the
condensate pumped by a laser beam. The coherence of the amplified wave was verified by observing
its interference with a reference matter wave, which was produced by applying a second (reference)
Bragg pulse to the condensate.  The interference signal was observed after 35~ms of ballistic
expansion. The fringes on the right side show the interference between the amplified input and the
reference matter wave.  Reprinted by permission from Nature, Ref.~\protect \cite{inou99mwatex},
copyright 1999 Macmillan Magazines Ltd.
    \label{MWAsetup}}
    \end{center}
\end{figure}

Our observation of superradiance can be regarded as the
observation of matter wave amplification of noise, i.e.\ of
spontaneously scattered atoms.  To examine this amplification
mechanism further, we performed an experiment in which, rather
than examine the amplification of noise, we amplify a
well--defined input signal, i.e.\ a matter wave of well--defined
momentum which was generated by Bragg scattering. By comparing the
input and output waves, we could characterize the amplification
process.

The input atoms were generated by exposing the condensate to a
pulsed optical standing wave which transferred a small fraction of
the atoms (between $10^{-4}$ and  $10^{-2}$) into a recoil mode by
Bragg diffraction \cite{kozu99bragg,sten99brag}. Both laser beams
were red-detuned by 1.7 GHz from the $3S_{1/2} , |F=1\rangle
\rightarrow 3P_{3/2} , |F = 0,1,2\rangle$ transition to suppress
normal Rayleigh scattering. The geometry of the light beams is
shown in Fig.~\ref{MWAsetup}. The beam which was perpendicular to
the long axis of the condensate (radial beam) was blue detuned by
50 kHz relative to the axial beam.  This detuning fulfilled the
Bragg resonance condition.

\begin{figure}
    \begin{center}
    \includegraphics[height=2.5in]{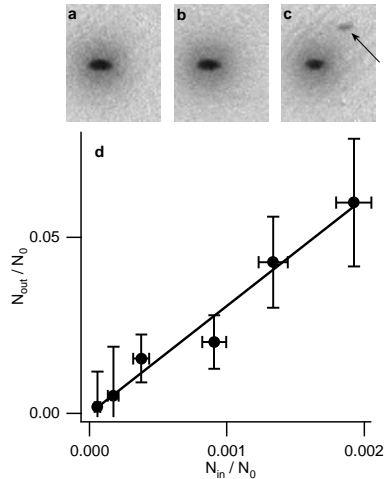}
    \caption{Input--output characteristic of the matter-wave amplifier.  ({\bf a-c}) Typical
time-of-flight absorption images demonstrating matter wave amplification. The output of the seeded
amplifier ({\bf c}) is clearly visible, whereas no recoiling atoms are discernible in the case
without amplification ({\bf a}) or amplification without the input ({\bf b}). The size of the
images is 2.8~mm $\times$ 2.3~mm. ({\bf d}) Output of the amplifier as a function of the number of
atoms at the input.  A straight line fit shows a number gain of 30.  Reprinted by permission from
Nature, Ref.~\protect \cite{inou99mwatex}, copyright 1999 Macmillan Magazines Ltd.
    \label{MWAgain}}
    \end{center}
\end{figure}

Amplification of the input matter wave was realized by applying an intense radial pump pulse for
the next 20 $\mu$s with a typical intensity of 40 mW/cm$^2$. The number of atoms in the recoil mode
was determined by suddenly switching off the trap and observing the ballistically expanding atoms
after 35 ms of time-of-flight using resonant absorption imaging. After the expansion, the
condensate and the recoiling atoms were fully separated (Fig.~\ref{MWAgain}c).

Fig.~\ref{MWAgain} shows the input-output characteristics of the amplifier. The number of input
atoms was below the detection limit of our absorption imaging (Fig.~\ref{MWAgain}a) and was
determined from a calibration of the Bragg process at high laser powers, where the diffracted atoms
were clearly visible in the images.  The amplification pulse alone, although above the threshold
for superradiance \cite{inou99super}, did not generate a discernible signal of atoms in the recoil
mode (Fig.~\ref{MWAgain}b). When the weak input matter wave was added, the amplified signal was
clearly visible (Fig.~\ref{MWAgain}c).  The gain was controlled by the intensity of the pump pulse
(see Eq.~(\ref{eq:gain-eq})) and typically varied between 10 and 100.  Fig.~\ref{MWAgain}d shows the
observed linear relationship between the atom numbers in the input and the amplified output with a
number gain of 30.

The phase of the amplified matter wave was determined with an interferometric technique
(Fig.~\ref{MWAsetup}). For this, a reference matter wave was split off the condensate in the same
way as the first (input) wave. The phase of the reference matter wave was scanned by shifting the
phase of the radio-frequency signal that drove the acousto-optic modulator generating the axial
Bragg beam. We then observed the interference between the reference and the amplified matter waves
by measuring the number of atoms in the recoil mode.

When the input was comparable in intensity to the reference matter
wave, high contrast fringes were observed even without
amplification. Fringes were barely visible when the input was
about 40 times weaker in population. After amplification, we
regained a large visibility (Fig.~\ref{MWAsetup}). This increase
in visibility proved the coherent nature of the matter wave
amplification process. The increase in visibility of the
interference fringes was a factor of two, less than the expected
square root of the total gain of thirty.  This might be due to a
distortion of the matter wave during the amplification, but this
effect requires further study.  A similar experiment with rubidium
atoms was done at the University of Tokyo \cite{kozu99amp}.

This experiment can  be regarded as a demonstration of an active atom interferometer. It realizes a
two-pulse atom interferometer with phase-coherent amplification in one of the arms. Such active
interferometers may be advantageous for precise measurements of phase shifts in highly absorptive
media, e.g. for measurements of the index of (matter wave) refraction when a condensate passes
through a gas of atoms or molecules \cite{schm95}. Since the most accurate optical gyroscopes are
active interferometers \cite{sted97}, atom amplification might also play a role in future
matter-wave gyroscopes \cite{gust97}.

\section{Spinor Bose--Einstein condensates}

\label{sec:spinor}

The experiments described in the previous sections explored the
nature of Bose--Einstein condensates of atomic gases in which all
the atoms were in the same internal state, in particular, the
$|F=1, m_F = -1\rangle$ hyperfine state of sodium.  The
Bose--Einstein condensation phase transition leads to a non--zero
value of a \emph{scalar} order parameter, the condensate
wavefunction $\psi(\vr) = \langle \bfo(\vr) \rangle$. As such, a
single--component gaseous condensate can be considered as a
simple, tractable model of the more complicated spinless
superfluid $^4$He. Many experiments in recent years have explored
aspects of this connection between the two quantum fluids
\cite{kett99var}.

However, unlike $^4$He, alkali atoms can have a non--zero spin and
therefore numerous internal hyperfine states which are stable
electronic ground states.  Thus, there exists the possibility of
creating a quantum fluid simultaneously composed of several,
distinguishable components by Bose condensing a gas of atoms in
several hyperfine states.

The study of multi--component superfluid systems has been a
tantalizing goal of low--temperature physics for decades. The
earliest discussion focused on $^{4}$He -- $^{6}$He mixtures.
$^{6}$He is radioactive with a half-life of 1 second. An ambitious
experiment by Guttman and Arnold~\cite{gutt53} in 1953 sought
evidence for the superfluid flow of $^{6}$He mixed with $^{4}$He
to no avail. Nevertheless, this pursuit touched off a series of
theoretical works on two--component superfluid
hydrodynamics~\cite[and others since]{khal57}. In 1978, Colson and
Fetter~\cite{cols78} considered such mixtures in the context of
mean--field theories which apply directly to current experiments,
and discussed the criterion for interactions between the
superfluids to cause miscibility or phase--separation. After
progress in the stabilization of a spin--polarized atomic hydrogen
gas, Siggia and Ruckenstein~\cite{sigg80} considered the use of
different hyperfine states to achieve a mixture of superfluids.
Since the observation of gaseous Bose condensates, the interest in
multi--component condensates has been revived with a flurry of
theoretical attention~\cite[for
example]{ho96bin,esry97hf,gold97,law97}.

For atomic gases in magnetic traps, the availability of hyperfine states is restricted by the
requirement that the trapped atoms remain in weak--field seeking states. For example, alkali atoms
with a nuclear spin of $I = 3/2$, such as $^{87}$Rb and $^{23}$Na, have three weak--field seeking
states at zero--field: one in the lower hyperfine manifold ($|F=1, m_F = -1\rangle$), and two in
the upper hyperfine manifold ($|F=2, m_F = 1, 2\rangle$). Generally, the simultaneous confinement
of more than one of these states is unstable against exothermic hyperfine state changing
collisions.

However, it was recently found that magnetically--trapped multi--component gases of $^{87}$Rb are
quite long lived due to a fortunate near--equality of the singlet and triplet scattering lengths
which greatly suppresses the spin exchange collision rate
\cite{myat97,burk97,kokk97coll,juli97stab}.   This allowed for the creation of multi--component
condensates by the simultaneous magnetic confinement of $^{87}$Rb atoms in the $|F = 1, m_F =
-1\rangle$ and $|F=2, m_F = 2\rangle$ \cite{myat97} (and, in later work, $\fmst{2}{1}$) states. The
Boulder group has used magnetically trapped multi--component condensates for a remarkable series of
experiments. Studies have probed the spatial separation of a two--component Bose--Einstein
condensate \cite{myat97,hall98dyn} and the stability of a relative phase between the two components
even in the presence of dissipation \cite{hall98phas}.  The Boulder group has also explored the
nature of multi--component condensates in the presence of continuous resonant and non--resonant rf
coupling between the components. Such coupling links the external center--of--mass and the internal
degrees of freedom of the condensed gas and gives rise to a rich variety of dynamical effects such
as Josephson--type oscillations \cite{will99jose} and spin--wave excitations
\cite{matt99twist,will00couple}, as well as new techniques for manipulating the phase of the
condensate wavefunction \cite{matt99vort,will99prep}. Some of this work is reviewed in Refs.\
\cite{corn98jltp,corn99var}.

In contrast, a far--off--resonant optical trap confines atoms
regardless of their hyperfine state \cite{stam98odt}. Thus, the
atomic spin is liberated from the requirements of magnetic
trapping and becomes a new degree of freedom.  In particular, all
atoms in the lower hyperfine manifold, for example the $F=1$
hyperfine manifold of sodium, can be stably trapped simultaneously
without suffering from hyperfine manifold changing collisions.
Such multi--component optically trapped condensates are
represented by an order parameter which is a \emph{vector} in
hyperfine spin space, and are thus called spinor Bose--Einstein
condensates. A variety of new phenomena are predicted for this new
quantum fluid such as spin textures, spin waves, and coupling
between atomic spin and superfluid
flow~\cite{ho98,ohmi98,law98spin2}.

Spinor Bose condensates differ from other multi--component Bose condensates, such as the
experimentally realized $^{87}$Rb mixtures or the proposed mixtures of several atomic species, in
important ways stemming from symmetries under rotations of the vectorial order parameter.
Furthermore, spin relaxation collisions within the lower $F=1$ hyperfine manifold
\begin{equation}\label{eq:spinrelax}
  \mst{0} + \mst{0} \leftrightarrow \mst{+1} + \mst{-1}
\end{equation}
allow for population mixing among the different hyperfine states without trap loss.  In contrast,
spin relaxation is the major limitation to the lifetime (about a second) of the $^{87}$Rb mixtures.

Since the realization of an optical trap for Bose--Einstein
condensates \cite{stam98odt}, our group has performed several
experimental studies of this new quantum fluid.  In three
different experiments we explored the ground--state spin structure
of spinor condensates in external magnetic fields
\cite{sten98spin}, the formation and persistence of metastable
spin domain configurations \cite{mies99meta}, and the transport
across spin domain boundaries by quantum tunneling
\cite{stam99tun}.  In this section, we summarize our current
understanding of this fluid as derived from our experiments and
from a growing number of theoretical works. While a portion of
this work has been reviewed in Refs.\ \cite{sten98odt,kett99var},
this section is the first comprehensive review on spinor
Bose--Einstein condensates.

\subsection{The implications of rotational symmetry}

\label{sec:rotsym}

An $F = 1$ spinor Bose--Einstein condensate is described by a three--component order parameter
\begin{equation}\label{eq:3comporder}
  \vec{\psi}(\vr) = \left( \begin{array}{c}
  \psi_{1}(\vr) \\
  \psi_{0}(\vr) \\
  \psi_{-1}(\vr) \end{array} \right)
\end{equation}
and can thus be regarded as a particular instance of a
multi--component condensate.  Here, the subscripts refer to the
spin projection on the quantization axis, $m_F = -1, 0, +1$.
However, the spinor Bose--Einstein condensate is distinguished
from a general, multi--component quantum fluid by the fact that
the order parameter $\vec{\psi}$ transforms as a vector.  The
vectorial character of the order parameter has a pronounced effect
on interatomic interactions, and defines important features of the
spinor condensate at zero magnetic field, where the rotational
symmetry of the system is preserved.

In second--quantized notation, the Hamiltonian for a multi--component gas has the general form
\begin{eqnarray}\label{eq:multigpe}
  \hat{\mathcal{H}} & = & \int d^3 \vr \, \left\{ \bfod_i(\vr) \left( - \frac{\hbar^2 \nabla^2}{2
  m} \, \delta_{i j} + U_{i j}(\vr) \right) \bfo_j(\vr) \right. \nonumber \\
  & & + \frac{g_{ij,kl}}{2} \int d^3 \vr_1 \, d^3 \vr_2 \,
   \bfod_i(\vr_1) \bfod_j(\vr_2)
    \bfo_k(\vr_2) \bfo_l(\vr_1) \, \delta(\vr_1 - \vr_2) \left. \right\}
\end{eqnarray}
where the indices $i,j,k,l$ correspond to the $N$ components of the gas, and repeated indices are
summed.  The general form of the external potential $U_{i j}(\vr)$ allows for a potential which is
not diagonal in the hyperfine spin basis, in which case it can represent a Josephson-type coupling
between spin components. The $U_{i j}(\vr)$ terms also contain the effects of magnetic fields which
are discussed in later sections.

The interatomic interaction (second line of the above expression) has been approximated as a
contact interaction in which the coefficients $g_{ij,kl}$ describe the strength of the various
elastic and inelastic (state converting) collisions.  These generally constitute a large number of
free parameters. Particle exchange ($g_{ij,kl} = g_{ji,kl}$) and time--reversal ($g_{ij,kl} =
g_{kl,ij}$) symmetries reduce the number of free parameters from $N^4$ to $[(N^2 + N + 2)(N^2 +
N)]/8$ but still a large number of interaction parameters remain\footnote{Due to particle exchange
symmetry, we may denote the interaction parameter as $g_{\mathcal{A}, \mathcal{B}}$ where
$\mathcal{A}$ and $\mathcal{B}$ are elements of the set $\mathbb{P}$ of distinct unordered pairs of
indices $i,j$.  There are $Z = N (N+1) / 2$ such pairs.  Due to time--reversal symmetry, the
interaction parameters are enumerated as $g_{\mathcal{Z}}$ where $\mathcal{Z}$ is an unordered pair
of elements from $\mathbb{P}$.  Thus, the number of free interaction parameters is $Z (Z+1) /2 =
[(N^2 + N + 2)(N^2 + N)]/8$.}.  For a three--component condensate, the number of interaction
parameters is 21.

This situation is greatly simplified in the case of spinor
condensates due to rotational symmetry. The rotationally symmetric
characterization of two--body collisions among atoms of hyperfine
spin $F_1$ and $F_2$ can only depend on their total spin $f = F_1
+ F_2$ and not on its orientation. Thus, in the $s$--wave limit,
the interatomic interaction $V\subrm{int}(\vr_1 - \vr_2)$ is
reduced to the form \cite{ho98,ohmi98}
\begin{equation}\label{eq:rotcontact}
  V\subrm{int}(\vr_1 - \vr_2) =
   \frac{4 \pi \hbar^2}{m} \, \delta(\vr_1 - \vr_2) \sum_{f}
  a_{f}  \tilde{\mathcal{P}}_{f}
\end{equation}
where $a_{f}$ is the scattering length for collisions between atoms with total spin $f$, and
$\tilde{\mathcal{P}}_{f}$ is the projection operator for the total spin.  For colliding bosons of
spin $F$, $f$ can take the values $f = {0,2,\ldots , 2F}$. Therefore, the interaction parameters
$g_{ij,kl}$ describing collisions among the $N = 2F+1$ different components of a spinor condensate
of spin $F$ depend only on $F+1 = (N+1)/2$ scattering lengths.

In particular, for the $F=1$ spinor system which was realized in gaseous sodium, interactions are
described fully by just two parameters, the scattering lengths $a_{f=0}$ and $a_{f=2}$. The
interaction potential can then be written as
\begin{equation}\label{eq:intf1f2}
 V\subrm{int}(\vr_1 - \vr_2) =
  \left( g_0 + g_2 \vec{F}_1 \cdot \vec{F}_2 \right) \, \delta(\vr_1 - \vr_2)
\end{equation}
where the parameters $g_0$ and $g_2$ are defined as
\begin{eqnarray}\label{eq:defineg0g2}
  g_0 & = & \frac{4 \pi \hbar^2}{m} \, \frac{2 a_{f=2} + a_{f=0}}{3} \\
  g_2 & = & \frac{4 \pi \hbar^2}{m} \, \frac{a_{f=2} - a_{f=0}}{3}
\end{eqnarray}
and $\vec{F}_1$ and $\vec{F}_2$ are the spin operators for the two colliding particles.  The
spin--dependence of the interatomic interaction is thereby isolated in the rotationally--symmetric
term $ g_2 \vec{F}_1 \cdot \vec{F}_2 \, \delta(\vr_1 - \vr_2)$. The Hamiltonian describing a
weakly--interacting Bose gas of spin $F=1$ now becomes \cite{ho98,ohmi98}
\begin{eqnarray}\label{eq:spinhami}
  \hat{\mathcal{H}} & = & \int d^3 \vr \, \left\{
  \bfod_i(\vr) \left( - \frac{\hbar^2 \nabla^2}{2
  m} \, \delta_{i j} + V_{i j}(\vr) \right) \bfo_j(\vr) \right.  \\
  & & + \frac{1}{2} \left. \left[
   g_0  \bfod_i(\vr) \bfod_j(\vr)  \bfo_i(\vr) \bfo_j(\vr) \right. \right. \nonumber \\
   & &  + \, g_2 \left. \left.  \left( \bfod_i(\vr) (F_\eta)_{i j} \bfo_j(\vr) \right)
   \cdot \left( \bfod_k(\vr) (F_\eta)_{k l} \bfo_l(\vr) \right)
   \right] \right\} \nonumber
\end{eqnarray}
where the index $\eta$ runs over the three coordinate axes $x, y, z$.

Equivalently, the Hamiltonian can be written in the form of Eq.\ \ref{eq:multigpe}, where all
non--zero $g_{i j, k l}$ are determined by $g_0$ and $g_2$ as shown in Table \ref{tab:spinorscats}.
Taking into account particle exchange symmetries and gathering similar terms, the interaction
Hamiltonian can be written as
\begin{eqnarray}\label{eq:hint}
  \mathcal{H}\subrm{int} & = & \frac{1}{2} \int d^3 \vr \, \left[
  (g_0 + g_2) \bfod_1 \bfod_1 \bfo_1 \bfo_1 + g_0 \bfod_0 \bfod_0
  \bfo_0 \bfo_0 \right. \nonumber \\
  & & \mbox{} + (g_0 + g_2) \bfod_{-1} \bfod_{-1} \bfo_{-1}
  \bfo_{-1} + 2 (g_0+g_2) \bfod_1 \bfod_0 \bfo_1 \bfo_0  \nonumber \\
  &  & \mbox{} + 2 (g_0+g_2) \bfod_{-1} \bfod_0 \bfo_{-1} \bfo_0
   + 2 (g_0 - g_2) \bfod_1 \bfod_{-1} \bfo_1 \bfo_{-1}  \nonumber \\
   & & \left. \mbox{} + 2 g_2 (\bfod_0 \bfod_0 \bfo_1 \bfo_{-1} + \bfod_1
   \bfod_{-1} \bfo_0 \bfo_0 ) \right]
\end{eqnarray}
The first three terms are the self--scattering terms, the following three the cross--scattering
terms, and the last line contains the spin relaxation terms.

\begin{table}
    \begin{center}
        \vskip \baselineskip
        \begin{tabular}{c|c|c|c}
        $g_{AB,AB}$ & $m_F = +1$ & $m_F = 0$ & $m_F = -1$ \\
        \hline
        $m_F = +1$ & $g_0 + g_2$ & $\frac{g_0 + g_2}{2}$ & $\frac{g_0
        - g_2}{2}$ \\
         $m_F = 0$ & $\frac{g_0 + g_2}{2}$ & $g_0$ & $\frac{g_0 + g_2}{2}$ \\
        $m_F = -1$ & $\frac{g_0
        - g_2}{2}$  & $\frac{g_0 + g_2}{2}$ &$g_0 + g_2$
        \end{tabular}
\caption[Interaction parameters for the $F=1$ spinor system]{
Interaction  parameters $g_{ij,kl}$ describing elastic collisions
in the $F=1$ spinor system. The interaction Hamiltonian is defined
as in Eq.\ \ref{eq:multigpe}.  The interaction parameter
responsible for spin relaxation is $g_{0 0, -1 +1} = g_2$.  The
cross--species scattering length discussed in Sec.\
\ref{sec:immisc} is given as $g_{AB,AB} = 2 \pi \hbar^2 a_{A B}
/m$.
        \label{tab:spinorscats}}
    \end{center}
\end{table}

It is important to note that this simplification of the
interaction term is only strictly valid in the absence of magnetic
fields.  As the magnetic field is increased, the hyperfine spin
$F$ is no longer a good quantum number for describing the atomic
states, and thus collisional properties deviate from the relations
imposed by rotational symmetry. Dramatic deviations from
rotational symmetry can be seen, for example, at a Feshbach
resonance where the scattering length for one particular
collisional channel is greatly varied.  Nonetheless, away from
Feshbach resonances and at low magnetic fields for which the
Zeeman shifts are much smaller than the hyperfine splitting, one
can expect the zero--field description of collisional interactions
to remain valid.

The properties of spinor Bose--Einstein condensates at
zero--magnetic field have been recently discussed by a number of
authors.  Ho \cite{ho98} and Ohmi and Machida \cite{ohmi98}
considered the ground state of a spinor condensate by an extension
of mean--field theory (see also \cite{huan99spinor}). The
$N$--particle condensate ground state is found by replacing the
field operators $\bfo_i$ in Eq.\ \ref{eq:spinhami} with
$c$--number order parameters $\psi_i$. It is convenient to express
the order parameter as $\vec{\psi} = \sqrt{n} \vec{\zeta}$ where
$n$ is the atomic density and $\vec{\zeta}$ is a three--component
spinor of normalization $|\vec{\zeta}| = 1$, obtaining the energy
functional
\begin{equation}\label{eq:spinefunc}
  E = \int d^3 \vr \, \left[ \psi_i^*(\vr) \left( - \frac{\hbar^2
  \nabla^2}{2m} \right) \psi_i(\vr) + (U(\vr) - \mu) n(\vr)
   + \frac{n^2}{2} \left( g_0 + g_2 \langle \vec{F} \rangle^2 \right)
  \right]
\end{equation}
Here it is assumed that the external potential is scalar, i.e.\ diagonal in the hyperfine spin
basis and equal for each of the spin components.  The chemical potential $\mu$ determines the
number of atoms in the condensate.

The ground--state spinor $\vec{\zeta}$ is determined by minimizing the spin--dependent interaction
energy, $n^2  g_2 \langle \vec{F} \rangle^2 / 2$, which is expressed in terms of the average
condensate spin $\langle \vec{F} \rangle_\eta = \zeta_i^* (F_\eta)_{i j} \zeta_j$.  There are two
distinct solutions depending on the sign of the spin--dependent interaction parameter $g_2$:
\begin{itemize}
  \item $g_2 > 0$: the collisional coupling is
anti--ferromagnetic as the condensate lowers its energy by minimizing its average spin, i.e.\ by
making $|\langle \vec{F} \rangle| = 0$. The ground state spinor is then one of a degenerate set of
spinors, the ``polar'' states, corresponding to all possible rotations of the hyperfine state $|m_F
= 0\rangle$.

  \item $g_2 < 0$:  the collisional coupling is ferromagnetic as the
condensate lowers its energy by maximizing its average spin, i.e.\ by making $|\langle \vec{F}
\rangle| = 1$.  In this case the ground state spinors correspond to all rotations of the hyperfine
state $|m_F = +1\rangle$.
\end{itemize}

Law, Pu, and Bigelow \cite{law98spin2} explicitly calculated the many--body state of a homogeneous
spinor Bose--Einstein condensate without assuming a Hartree form.  They describe an elegant
transformation of the spinor Hamiltonian in terms of operators which obey an angular momentum
algebra, and thus immediately yield eigenstates of the many--body Hamiltonian and their energy
spectrum.  Anti--ferromagnetic coupling leads to a unique ground--state which cannot be represented
in the Hartree form and which has super--Poissonian fluctuations in the population of each spin
state. Ferromagnetic coupling yields a degenerate family of ground states with sub--Poissonian
fluctuations. The Hartree solution with all atoms in the $|m_F = +1\rangle$ state is included in
the family of ferromagnetic ground states obtained by Law, Pu and Bigelow.

The highly entangled ground state of Law \emph{et al.}\ for the anti--ferromagnetic case differs in
energy from the ground state in the mean--field formalism (all atoms in the $|m_F = 0\rangle$
state) only by a term which is proportional to  $N$, the number of atoms.  This energy is $N$ times
smaller than the mean-field energy which scales as $N^2$.  Therefore, it will be difficult to
observe this highly correlated state. Several authors have extended the treatment to finite
magnetic fields and shown that the singlet state of Ref.\ \cite{law98spin2} is only the ground
state at very small magnetic fields; otherwise the ground state approaches the Hartree form
\cite{koas00, ho99frag}.

The scattering lengths $a_{f = 0}$ and $a_{f = 2}$ can be
calculated by incorporating data from a variety of experiments
into models of the interparticle potentials. For sodium, it is
predicted that ${a_{f = 2}} = 2.75$ nm \cite{ties96} and ${a_{f =
2}} - {a_{f = 0}} = 0.29$ nm \cite{burk98}. Thus, an $F = 1$
spinor condensate of sodium should be anti--ferromagnetic.  For
$^{87}$Rb, it appears that ${a_{f = 2}} - {a_{f = 0}} < 0$ and
thus an $F=1$ spinor condensate of $^{87}$Rb should be
ferromagnetic \cite{ho98}.

\subsection{Tailoring the ground--state structure with magnetic fields}

These theoretical considerations differ from the conditions in our experiment in two major regards:
the imposition of spin conservation, and the effects of field inhomogeneities and quadratic Zeeman
shifts. These differences are exemplified by considering the outcome of an experiment we performed,
in which an optically--trapped Bose condensed cloud was prepared with all atoms in the $|m_F =
0\rangle$ state.  The cloud was then allowed to equilibrate to the ground state by spin relaxation
and by spatial redistribution \cite{sten98spin}.  For simplicity, consider only variations in the
condensate wavefunction in one dimension, the $\hat{z}$--direction, and define the coordinate $z$
to give the distance from the center of the condensate. This corresponds to the experimental
situation in which the spinor condensates were held in the highly anisotropic, cigar--shaped
potential of an optical trap.

\begin{figure}
    \begin{center}
    \includegraphics[height=1.5in]{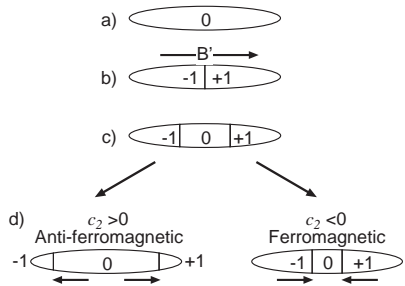}
     \caption[Constructing the ground--state $\langle F_{z}  \rangle = 0$ spinor in a trap]{
Constructing the ground--state $\langle F_{z}\rangle = 0$ spinor in a trap. In each step (a -- d),
another contribution to the Hamiltonian is considered (see text): a) linear Zeeman shift from
homogeneous field, b) linear Zeeman shift from field gradient, c) quadratic Zeeman shift, d)
spin-dependent mean-field interaction.} \label{fig:spinorground}
    \end{center}
\end{figure}

Let us consider the evolution of the pure $m_{F} = 0$ condensate as we slowly add the effects of
magnetic fields and mean--field interactions and construct the relevant energy functional (Fig.\
\ref{fig:spinorground}).

\begin{itemize}
    \item[a)] {\it Linear Zeeman shift from homogeneous field.}
    In a homogeneous magnetic field of strength $B_0$, the linear Zeeman energy
\begin{equation}\label{eq:elz}
  E\subrm{lin} = - p \int d^3 r \, n  \langle F_z \rangle
\end{equation}
    is minimized by placing all atoms in the strong--field seeking $|m_{F} = +1\rangle$ state (a
    ferromagnetic state)\footnote{ The operator $F_z$ gives the component of the spin along the
    direction of the magnetic field.  This direction need not be the same as the long axis of the
    trapped condensate.}.  Here, $p = g \mu_B B_0$ where $g$ is the Land\'{e} $g$--factor and $\mu_B$
    is the Bohr magneton.

    However, spin non-conserving collisions (dipolar relaxation), which would be necessary to transform
    the $m_F = 0$ condensate to the $m_F = +1$ ground state, are negligible over the lifetime of the
    condensate.  Thus, the total spin is a conserved quantity, and rather than considering the global
    ground state of the system, one must consider the lowest energy state under the restriction of spin
    conservation.  Formally, one minimizes the restricted energy functional
\begin{equation}\label{eq:k}
    {K}\subrm{tot} = E\subrm{tot} + \tilde{p} \int d^3 \vr \, n \langle F_z \rangle
\end{equation}
    where $\tilde{p}$ is a Lagrange parameter determined by the given value of the total spin. For the
    case where $\int \langle F_z \rangle \, n \, d^3 \vr = 0$, the linear Zeeman shift of a homogeneous
    magnetic field ($E_{lZ}$) is exactly canceled ($p_0 = \tilde{p}$).

    This is an important point: spin conservation allows one to study the effects of the small
    spin--dependent interaction energies even at magnetic fields for which the linear Zeeman energy
    would otherwise be dominant.  For example, in the case of sodium, the spin--dependent mean--field
    energy is just $c_2 n \simeq h \times 50$ Hz for a typical density of $n = 3 \times 10^{14} \,
    \mbox{cm}^{-3}$.  Thus, without the restriction of spin conservation, the ground--state spinor
    would trivially consist of all atoms in the $|m_F = +1\rangle$ state at a magnetic field of just 70
    $\mu$G, and any interesting structure, correlations or dynamics due to the anti--ferromagnetic
    coupling would be obscured.  However, due to spin conservation, many of these effects can be
    studied even in the absence of such demanding field stability.

    \item[b)] {\it Linear Zeeman shift from field gradient.}
    A field gradient $B'$ along the long axis of the condensate
    introduces an energy term
\begin{equation}\label{eq:egrad}
  E\subrm{grad} = - \int d^3 \vr \, p(z) n \langle F_z \rangle
\end{equation}
    with $p(z) = g \mu_B B' z$, which makes it energetically favorable for two $m_{F} = 0$
    atoms to collide and produce a $m_{F} = +1$
    atom on the high--field end of the cloud, and a $m_{F} = -1$ atom on
    the low--field end.  Thus, the condensate is magnetically polarized
    into two pure spin domains.

    \item[c)] {\it Quadratic Zeeman shift from homogeneous field.}
    The quadratic Zeeman shift at a field $B_{0}$
    introduces
    an energy term of the form
\begin{equation}\label{eq:quad}
  E\subrm{quad} =  q \int d^3 \vr \, n \langle F_z^2 \rangle
\end{equation}
    which causes the energy of a $m_{F} = 0$ atom to
    be lower than the average energy of a $m_{F} = +1$ and $m_{F} = -1$
    atom by an amount $q = \hat{q} B_{0}^{2}$.
    For sodium, $\hat{q} = h \times  390 \, \mbox{Hz}/\mbox{G}^{2}$.
    The quadratic Zeeman energy favors population in the $|m_F = 0\rangle$
    state,
    and thus introduces an $m_{F} = 0$ domain at the center of
    the cloud with boundaries at $q = |p(z)|$ (see Fig.\ \ref{fig:spinorground}c).

    \item[d)] {\it Spin-dependent mean--field interaction.}
    As discussed previously, the collisional interactions give
    a spin--dependent energy term of the form
\begin{equation}\label{eq:espin}
  E\subrm{int} = \frac{1}{2} \int d^3 \vr \, g_2 n^2 \langle \vec{F}
  \rangle^2
\end{equation}
    Anti--ferromagnetic coupling ($g_2 > 0$) favors the polar
    state, and thus makes the central $m_F = 0$ domain larger.
    Ferromagnetic coupling ($g_2 < 0$) favors the ferromagnetic
    states $|m_f = +1\rangle$ and $|m_F = -1 \rangle$ at the ends
    of the cloud, and thus the $m_F = 0$ domain shrinks (see Fig.\ \ref{fig:spinorground}d).
\end{itemize}

Collecting spin--dependent energy terms, we now have the restricted energy functional
\begin{eqnarray}\label{eq:kspin}
  {K}\subrm{tot} & = & \int d^3 \vr \, \left( {K}_0 + K\subrm{spin} \right) \\
  K_0 & = & \vec{\psi}^\dagger \left( - \frac{\hbar^2 \nabla^2}{2 m} \right) \vec{\psi} +
  (U(\vr) - \mu + \frac{g_0}{2} n) n \\
  K\subrm{spin} & = & \left( - p(z) \langle
  F_z \rangle + q \langle F_z^2 \rangle + c \langle
  \vec{F} \rangle^2 \right) n
\end{eqnarray}
where $c = g_2 n / 2$.  The ground--state spin structure of a spinor condensate is found by
minimizing $K\subrm{tot}$.

\subsection{Spin--domain diagrams: a local density approximation to
the spin structure of spinor condensates}

\label{sec:ldaspinor}

\begin{figure}
    \begin{center}
    \includegraphics[height=2in]{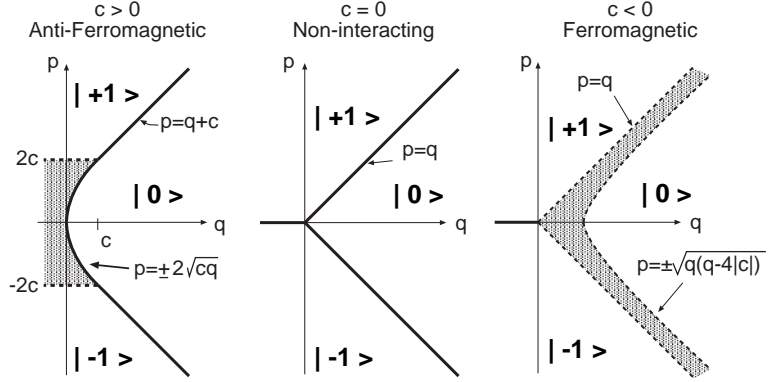}
    \caption[Spin--domain diagrams for $F = 1$ spinor condensates]{
Spin--domain diagrams for $F = 1$ spinor condensates for the three
cases $c>0$ (anti--ferromagnetic collisional coupling), $c=0$ (no
spin--dependent interactions), and $c<0$ (ferromagnetic coupling).
The ground--state spinors which minimize $K\subrm{spin}$ for the
given values of $p$, $q$, and $c$ are shown.  Regions in which the
ground state contains just one component are labeled by the
magnetic hyperfine sublevel.  Regions of mixed components are
shaded. The ground--state spin--domain structure of a spinor
condensate in the presence of a magnetic bias field and field
gradient is indicated by a vertical line across the relevant
domain diagram. Reprinted by permission from Nature,
\cite{sten98spin}, copyright 1999 Macmillan Magazines Ltd.
    \label{fig:spindiagram}}
    \end{center}
\end{figure}

Focusing on the spin--dependent part, $K\subrm{spin}$ has a simple and elegant form.  Let us step
away from the specific experimental considerations and consider a homogeneous spinor condensate
with some arbitrary, uniform values of the parameters $p$, $q$ and $c$.  The ground--state spinors
obtained by minimizing the energy functional $K\subrm{spin}$ are indicated in the three diagrams of
Fig.\ \ref{fig:spindiagram} for three conditions on the parameter $c$ \cite{sten98spin}.  For a
non--interacting gas ($c=0$), the ground--state spinor is determined by the effects of magnetic
fields alone.

In the case of anti--ferromagnetic coupling ($c > 0$), the spinor diagram changes in two
significant ways.  First, since the anti--ferromagnetic energy favors the $|m_F = 0\rangle$ polar
state, the region in which the $|m_F = 0\rangle$ state is the ground state is enlarged.  Second, at
low values of $q$, in addition to the region of pure hyperfine states, a region is introduced in
which the ground--state spinor is a superposition of the $|m_F = \pm 1\rangle$ states.  While
admixing the $|m_F = -1\rangle$ ($|m_F = +1\rangle$) state for $p > 0$ ($p<0$) increases the linear
Zeeman energy, it reduces the interaction energy by making the superposition state more polar in
character. Explicitly, in the shaded region, $\langle F_z \rangle = p / 2c$ independent of $q$.
The effect of the linear Zeeman energy on anti--ferromagnetic condensates was considered also by
Ohmi and Machida \cite{ohmi98} whose work did not include the quadratic Zeeman energy, and thus
concerned the $q = 0$ axis of the spin--domain diagram.

In the case of ferromagnetic coupling ($c < 0$), the situation is different.  The higher
interaction energy for the polar state diminishes the region in which the ground state is the $|m_F
= 0 \rangle$ state.  In between the regions of single--component ground--state spinors there is a
region in which {\it all three} hyperfine states are generally mixed.  In these regions, the $|m_F
= 0\rangle$ state is mixed predominantly with a large population of only one of the $|m_F = \pm
1\rangle$ states and with  a small population of the other.

Returning now to the experimental situation, the spin--domain diagram is used to describe the
ground--state spin structure of a spinor condensate through a local density approximation.  The
values of $q$ and $c$ (but not its sign) are determined by the magnetic bias field and by the
condensate density, which we assume for now to be constant across the condensate.  The coefficient
$p$ varies across the condensate due to the presence of a magnetic field gradient; thus, the
variations in the condensate spin structure across the length of the condensate are determined by
scanning along a vertical line in the spin--domain diagrams.  The length of this line is determined
by the condensate length and the field gradient $B'$.  The center of this vertical line is
determined by the total spin in the condensate: it moves upwards (larger $p$) as $\langle F_z
\rangle$ is increased and moves downwards (smaller $p$) as $\langle F_z \rangle$ is decreased.
Thus, by adjusting the condensate density, the magnetic bias field, and the total spin of the
cloud, all regions of the spin--domain diagrams are accessible.

\subsection{Experimental methods for the study of spinor condensates}

Having introduced this new quantum fluid, let us describe how we
made it, and how we probed it. First, magnetically--trapped
Bose--Einstein condensates were produced in the $|F = 1, m_{F} =
-1\rangle$ hyperfine state and transferred to an optical trap
\cite{stam98odt}. Then, we pulsed on rf fields of variable
strength which were swept in frequency to distribute the
optically--trapped atoms among the $F=1$ hyperfine sublevels by
the method of adiabatic rapid passage \cite{mewe97}. High
amplitudes or slow sweep rates transferred all the atoms from one
hyperfine state to another, while low amplitudes or fast sweep
rates transferred just a fraction of the atoms. To achieve an
arbitrary hyperfine distribution, it was necessary to make these
rf--transitions at large (15 -- 30 G) bias fields, separating the
$|m_{F} = +1\rangle \rightarrow |m_{F} = 0\rangle$ and $|m_{F} =
0\rangle \rightarrow |m_{F} = -1\rangle$ transition frequencies by
about 1 MHz due to the quadratic Zeeman shift. Otherwise, at low
fields where level spacings between the hyperfine sublevels are
equal, rf fields can only be used to rotate the atomic spin vector
and cannot, for example, change the atoms from the ferromagnetic
$|m_F = +1\rangle$ state to the polar $|m_F = 0\rangle$ state.

\begin{figure}
    \begin{center}
    \includegraphics[height=1.7 in]{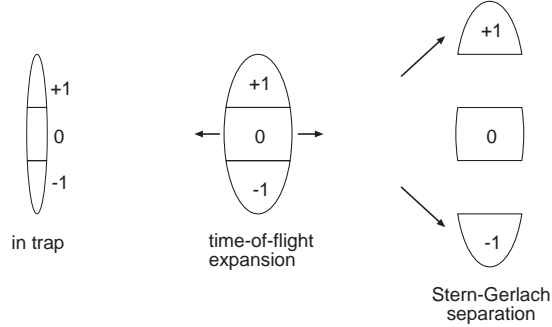}
    \caption[Probing spinor condensates]{
Probing spinor condensates. After release from the elongated optical trap, the trapped spinor
condensate expands primarily radially while maintaining the axial hyperfine distribution.  A
magnetic field gradient is then used to separate out the different components while preserving
their shape. A subsequent absorption probe reveals the spatial and hyperfine distributions in the
trap.\label{fig:spinorprobe}}
    \end{center}
\end{figure}

After state preparation, the optically--trapped spinor condensates
were allowed to evolve in the presence of variable magnetic bias
fields and field gradients.  After a variable dwell time in the
optical trap, the spinor condensates were probed by
time--of--flight imaging combined with a Stern--Gerlach spin
separation (Fig.\ \ref{fig:spinorprobe}). The optical trap was
suddenly switched off, allowing the atoms to expand primarily
radially from the highly anisotropic optical trap. Then, after
allowing about 5 ms for the interaction energy to be completely
converted to kinetic energy, a magnetic field gradient was applied
which separated the spin state populations without distorting
them. Finally, after 15 -- 30 ms, the atoms were optically pumped
to the $|F = 2\rangle$ hyperfine manifold. This gave the same
optical cross--section for all the atoms in the subsequent
absorption probing on the $|F=2, m_{F} = 2\rangle \rightarrow
|F'=3, m_{F'} = 3\rangle$ cycling transition. This probing method
obtained both the spatial and hyperfine distributions along the
axis of the optical trap in a single image.

\subsection{The formation of ground--state spin domains}

\label{sec:spinorground}

\begin{figure}
    \begin{center}
    \includegraphics[height=2in]{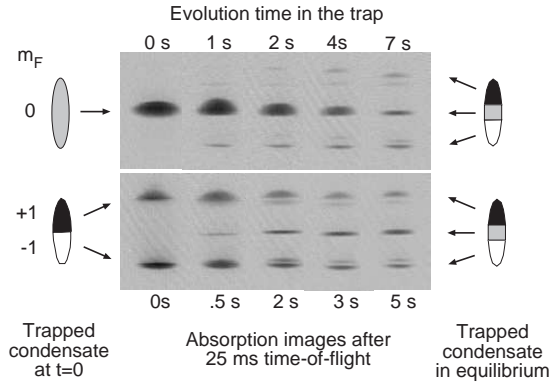}
        \caption[Formation of ground state spin domains]{
Formation of ground state spin domains. Absorption images of
ballistically expanding spinor condensates show both the spatial
and hyperfine distributions. The images of clouds with various
dwell times in the trap show the evolution to the same equilibrium
for condensates prepared in either a pure $|m_F =0\rangle$ state
(upper row) or in equally populated $|m_F =\pm 1\rangle$ states
(lower row). The bias field and field gradient during the dwell
time was $B_0=20\,{\rm mG}$ and $B^\prime=11\,{\rm mG/cm}$. The
image size for each spinor condensate is 1.7 $\times$ 2.7 mm.
    \label{fig:spinor-eq}}
    \end{center}
\end{figure}

In Ref.\ \cite{sten98spin} we explored the ground--state structure of spinor Bose--Einstein
condensates with an average spin of $\langle F_z \rangle = 0$. Condensates were prepared either
with all atoms in the $\mst{0}$ state or in an equal mixture of the $\mst{\pm 1}$ states.  The
atoms were then allowed to relax to their equilibrium distribution in the presence of a variable
magnetic bias field and field gradient.   Probing at variable times after the state preparation
revealed that the condensates relaxed to the same spin structure from either of the initial
conditions, and remained thereafter in the same equilibrium state (Fig.\ \ref{fig:spinor-eq}).

\begin{figure}
    \begin{center}
    \includegraphics[height=2.5 in]{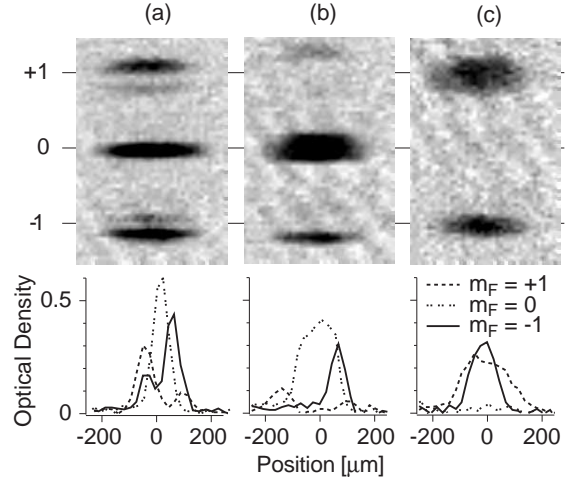}
    \caption[Ground state spin domains in $F=1$ spinor Bose-Einstein condensates]{
Ground state spin domains in $F=1$ spinor Bose-Einstein condensates.  Time-of-flight images after
Stern-Gerlach separation are shown, along with the indicated axial density profiles in the optical
trap, for which the Stern-Gerlach separation was ``undone.'' Images (a) and (b) show spin domains
of all three components. Image (c) shows a miscible $m_{F} = \pm 1$ component condensate.
Conditions are: (a) $B = 20 \, \mbox{mG}$, $B' = 11 \, \mbox{mG/cm}$; (b) $B = 100 \, \mbox{mG}$,
$B' = 11 \, \mbox{mG/cm}$; (c) $B = 20 \, \mbox{mG}$, $B' \simeq 0$, $\langle F_{z} \rangle >
0$.\label{fig:spinorcurves}}
    \end{center}
\end{figure}

\begin{figure}
    \begin{center}
    \includegraphics[height=2in]{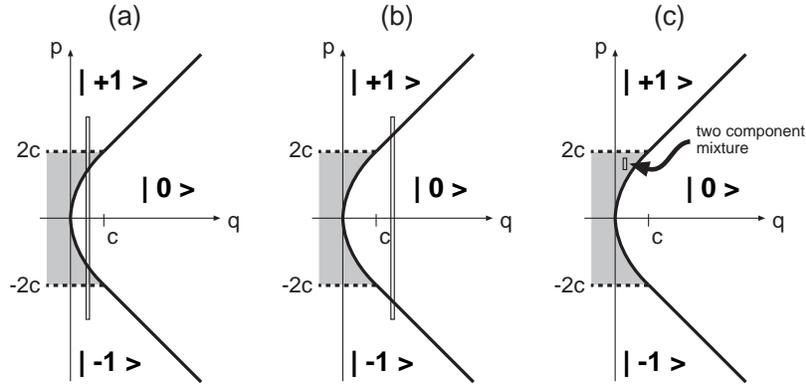}
    \caption[Representation of ground--state spin--domain structures on the spin--domain diagram]{
Representation of ground--state spin--domain structures of Fig.\ \ref{fig:spinorcurves} on the
spin--domain diagram.  The spin structures shown in Fig.\ \ref{fig:spinorcurves}a and b correspond
to long vertical lines through the spin--domain diagram, centered at $p = 0$ to correspond to an
average spin $\langle F_z \rangle = 0$.  (a) At low magnetic field $q < c$ and mixed--spin domains
are introduced.  (b) At higher fields, they are absent. (c) Lowering the gradient focuses on a
small portion of the diagram in which a cloud of non--zero spin consists of overlapping $m_F = \pm
1$ components.
    \label{fig:abcdiagrams}}
    \end{center}
\end{figure}

Figure \ref{fig:spinorcurves} shows three examples of the spin
structures which were observed.  The corresponding representations
of these structures in the anti--ferromagnetic spin--domain
diagram are indicated in Fig.\ \ref{fig:abcdiagrams}.

Fig.\ \ref{fig:spinorcurves}b shows the equilibrium structure of a
spinor condensate for which the quadratic Zeeman energy is larger
than the interaction energy ($q > c$).   The magnetic field
gradient is sufficiently strong so that $|p| > q + c$ at the ends
of the cloud, and thus the condensate consists of three pure spin
domains with a $m_F = +1$ domain on the high--field end of the
cloud, a $m_F = 0$ domain in the center, and a $m_F = -1$ domain
at the low--field end.

In Fig.\ \ref{fig:spinorcurves}a, the magnetic field is weaker,
and thus the central $m_F= 0$ domain is now flanked by regions in
which the $m_F = \pm 1$ components are mixed.  The appearance of a
large population of the $m_F = 1$ ($m_F = -1$) component on the
low--field (high--field) end of the cloud provides a qualitative
confirmation of the anti--ferromagnetic collisional coupling in
the $F = 1$ hyperfine manifold of sodium.  The division between
the $m_F= 0$ domain and the domains containing the $m_F = \pm 1$
components indicates the immiscibility of the $m_F = 0$ component
with the others.

Finally, in Fig.\ \ref{fig:spinorcurves}c, a condensate is shown
for which the total spin is greater than zero. The magnetic field
gradient is about zero, and thus the condensate corresponds to a
point on the spin--domain diagram in which the $m_F = \pm 1$
components are mixed.  This situation nicely demonstrates the
miscibility of the $m_F = \pm 1$ components with each other.  The
different widths of the two components may be due to the curvature
of stray magnetic fields, which has opposite effects on the
$\mst{\pm 1}$ states due to their different magnetic moments.
Another explanation was recently given by Huang and Gou
\cite{huan99field} who ascribe the different widths to the
inhomogeneous density of the condensate.

There were some apparent discrepancies between our observations
and the local density description (Sec.\ \ref{sec:ldaspinor}). For
instance, as shown in Figs.\ \ref{fig:spinorcurves}a and b, the
separation between the central $m_F = 0$ domain and the
neighboring $m_F = \pm 1$ domains was not sharp, as would have
been predicted by the spin--domain diagram which neglects the
kinetic energy.  Including the kinetic energy requires that the
the condensate spinor vary gradually between the domains.  As
discussed in Secs.\ \ref{sec:metastability} and
\ref{sec:tunneling}, the condensate spinor should change over a
length scale given by $\xi_s = \sqrt{\hbar^2 / 2 m \Delta E}$,
where
\begin{equation}\label{eq:barriersize}
  \Delta E = \mu_0 \left(
\sqrt{\frac{g_0 + g_2}{g_0}} - 1 \right) \simeq 0.018 \mu_0
\end{equation}
is the height of the interaction energy barrier which expels $m_F
= 0$ atoms from $m_F = 1$ spin domains, and $\mu_0 = g_0 n$ is the
chemical potential of the $m_F = 0$ atoms at a condensate density
of $n$. This width $\xi_s$ can be called a ``spin healing length''
in analogy with the healing length $\xi = \sqrt{\hbar^2 / 2 m
\mu}$ which is the minimum length for density variations. For
typical conditions, $\xi_s \simeq 1$ -- $2  \, \mu$m.

In our time--of--flight images (Fig.\ \ref{fig:spinorcurves}), the overlap between the components
appears to be much larger, on the order of tens of microns. This discrepancy may just be an
artifact of the indirect probing technique from which the spin structure of the trapped condensate
is inferred.  During the expansion of the condensate, the kinetic energy at the spin--domain
boundaries is released axially, imparting velocities of $\sqrt{2 \Delta E / m} \simeq 2$ mm/s.
During the 25 ms time of flight, this would cause a sharp boundary between spin components to be
smeared out by $\simeq 50 \, \mu$m, consistent with the observed width of the overlap between the
$m_F = 0$ and $m_F = \pm 1$ components. Thus, our time--of--flight imaging technique cannot
properly characterize the boundary between neighboring spin domains.  In future work, it would be
interesting to examine such boundaries with an {\it in situ} imaging technique, perhaps to observe
the spatial structures recently predicted by Isoshima, Machida and Ohmi \cite{isos99spin}. It is
interesting to note that the radial expansion of the cigar--shaped condensate occurs at a velocity
near the speed of Bogoliubov sound, which describes the propagation of density waves, while the
axial expansion of a spin domain boundary occurs at a ``spin sound velocity'' which would describe
the propagation of spin waves.

\subsection{Miscibility and immiscibility of spinor condensate components}
\label{sec:immisc}

As discussed above, the spin--domain diagram and the observed
ground--state spin structures  showed evidence for the miscibility
of the $m_F=-1$ and $m_F =+1$ components and the immiscibility of
$m_F =\pm 1$ and $m_F =0$ components.  The bulk miscibility or
immiscibility of two--component condensate mixtures is predicted
by mean--field theory \cite{cols78,ho96bin,esry97hf,ao98,gold97}.
The interaction energy density of such condensates is given by
\begin{equation}\label{eq:intenergy2comp}
  E = \frac{1}{2}
(n_{a}^{2} g_{a} + n_{b}^{2} g_{b} + 2 n_{a} n_{b} g_{ab})
\end{equation}
where $m$ is the common atomic mass, and $n_{a}$ and $n_{b}$ are the densities of each of the
components.  The interaction parameters are given generally as $g = 4 \pi \hbar^2 a / m$ where
$a_{a}$ and $a_{b}$ are the same--species scattering lengths, and $a_{ab}$ is the scattering length
for interspecies collisions. Consider a two--component mixture in a box of volume $V$ with $N$
atoms in each component. If the condensates overlap, their total mean--field energy is
\begin{equation}
    E_{O} = \frac{N^{2}}{2 V}
    \left(g_{a} + g_{b} + 2 g_{ab}\right)
\end{equation}
If they phase separate, their energy is
\begin{equation}
    E_{S} = \frac{N^2}{2} \left( \frac{g_{a}}{V_{a}}
     + \frac{g_{b}}{V_{b}} \right)
\end{equation}
The volumes $V_{a}$ and $V_{b}$ occupied by each of the separated
condensates are determined by the condition of equal pressure:
\begin{equation}
    g_{a} \left( \frac{N}{V_{a}}\right)^{2} = g_{b} \left(
    \frac{N}{V_{b}}\right)^{2}
\end{equation}
Comparing the energies $E_{O}$ and $E_{S}$ the condensates will
phase--separate if $g_{ab} > \sqrt{g_{a} g_{b}}$, and will mix if
$g_{ab} < \sqrt{g_{a} g_{b}}$.

In the $F = 1$ three--component spinor system, the scattering lengths are determined by $a_{f = 0}$
and $a_{f = 2}$. Defining $\bar{a} = (2 a_{f=2} + a_{f=0}) / 3$ and $\Delta a = (a_{f=2} - a_{f=0})
/ 3$, the scattering lengths for the $m_{F} = 1 , 0$ two--component system (or equivalently the
$m_{F} = -1, 0$ system) are given by $a_{0} = \bar{a}$, and $a_{1} = a_{0,1} = \bar{a} + \Delta a$.
Since $\Delta a$ is positive for sodium, the condition $a_{0,1} > \sqrt{a_{0}a_{1}}$ applies and
the components should phase-separate, as we have observed~\cite{sten98spin,mies99meta}.
Interestingly, this phase-separation should not occur in the non-condensed cloud because the
same-species mean--field interaction energies are doubled due to exchange terms.

In the $m_{F} = \pm 1$ two component system, the scattering
lengths are $a_{1} = a_{-1} = \bar{a} + \Delta a$ and $a_{1,-1} =
\bar{a} - \Delta a$.  Thus, $a_{1,-1} < \sqrt{a_{1} a_{-1}}$, and
these two components should mix. Indeed, as shown in Fig.\
~\ref{fig:spinorcurves}c, an equilibrium spinor condensate with
$\langle F_{z} \rangle \neq 0$, small field gradient, and
near-zero field consists of an overlapping mixture of atoms in the
$m_{F} = \pm 1$ states. This particular miscible two--component
system has an important advantage. If the trapping potential
varies across a two--component condensate, the lowest energy state
may be a phase--separated state if $a_{a} \neq a_{b}$ even though
the condition $a_{ab} < \sqrt{a_{a} a_{b}}$ is
fulfilled~\cite{pu98two}.  In this case, the atoms with the
smaller scattering length concentrate near the trap center, making
it harder to observe miscibility. However, in the $m_{F} = \pm 1$
system, the two scattering lengths $a_{1}$ and $a_{-1}$ are equal
by rotational symmetry, so the components mix completely even in a
trapping potential.

\subsection{Metastable states of spinor Bose--Einstein condensates}
\label{sec:metastability}

Having observed and explained ground state spin--domain
structures, we began to explore dynamical properties of spinor
Bose--Einstein condensates. Ref.\ \cite{mies99meta} discusses the
observation of long--lived excited states of spinor Bose--Einstein
condensates.  We observed two complementary types of metastable
states: one type in which a two--component condensate was stable
in spin composition but which persisted in a non--equilibrium
structure of spin domains, and another in which a spatially
uniform condensate was metastable with respect to spin relaxation
to an equilibrium spin composition.  In each case, the energy
barriers to relaxation to the ground state (the activation energy)
were identified and found to be much smaller than the thermal
energies of the metastable gases; thus, such states would not be
metastable in a non--condensed cloud.  However, in a Bose
condensed cloud with a large condensate fraction, the thermal
energy is only available to the scarce thermal component and thus
thermal relaxation is considerably slowed.  In other words,
Bose--Einstein condensation allows the study of weak effects in an
energy regime which is much lower than the temperature of the gas.

\subsubsection{Metastable spin--domain structures}
\label{sec:metabubbles}

The first type of metastability was observed in spinor Bose--Einstein condensates in a high
magnetic field (15 G) which caused a two--component $m_F = 0, 1$ cloud to be stable in spin
composition.  This occurs because the large quadratic Zeeman shift makes $\mst{0} + \mst{0}
\rightarrow \mst{+1} + \mst{-1}$ collisions endothermic, even if the $\mst{\pm 1}$ atoms would move
to the ends of the condensate.  We prepared clouds in an equal superposition of the two hyperfine
states using a brief rf pulse, and then allowed the system to equilibrate.

\begin{figure}
  \begin{center}
    \includegraphics[width=4in]{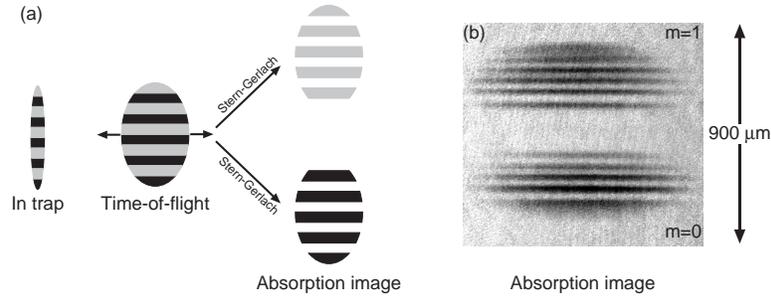}
    \caption[Quantum ``bubbles'': a metastable arrangement of alternating
    spin domains in a two--component condensate]{
Quantum ``bubbles'': a metastable arrangement of alternating spin domains in a two--component
condensate.  (a) A two--component optically--trapped condensate composed of atoms in the $\mst{0}$
(black) and $\mst{1}$ (grey) hyperfine states expands primarily radially once the trap is switched
off.  A magnetic field gradient is used to separate the two spin states before imaging, allowing a
determination of the axial distribution of the two components in the trap.  (b) The observed
density striations indicate that the trapped condensate was composed of alternating $m_F = 0$ and
$m_F = 1$ spin domains.
    \label{fig:bubbleimage}}
  \end{center}
\end{figure}

Because of the immiscibility of the $m_F = 0$ and $m_F = 1$ components, the ground--state spin
structure in this case consists of two phase--separated spin domains, one for each of the
components, on opposite sides of the elongated trap with a domain boundary in the middle.  What we
observed was dramatically different: a spontaneously formed, metastable arrangement of alternating
$m_F = 0$ and $m_F = 1$ spin domains (Fig.\ \ref{fig:bubbleimage}).

\begin{figure}
    \begin{center}
    \includegraphics[height=1.6 in]{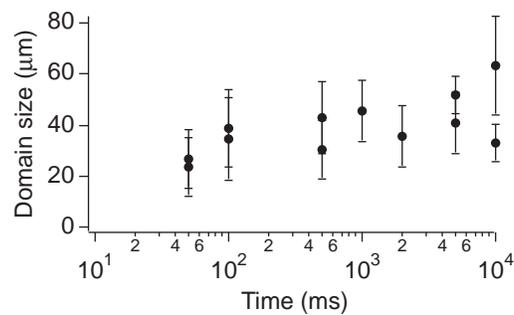}
    \caption[Evolution of metastable domain sizes after state preparation]{
Evolution of metastable domain sizes. Shown is the average size (points) and standard deviation
(one sigma error bars) of the spin domains observed at various time after the initial state
preparation at a bias field of 15 G.  Each point corresponds to measurements made on a different
condensate.
    \label{fig:bubblesize}}
    \end{center}
\end{figure}

These spin striations began forming within about 50 ms of the
initial preparation of two overlapping, immiscible components. The
striations were initially angled due to radial excitations in the
narrow spinor condensates which soon damped out, leaving strictly
horizontal striations. The observed width of the spin domains grew
to an equilibrium value of about 40 $\mu$m within about 100 ms
(Fig.\ \ref{fig:bubblesize}). Thereafter, the clouds were
essentially unchanged, remaining in the metastable state for 10 or
more seconds as the number of trapped atoms slowly decayed due to
three--body trap losses.

\begin{figure}
    \begin{center}
    \includegraphics[height=3in]{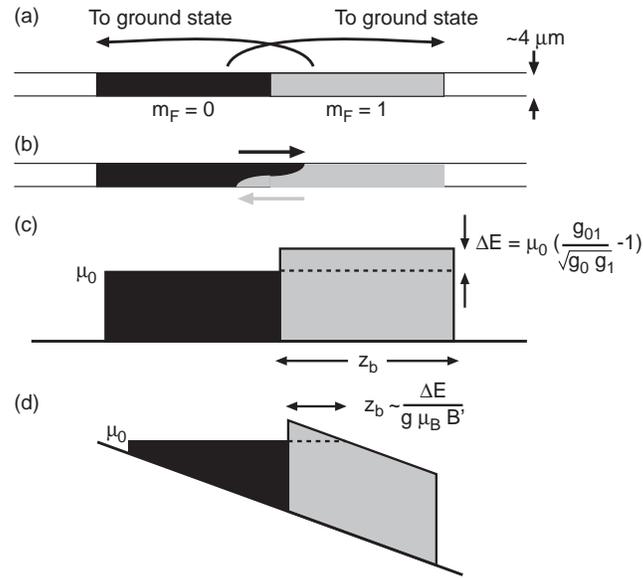}
    \caption[Energy barriers responsible for the metastability of spin domains]{
Energy barriers responsible for the metastability of spin domains in a two--component $m_F = 0, 1$
condensate.  (a) Metastable domains of the $m_F = 0$ (black) and $m_F = 1$ (gray) components are
held in a narrow optical trap.  For the population in these domains to decay to the ground state
they must either pass each other without overlapping, or pass through each other. (b) The former is
prohibited due to the kinetic energy barrier of modifying the condensate wavefunction on a radial
length scale.  (c) The latter is prohibited due to an interaction energy barrier $\Delta E$, shown
here for the passage of an $m_F = 0$ domain of chemical potential $\mu_0$ through the neighboring
$m_F = 1$ domain. (d) If a magnetic field gradient $B'$ is imposed, the width of the classically
forbidden region for the passage of $m_F = 0$ atoms through the $m_F = 1$ domain is reduced to $z_b
= \Delta E / g \mu_B B'$, and the tunneling rate which governs the decay of the metastable domains
is increased.
    \label{fig:explainbubbles}}
    \end{center}
\end{figure}

To understand the reason for the metastability, let us consider two adjacent spin domains of the
$m_F = 0$ and $m_F = 1$ components, as shown in Fig.\ \ref{fig:explainbubbles}. Suppose that, in
order for the condensate to decay to its ground state, the atoms of each spin domain must somehow
be transported across the spin--domain boundary.  For this to occur, the atoms of one component
must either pass around or through the other component.

If one component passes around the other, the condensate wavefunction must be varied spatially in
the radial direction. This gives a kinetic energy barrier to decay of about $\hbar^2 / 2 m r^2
\simeq k_B \times 3$ nK where $r \simeq 2 \, \mu$m is the condensate radius. The passage of one
component through the other is limited by an interaction barrier since the components are
immiscible.  As explained in Ref.\ \cite{stam99tun}, the interaction energy barrier for atoms in
the $\mst{0}$ state to pass through a $m_F = 1$ domain is (see Eq.\ \ref{eq:barriersize}) $\Delta E
\simeq g_2 / 2 g_0 \times \mu_0 = c$ where $\mu_0$ is the chemical potential of atoms in the $m_F =
0$ spin domain, and $c$ is the spin--dependent interaction energy introduced in Sec.\
\ref{sec:rotsym}. At a chemical potential $\mu_0 = k_B \times 300$ nK, the interaction energy
barrier height is $c \simeq k_B \times 5$ nK. The probability of tunneling through the spin domains
is exceedingly small because the axial length of the domains $z_B \simeq 40 \, \mu$m is much larger
than the spin healing length $\xi_s \simeq 1.4 \, \mu$m. Thus, because of these two energy
barriers, the non--equilibrium arrangement of spin--domains is metastable.

The formation of these metastable spin--domains was considered
recently by Chui and Ao \cite{chui99spin} as a spinodal
decomposition in a binary system. Equilibration occurs on two time
scales.  First, on a short time scale determined by the
spin--dependent interaction energy $c$ as $t \simeq \hbar / c$,
the homogeneous initial state begins to phase separate into small
domains of length scale $\xi_s = \sqrt{\hbar^2 / 2 m \Delta E}$.
Thereafter, the domains grow by the coalescence of small spin
domains on the long time scales required for quantum tunneling.
This work supports the physical picture which we suggested in
Ref.\ \cite{mies99meta}.  A similar picture emerges from the work
of Pu {\it et al.} \cite{pu99dyn} who consider the quantum
dynamics (without dissipation) of a trapped condensate composed of
two overlapping components which tend to phase separate.  In their
calculations, they observe the evolution of fine spatial features
which exemplify the instability of such a system against spin
domain formation.

\subsubsection{Metastable spin composition}

\begin{figure}
    \begin{center}
    \includegraphics[height=1.5 in]{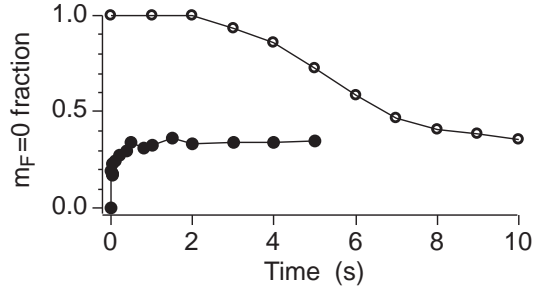}
    \caption{
Metastability of the pure $|m_{F} = 0\rangle$ state in the presence of a magnetic bias field (250
mG), and gradient (44 mG/cm). The evolution toward equilibrium of an initially pure $|m_{F} = 0
\rangle$ condensate (open symbols), and a mixture of $|m_{F} = 1\rangle$ and $|m_{F} = -1 \rangle$
(closed symbols) is shown by plotting the fraction of atoms in the $|m_{F} = 0 \rangle$ state vs.\
dwell time in the optical trap.  Figure taken from Ref.\ \protect \cite{mies99meta}.
    \label{metastable-m=0.eps}}
    \end{center}
\end{figure}

Another type of metastability was discovered in the studies of ground--state spin domains of a
$\langle F_z \rangle = 0$ condensate discussed in Sec.\ \ref{sec:spinorground}.  Condensates were
prepared either with all atoms in the $\mst{0}$ state or in an equal mixture of the $\mst{\pm 1}$
states.  While the ground state reached from either starting condition was the same (Fig.\
\ref{fig:spinor-eq}), equilibration occurred on much different time scales. When starting from the
$|m_{F} = 0\rangle$ state, the condensate remained unchanged for several seconds before evolving
over the next few seconds to the ground state.  When starting from the $|m_{F} = \pm 1\rangle$
superposition, the fraction of atoms in the $|m_{F} = 0\rangle$ state grew without delay, arriving
at equilibrium within less than a second (Fig.\ \ref{metastable-m=0.eps}).

This difference can be understood by considering a spin-relaxation
collision, in which two \mst{0} atoms collide to produce a \mst{1}
and a \mst{-1} atom. In the presence of a magnetic field $B_{0}$,
quadratic Zeeman shifts cause the energy of the two \mst{0} atoms
to be lower than that of the \mst{1} and \mst{-1} atoms.  Due to
this activation energy, condensate atoms in the  \mst{0} state
could not undergo spin-relaxation collisions. Thus, even though
the creation of $m_F = 1$ and $m_F = -1$ spin domains at the ends
of the condensate is energetically favored {\it globally} in the
presence of a magnetic field gradient, the $m_F = 0$ condensate
cannot overcome the {\it local} energy barrier for
spin-relaxation. In contrast, condensate atoms in the  \mst{1} and
\mst{-1} states can directly lower their energy through such
collisions, and equilibrate quickly.

In support of this explanation, the metastability time was found to depend strongly on the
quadratic Zeeman energy which was varied by changing the magnetic bias field.  Importantly, the
equilibration time changed significantly when $q$ was varied by less than a nanokelvin; this
dependence excludes thermal spin relaxation (in a gas at a temperature $T \sim 100$ nK $\gg q /
k_B$) as the equilibration mechanism, and suggests that the metastable condensate decays to the
ground state via quantum tunneling.  In such tunneling, a pair of atoms in the $\mst{\pm 1}$ states
would be produced in a classically forbidden collision, and would then tunnel to opposite ends of
the cloud where their energy is lowered due to the magnetic field gradient. It would be interesting
to study this process further in the future.

\subsection{Quantum tunneling}
\label{sec:tunneling}

Metastable states can generally overcome the activation energy barrier for decay to the ground
state in two ways.  Classically, the system can decay by acquiring thermal energy larger than the
activation energy.  However, even without this thermal energy, the system can decay to the ground
state by quantum tunneling. Tunneling describes a wide range of phenomena such as nuclear decay,
field ionization of neutral atoms, and scanning tunneling microscopy.  In macroscopic quantum
systems, coherent tunneling can lead to a variety of Josephson effects \cite{jose62} which have
been observed in superconductors and quantum fluids.  As exemplified by the observation of
metastable states which persist in spite of temperatures higher than the activation energy, gaseous
Bose--Einstein condensates are an appealing new system to study tunneling and Josephson
oscillations \cite{java86,smer97,zapa98,ande98atla}.

In our study of optically trapped spinor Bose--Einstein condensates, we looked at the decay of
metastable spin domains via quantum tunneling through the spin domain boundaries \cite{stam99tun}.
Tunneling barriers were formed not by an external potential, but rather by the intrinsic repulsion
between two immiscible components of a quantum fluid. Tunneling across spin domain boundaries is a
spin transport mechanism inherent to such a fluid (emphasized in \cite{chui99spin}), and the
tunneling rates are sensitive probes of the structure of the domain boundaries.  From a practical
viewpoint, the use of phase--separated spin domains rather than externally imposed potentials is an
attractive option for future studies of tunneling in Bose condensates since the energy barriers for
tunneling are naturally of nanokelvin--scale height and micron--scale width.

\begin{figure}
  \begin{center}
    \includegraphics[height=2.6 in]{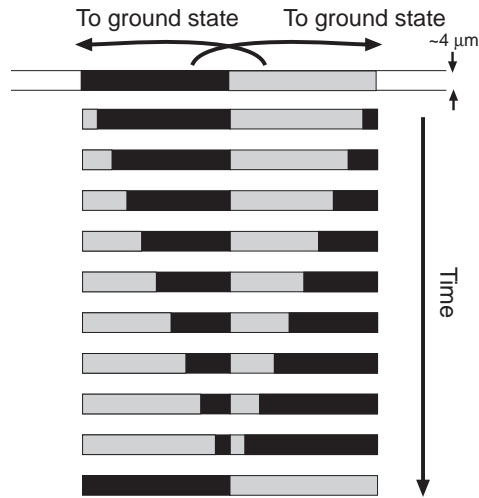}
    \caption[Evolution from a metastable state to the ground state]{
Evolution from a metastable state to the ground state.  The ground state domain structure is formed
by tunneling of the two components through each other.  The intermediate state is characterized by
outer ground state domains and inner metastable domains.
    \label{fig:tunnelevolution}}
  \end{center}
\end{figure}

The system we chose for our study was a simple, well--characterized metastable arrangement of spin
domains in a spinor condensate composed of the $m_F = 1$ and $m_F = 0$ components.  Such a state
was obtained by first preparing a spinor condensate in a superposition of the $\mst{0, 1}$ states.
A strong field gradient (several G/cm) was applied to break up the many--domain metastable state
(discussed in the previous section) and separate the spin components into the two--domain ground
state.  Then, a weak gradient was applied in the opposite direction, which energetically favored
the rearrangement of the spin domains on opposite ends of the optical trap, and thus yielded a
two--domain metastable state. This simple system allowed for the characterization of the decay of
metastable states by the easy identification of atoms in the metastable and ground--state spin
domains of each component (Fig.\ \ref{fig:tunnelevolution}).

\begin{figure}
  \begin{center}
    \includegraphics[width=3.5 in]{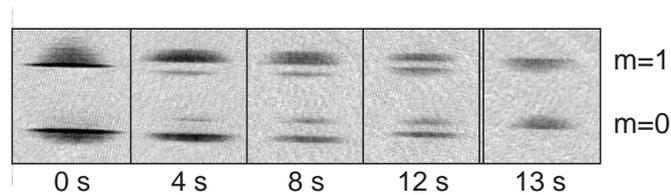}
    \caption[Decay of a metastable state by thermal relaxation and
    quantum tunneling]{
Decay of a metastable state by thermal relaxation and quantum tunneling.  A two--domain metastable
state was prepared at a 15 G axial bias field and a 0.1 G/cm gradient. Images show metastable
(outer parts in the time-of-flight picture) and ground state (inner parts) spin domains of $|m_{F}
= 1\rangle$ and $|m_{F} = 0\rangle$ atoms, probed at various times after state preparation. The
height of each image is 1.3 mm.  The number of atoms in the ground state domains grew slowly over
the first 12 s, after which the atoms tunneled quickly to the ground state. The total number of
atoms (and thus the condensate density) decreased during the dwell time due to inelastic
three--body collisions. Thus, the slow thermal relaxation gave way to a rapid tunneling once the
density fell below a threshold value.  Figure taken from Ref.\ \protect \cite{mies99meta}.
    \label{fig:creep}}
  \end{center}
\end{figure}

To ascribe the decay of these metastable spin domains to quantum
tunneling, it was first necessary to rule out thermal relaxation
as a decay mechanism. Figure \ref{fig:creep} shows a series of
time--of--flight images reflecting the state of the two--domain
metastable state at various times after the state was initially
prepared and held under a 0.1 G/cm gradient.  One can identify two
stages in the decay of this metastable cloud to the ground state:
a slow decay over the first 12 seconds, followed by a rapid decay
to the ground state within less than one second. The slow decay
was found to be rather insensitive to changes in the condensate
density and field gradient, and was thus consistent with thermal
relaxation wherein condensate atoms from the metastable state are
thermally excited, and then re-condense into the ground state
domains.  The number of atoms which accumulated in the
ground--state domains reached a nearly constant value of about $5
\times 10^4$, perhaps due to a dynamic equilibrium between the
growth of the domain via re--condensation and its depletion via
inelastic losses.  We also observed that the total population in
the $\mst{0}$ spin state decreases more rapidly than that in the
$\mst{1}$ state, indicating higher inelastic collision rates for
the $\mst{0}$ state.

The rapid decay which ensued (after 12 seconds) was due to quantum tunneling. As discussed below,
the tunneling rate is acutely sensitive to the condensate density.  While the condensate is stored
in the optical trap, its density decreases as atoms are lost from the trap due to three--body
inelastic collisions.  Thus, as time progressed, the density of the metastable condensate in Fig.\
\ref{fig:creep} decreased until the tunneling rate was fast enough to cause a rapid (within about 1
second) relaxation to the ground state.

\begin{figure}
  \begin{center}
    \includegraphics[height=2.2 in]{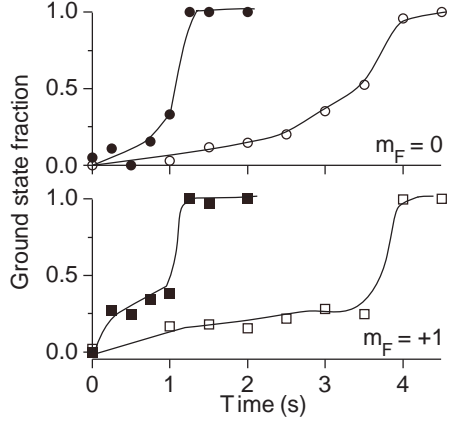}
    \caption[A comparison of the decay of metastable states at different densities]{
A comparison of the decay of metastable states at different densities.  Shown is the fraction of
$m_F = 0$ and $m_F = 1$ atoms found in the ground--state domains after a variable delay time at a
constant field gradient of 0.35 G/cm and a 2 G bias field.  The evolution is shown for condensates
prepared with different populations in the initial state, either $1 \times 10^6$  (open symbols) or
$2.5 \times 10^5$ atoms (closed symbols). The rapid tunneling in the denser condensate occurred at
a later time.  The solid lines are guides to the eye.
    \label{fig:tunnelseries}}
  \end{center}
\end{figure}

The dependence of the relaxation time on the condensate density is also shown in Fig.\
\ref{fig:tunnelseries}. Metastable condensates were prepared with different initial numbers of
atoms, and held in a constant magnetic field gradient. The metastable lifetimes for the two
starting conditions were different, with the denser condensate decaying to the ground state at a
later time.

A mean--field description of the tunneling rates was developed \cite{stam99tun}. We considered the
one--dimensional motion of a Bose--Einstein condensate composed of two immiscible components of
atoms in states $|a\rangle$ and $|b\rangle$ and atomic mass $m$, as in Fig.\
\ref{fig:explainbubbles} for the experimental realization of the $m_F = 1$ and $m_F = 0$
components.  The chemical potentials of the two components $\mu_i = g_i n_i$ are related by the
condition of constant pressure $\mu_a^2 / g_a = \mu_b^2 / g_b$ where $g_i = 4 \pi \hbar^2 a_i / m$,
$a_i$ is the same--species scattering length, and $i \in {a,b}$ labels the component.  The
mean--field interaction energy for each component is $U_i = g_i n_i + g_{ab} n_j$ ($i \neq j$)
where $g_{ab}$ is determined by the cross--species scattering length $a_{ab}$.

If we assume the boundary between the spin domains to be sharp, component $a$ is excluded from the
domain of component $b$ by an energy barrier $\Delta E$ given by
\begin{equation}\label{eq:energybarrier}
  \Delta E = g_{ab} n_b - g_a n_a = \left(
  \frac{g_{ab}}{\sqrt{g_a g_b}} - 1 \right) \mu_a
\end{equation}
This energy barrier is responsible for the metastability of spin domains which have an axial length
$z_b$ which is much larger than the spin healing length $\xi_s = \sqrt{\hbar^2 / 2 m\Delta E}$.

In the presence of a state--selective force $F$, which was experimentally applied in our
experiments using a magnetic field gradient $B'$, the energy barrier decreases away from the domain
boundary as $\Delta E(z) = \Delta E - F z$. Thus the width of the energy barrier is reduced to $z_b
= \Delta E / F$, and tunneling can occur when $z_b \sim \xi_s$. More precisely, the tunneling rate
across the barrier is given by the Fowler--Nordheim (WKB) equation
\begin{eqnarray}\label{eq:fowlnord}
    \frac{d N_a}{d t} & = &  \gamma \exp\left( - 2 \sqrt{\frac{2
    m}{\hbar^2}} \int_0^{z_b} \sqrt{\Delta E(z)} dz \right) \\
    & = & \gamma \exp\left( -\frac{4}{3} \sqrt{\frac{2 m}{\hbar^2}}
    \frac{\Delta E^{3/2}}{F}\right) \\
    & = & \gamma \exp \left( - \frac{4}{3}
    \frac{z_b}{\xi_s}\right)
\end{eqnarray}
which also describes the analogous phenomenon of the field emission of electrons from cold metals
\cite{fowl28}.  Here $\gamma$ is the total tunneling attempt rate (i.e.\ not the rate per
particle), and the exponential is the tunneling probability. This relation explains the strong
dependence of the tunneling rate on the condensate density $n$, which causes the density threshold
behaviour observed in Fig.\ \ref{fig:creep}.

\begin{figure}
    \begin{center}
    \includegraphics[width=4.5 in]{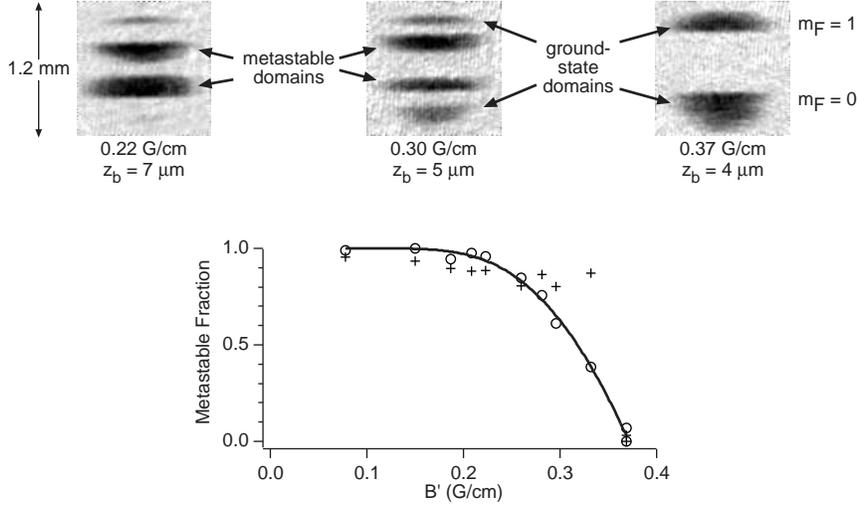}
    \caption[Tunneling across barriers of constant height and variable width]{
Tunneling across barriers of constant height and variable width $z_b$. Condensates at constant
density were probed after 2 seconds of tunneling at a variable field gradient $B'$.  The population
of atoms in the metastable and ground--state domains of each spin state were easily identified in
time--of--flight images (see Fig.\ \ref{fig:tunnelevolution}).  Also shown is the fraction of atoms
of each spin state in their metastable domain. Circles represent the $m_{F} = 0$ atoms, and pluses
the $m_{F} = 1$ atoms. The energy barrier height was $\approx 5$ nK at the chemical potential of
$\mu_0 \simeq 300$ nK. The tunneling rate depended strongly on the width of the energy barrier $z_b
= \Delta E / g \mu_B B'$: at $z_b = 7 \, \mu$m ($B' = 0.22$ G/cm) little tunneling was observed,
while at $z_b = 4 \, \mu$m ($B' = 0.37$ G/cm) the atoms had completely tunneled to the ground state
in 2 s.  The barrier attempt rate and tunneling probability were determined by a fit to the $m_{F}
= 0$ data (solid line), and found to agree with a mean--field model. The data indicate that the
tunneling rate for $m_F=0$ atoms is larger than that for $m_F=1$ atoms.
    \label{fig:metafrac}}
    \end{center}
\end{figure}

This mean--field description was tested experimentally by measuring the tunneling rate across
barriers of constant height and variable width.  For this, metastable condensates were prepared at
a constant density (giving a constant barrier height of $\approx 5$ nK), exposed to a variable
gradient (giving barrier widths between 4 and 20 $\mu$m), and allowed to decay for a period $\tau =
2$ s which was short enough that the condensate density did not vary appreciably due to trap
losses.  We then measured the number of atoms of each component in the metastable and ground--state
spin domains by time--of--flight absorption imaging (Fig.\ \ref{fig:metafrac}).  When the barrier
was wide (small field gradients), the tunneling rate was small and only a small fraction of atoms
were observed in the ground--state domains. When the barrier was narrow (large field gradients),
the tunneling rate was large and a large fraction of atoms were observed in the ground--state
domains.  The data were in quantitative agreement with our mean--field approach using the
scattering lengths calculated by Burke {\it et al.}\ \cite{burk98}.

\subsection{Magnetic field dependence of spin--domain boundaries}
\label{sec:fielddep}

\begin{figure}
    \begin{center}
    \includegraphics[height=1.8 in]{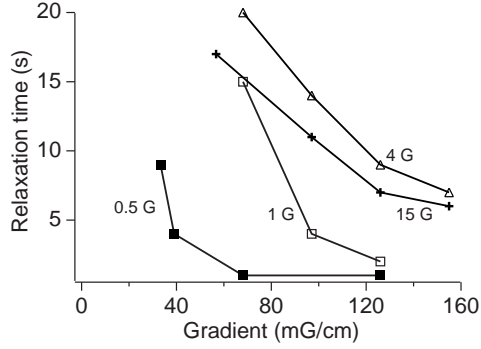}
    \caption[Variation of the metastable state lifetime with magnetic field]{
Variation of the metastable state lifetime with magnetic field and field gradient.  Metastable
condensates at a constant initial chemical potential of $\mu_0 \simeq 600$ nK were allowed to decay
at a variable bias field and field gradient.  Shown is the time at which the condensates were
observed to have relaxed completely to the ground state.  The dependence of the lifetime on the
field gradient at 15 G and 4 G was similar.  At fields of 1 G or lower, the lifetime was
dramatically shortened, indicating a substantially higher tunneling rate at low magnetic fields.
    \label{fig:timevsfield}}
    \end{center}
\end{figure}

While the tunneling rates measured at high magnetic fields (15 G)
agreed with a model of tunneling in a two--component Bose
condensate, at lower magnetic fields, a dramatic increase in the
tunneling rate was observed indicating the breakdown of the
two--component description. As described in our paper (see Fig.\ 4
of Ref.\ \cite{stam99tun}), the threshold chemical potential for
tunneling at a constant field gradient dramatically increased at
magnetic fields below about 1 G. Fig.\ \ref{fig:timevsfield} shows
similar data.  This strong increase in the tunneling rate at low
fields reveals that the structure of the domain boundary is
changed by the presence of the third spin component, the
$\mst{-1}$ state.

The introduction of the third spin component to the barrier has
the effect of reducing the energy barrier for the tunneling of
$m_F = 0$ atoms.  This can be seen by considering a mixture of
components $a$ and $b$ where $| a \rangle = \mst{0}$ and $|b
\rangle = \cos \theta \mst{1} - \sin \theta \mst{-1}$ ($0 \leq
\theta \leq \pi / 2$).  The interaction energy density of this
system can be written in the form of Eq.\ \ref{eq:intenergy2comp}
with the definitions
\begin{eqnarray}
  g_a & = & g_0 \\
  g_b & = & g_0 + \Delta g \cos^2 2 \theta \\
  g_{ab} & = & g_0 + \Delta g (1 - \sin 2 \theta)
\end{eqnarray}
where $\Delta g = g_2$.  Using these interaction parameters, the
energy barrier height which governs the tunneling of the $m_F = 0$
component becomes
\begin{equation}\label{eq:deltaetheta}
  \Delta E(\theta) = \left( \frac{g_0 + \Delta g (1 - \sin 2
  \theta)}{\sqrt{g_0 ( g_0 + \Delta g \cos^2 2 \theta)}} - 1
  \right) \mu_0 \simeq \frac{\Delta g}{g_0} \left(1 - \sin 2
  \theta - \frac{\cos^2 2 \theta}{2} \right) \mu_0
\end{equation}
for $\Delta g \ll g_0$.  For $\theta = 0$ or $\theta = \pi / 2$,
one recovers the energy barrier for tunneling through a pure $m_F
= 1$ or $m_F = -1$ barrier. For intermediate values, the height of
the energy barrier is reduced. Indeed, at $\theta = \pi / 4$, the
energy barrier disappears completely.  In this case, $|b \rangle =
\frac{1}{\sqrt{2}} (\mst{1} - \mst{-1})$ describes a polar state
obtained by rotating the $\mst{0}$ state.  The condensate spinor
can thus evolve continuously from state $|a\rangle$ to $|b
\rangle$ by a gradual rotation which is associated with a
gap--less (i.e.\ no barrier) Goldstone excitation mode.  Recent
calculations by Isoshima, Machida and Ohmi \cite{isos99spin}
confirm the disappearance of the energy barrier when the
populations in the $\mst{\pm 1}$ states become equal.    These
changes of effective interaction parameters in the presence of the
$\mst{-1}$ state provide a mechanism for the variation of the
tunneling rates with magnetic fields.

The local density approximation discussed in Sec.\
\ref{sec:ldaspinor} indicates that, indeed, the $m_F = -1$
component should be introduced to the spin domain boundaries at
low magnetic fields. According to this approximation, the
boundaries studied in our experiment between an $m_F = 0$ domain
and a region of $\langle F_z \rangle > 0$ are described in the
spin domain diagram (Fig.\ \ref{fig:abcdiagrams}) by a short
vertical line at constant $q$ (determined by the magnetic field)
which straddles the border of the $m_F = 0$ region in the $p > 0$
half--plane. The diagram thus suggests that a pure two--component
description of the boundary is correct for magnetic fields at
which the quadratic Zeeman energy is larger than the
spin--dependent interaction energy ($q > c$). Taking a typical
value of the chemical potential of $\mu_0 = 300$ nK, this
condition indicates that the two--component description of
tunneling should be valid for magnetic fields greater than 500
mG\footnote{Unlike in Ref.\ \cite{sten98spin}, we do not apply
here a one--dimensional approximation to reduce the effective
interaction energy by averaging over the radial directions.}.
Below this magnetic field, the $m_F = 0$ component is bordered not
by a pure $m_F = 1$ domain, but rather by a region in which the
$\mst{1}$ and $\mst{-1}$ states are mixed.

However, the local density approach fails to explain why the
tunneling rates should increase even at magnetic fields for which
$q > c$, as we observed. This failure stems from the incomplete
description of the spin domain boundary due to the neglect of
kinetic energy in the local density approximation.  As discussed
above, $\langle F_z \rangle$ must vary continuously  at the
boundary between spin domains.  One may approximate the spin
composition of the boundary region by explicitly minimizing the
energy of a spinor with a given value of $\langle F_z \rangle$,
and then applying a local density approximation in which we assume
that the spinor at each location in the boundary is determined by
the local value of $\langle F_z \rangle$.  We can write a spinor
for which $\langle F_z \rangle > 0$ as
\begin{equation}\label{eq:spinorfz}
  \vec{\zeta} = \left( \begin{array}{c}
  \sqrt{\afz + \epsilon^2} \\
  \sqrt{1 - \afz - 2 \epsilon^2} \\
  -\epsilon
  \end{array} \right)
\end{equation}
where a choice of complex phases has already been made which reduces the anti--ferromagnetic
interaction energy.  The spin--dependent energy $K\subrm{spin}$ is then
\begin{equation}\label{eq:ksptominimize}
  c \left( \afz^2 + 2 (1 - 2 \epsilon^2 - \afz)
  (\sqrt{\epsilon^2 + \afz} - \epsilon)^2 \right) + q (2 \epsilon^2 + \afz)
\end{equation}
which is minimized to determine the fractional population
$\epsilon^2$ in the $\mst{-1}$ state.

At high magnetic fields ($q \gg c$) one finds the approximate
solution
\begin{equation}\label{eq:zeroathighb}
  \epsilon^2 \simeq \frac{1}{(q/c)^2} \, \afz (1 - \afz)^2
\end{equation}
which indicates that atoms in the $\mst{-1}$ state are {\it always} energetically favored to reside
in the domain boundary\footnote{In our paper, an incorrect solution was given which indicated a
high--field scaling of $\epsilon^2 \propto B_0^{-2}$ \cite{stam99tun}.}.  Their population scales
with the magnetic field $B_0$ as $q^{-2} \propto B_0^{-4}$.  At a magnetic field of 15 G where $q /
c \approx 400$, the fraction of atoms in the $\mst{-1}$ state is exceedingly small and thus the
two--component approximation to the domain boundary should be quite accurate, as indicated by our
data.  At lower magnetic fields, the population in the $\mst{-1}$ state plays an increasingly
important role.

\begin{figure}
    \begin{center}
    \includegraphics[height=5in]{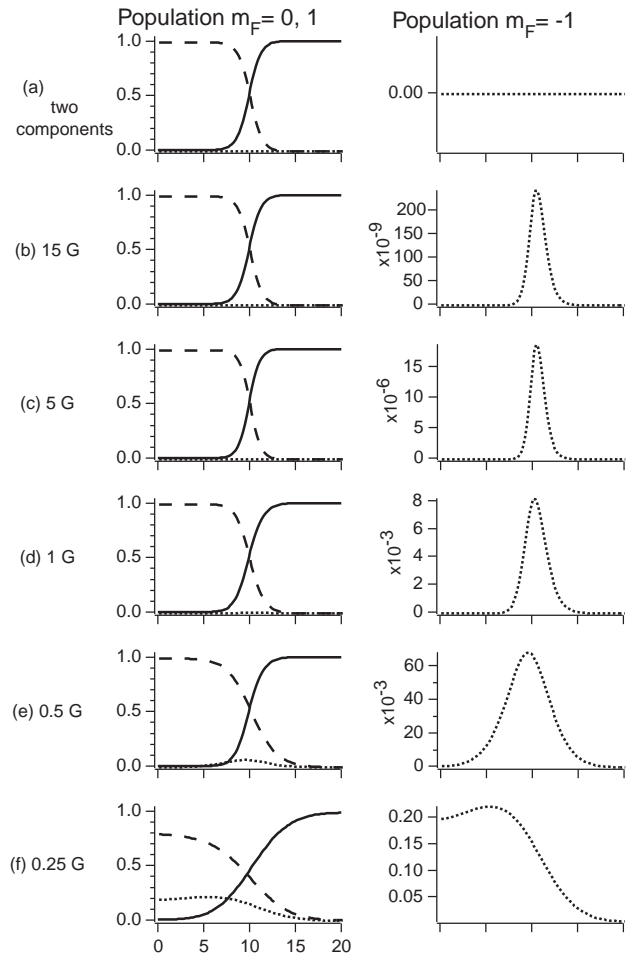}
    \caption[Numerical calculation of the spin structure at the domain boundaries at
    different magnetic fields]{
Numerical calculation of the spin structure at the domain boundaries at different magnetic fields.
Shown are the fractional population in the $\mst{1}$ (dashed line), $\mst{0}$ (solid line), and
$\mst{-1}$ (dotted line) states, with the $\mst{-1}$ population shown at right on an expanded
scale.  In (a), the calculation was restricted to just the $m_F = 0$ and $m_F = 1$ components, and
hence the results are magnetic field independent.  In the remaining graphs, all three components
were considered at magnetic fields of (b) 15 G, (c) 5 G, (d) 1 G, (e) 0.5 G, and (f) 0.25 G.  As
the magnetic field is decreased, the population in the $\mst{-1}$ state increases, and the boundary
region becomes wider.  Further details are found in Ref.\ \cite{stam99thesis}.
    \label{fig:allboundariesw}}
    \end{center}
\end{figure}

\begin{figure}
    \begin{center}
    \includegraphics[height=2in]{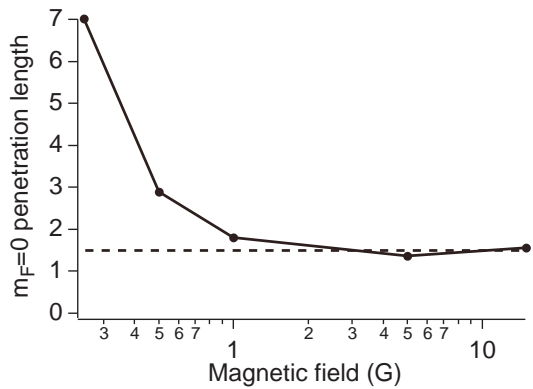}
    \caption[Variation of the spin--domain boundary width with magnetic field]{
Variation of the spin--domain boundary width with magnetic field. The calculated spatial variation
of the fractional population of atoms in the $\mst{0}$ state was fitted to the function $f(z) = [1
- \mbox{Erf}\left( (z - z_0) / \xi_0\right)] / 2$, where $\xi_0$ characterizes the penetration
length of the $m_F = 0$ component in the boundary region.  The points correspond to graphs (b -- f)
of Fig.\ \ref{fig:allboundariesw}, while the dashed line shows the penetration length calculated
for the two--component case. The penetration length is close to the two--component limit at high
magnetic fields, and then increases sharply at lower fields beginning at about 1 G, in agreement
with our experimental observation of an increased tunneling rate below this magnetic field value.
    \label{fig:bwidth}}
    \end{center}
\end{figure}

Guided by these simple approximate treatment, one can also explicitly calculate the spinor
wavefunction at the domain boundary which minimizes the total energy.  Such calculations were
presented in Ref.\ \cite{stam99thesis}, and the results are summarized in Figs.\
\ref{fig:allboundariesw} and \ref{fig:bwidth}.  One finds indeed that the fractional population of
atoms in the $\mst{-1}$ state is non--zero at all magnetic fields and scales as $B_0^{-4}$ at high
magnetic fields. The introduction of $m_F = -1$ atoms in the boundary layer increases the
penetration depth of $m_F = 0$ atoms in the boundary region, and thus should increase the rate of
tunneling across the domain boundary.  The magnetic field at which this effect becomes significant,
according to these calculations, is about 1 G for our experimental conditions and agrees with the
magnetic field value below which an increased tunneling rate was observed.

\section*{Conclusions}

These notes have reviewed advances in atomic BEC in two areas, spinor condensates and light
scattering.  Many further advances are described in other contributions to this volume.  This rapid
pace of developments during the last few years has taken the community by surprise. After decades
of an elusive search nobody expected that condensates would be so robust and relatively easy to
manipulate. Also, nobody imagined that such a simple system would pose so many challenges, not only
to experimentalists, but also to our fundamental understanding of physics. The list of future
challenges is long and includes the complete characterization of elastic and inelastic collisions
at ultralow temperatures, the exploration of superfluidity, vortices, and second sound in Bose
gases, the study of quantum-degenerate molecules and Fermi gases, the development of practical
``high-power'' atom lasers, and their application in atom optics and precision measurements.

\section*{Acknowledgments}
We are grateful to Chris Westbrook and Robin Kaiser for organizing a stimulating summer school, to
Shin Inouye for valuable discussions, to Chandra Raman and Ananth Chikkatur for comments on the
manuscript. The BEC work at MIT was done in collaboration with M.R. Andrews, A.P. Chikkatur, K.B.
Davis, D.S. Durfee, A. Görlitz, S. Gupta, Z. Hadzibabic, S. Inouye, M. Köhl, C.E. Kuklewicz, M.-O.
Mewes, H.-J. Miesner, R. Onofrio, T. Pfau, D.E. Pritchard, C. Raman, D.M. Stamper-Kurn, J. Stenger,
C.G. Townsend, N.J. van Druten, and J. Vogels.   This work was supported by NSF, ONR, JSEP, ARO,
NASA, and the David and Lucile Packard Foundation. D.M.S.-K.\ acknowledges additional support from
JSEP, and is presently supported by a Millikan Prize Postdoctoral Fellowship.


\end{document}